\documentclass[aps,nofootinbib,preprintnumbers,twocolumn,superscriptaddress,prd]{revtex4-1}

\usepackage{amsmath}
\usepackage{amssymb}
\usepackage{graphics}
\usepackage{epsfig}
\usepackage{bbm,bm}
\usepackage{psfrag}
\usepackage{hyperref}
\usepackage{color}
\usepackage{comment}
\usepackage{amsfonts}
\usepackage{dsfont}
\usepackage{booktabs}
\usepackage[markup=nocolor]{changes}
\usepackage{slashed}
\usepackage{cleveref}

\graphicspath{{Pictures/}}
\usepackage[utf8]{inputenc}

\usepackage[english]{babel}

\newcommand{\gs}{g_\mathrm{s}}
\newcommand{\gresc}{\hat{g}}
\newcommand{\bargs}{\bar{g}_\mathrm{s}}

\newcommand{\I}{\mathrm{i}}
\newcommand{\E}{\mathrm{e}}
\newcommand{\Nc}{N_{\mathrm{c}}}
\newcommand{\NL}{N_{\mathrm{L}}}

\newcommand{\NfL}{N_{\mathrm{f}}^{\mathrm{L}}}
\newcommand{\Nfc}{N_{\mathrm{f}}^{\mathrm{c}}}
\newcommand{\dgammaL}{d_\gamma^{\mathrm{L}}}
\newcommand{\dgammac}{d_\gamma^{\mathrm{c}}}
\newcommand{\SU}{\mathrm{SU}}
\newcommand{\ABL}{\mathcal{A}_\theta}
\newcommand{\ABR}{\mathcal{A}_\mathrm{H}}
\newcommand{\AW}{\mathcal{A}_\mathrm{W}}
\newcommand{\AF}{\mathcal{A}_\mathrm{F}}

\newcommand{\omegaBl}{\omega_\theta}
\newcommand{\omegaBr}{\omega_\mathrm{H}}
\newcommand{\omegaF}{\omega_\mathrm{F}}

\newcommand{\omegafBl}{z_\theta}
\newcommand{\omegafBr}{z_\mathrm{H}}
\newcommand{\omegafF}{z_{\mathrm{F}}}
\newcommand{\omegafW}{z_{\mathrm{W}}}

\newcommand{\omegaW}{\omega_{\mathrm{W}}}
\newcommand{\BG}{\mathcal{B}_\mathrm{W}}
\newcommand{\BF}{\mathcal{B}_\mathrm{F}}

\newcommand{\psiL}{\psi_{\mathrm{L}}}
\newcommand{\psiR}{\psi_{\mathrm{R}}}
\newcommand{\barpsiL}{\bar{\psi}_{\mathrm{L}}}
\newcommand{\barpsiR}{\bar{\psi}_{\mathrm{R}}}

\newcommand{\Nsc}{N_{\mathrm{sc}}}
\newcommand{\ZL}{Z_{\mathrm{L}}}
\newcommand{\ZR}{Z_{\mathrm{R}}}
\newcommand{\etaL}{\eta_{\mathrm{L}}}
\newcommand{\etaR}{\eta_{\mathrm{R}}}
\newcommand{\etaW}{\eta_{\mathrm{W}}}
\newcommand{\etaG}{\eta_{\mathrm{G}}}
\newcommand{\hatetaW}{\hat{\eta}_{\mathrm{W}}}
\newcommand{\hatetaG}{\hat{\eta}_{\mathrm{G}}}
\newcommand{\hatetaphi}{\hat{\eta}_{\phi}}
\newcommand{\ZW}{Z_{\mathrm{W}}}
\newcommand{\ZG}{Z_{\mathrm{G}}}
\newcommand{\trho}{\tilde{\rho}}
\newcommand{\MSbar}{\overline{\text{MS}}}

\newcommand{\be}{\begin{eqnarray}}
\newcommand{\de}{\partial}
\newcommand{\ee}{\end{eqnarray}}
\newcommand{\Eqref}[1]{Eq.~\eqref{#1}}
\newcommand{\Figref}[1]{Fig.~\ref{#1}}
\newcommand{\Secref}[1]{Sec.~\ref{#1}}
\newcommand{\Appref}[1]{App.~\ref{#1}}

  {\left\lbrace\begin{array}{@{}l@{}}}%
  {\end{array}\right.}

\newcommand{\widesim}[2][1.5]{
	\mathrel{\underset{#2}{\scalebox{#1}[1]{$\sim$}}}
}

\newcommand{\kap}{\hat{\kappa}}

\newcommand{\hresc}{\hat{h}}
\newcommand{\hrescs}{\hat{h}}
\newcommand{\hrescg}{\breve{h}}
\newcommand{\kapg}{\breve{k}}
\newcommand{\lrescg}{\breve{\lambda}}
\newcommand{\chis}{\chi_\mathrm{s}}
\newcommand{\chig}{\chi_g}
\newcommand{\lresc}{\hat{\lambda}}

\newcommand{\HTQCD}{$\mathbbm{Z}_2$-Yukawa-QCD}
\newcommand{\NAH}{non-Abelian Higgs}
\newcommand{\SULSUc}{$\SU(2)_\mathrm{L}\times\SU(3)_\mathrm{c}$}
\newcommand{\as}{a_\mathrm{s}}

\definecolor{ao(english)}{rgb}{0.0, 0.5, 0.0}
\definecolor{hg}{rgb}{0.8, 0, 0.5}

\begin{document}

\title{Scheme dependence of asymptotically free solutions}


\author{Holger Gies}
\email{holger.gies@uni-jena.de}
\affiliation{\mbox{\it Theoretisch-Physikalisches Institut, Friedrich-Schiller-Universit{\"a}t Jena,}
	\mbox{\it D-07743 Jena, Germany}}
\affiliation{\mbox{\it Abbe Center of Photonics, Friedrich-Schiller-Universit{\"a}t Jena,}
	\mbox{\it D-07743 Jena, Germany}}
\affiliation{Helmholtz-Institut Jena, Fr\"obelstieg 3, D-07743 Jena, Germany}

\author{Ren\'e Sondenheimer}
\email{rene.sondenheimer@uni-graz.at}
\affiliation{\mbox{\it Theoretisch-Physikalisches Institut, Friedrich-Schiller-Universit{\"a}t Jena,}
	\mbox{\it D-07743 Jena, Germany}}
\affiliation{\mbox{\it Institute of Physics, NAWI Graz, University of Graz, Universit\"atsplatz 5, A-8010 Graz, Austria}}

\author{Alessandro Ugolotti}
\email{alessandro.ugolotti@uni-jena.de}
\affiliation{\mbox{\it Theoretisch-Physikalisches Institut, Friedrich-Schiller-Universit{\"a}t Jena,}
	\mbox{\it D-07743 Jena, Germany}}

\author{Luca Zambelli}
\email{luca.zambelli@uni-jena.de}
\affiliation{\mbox{\it Theoretisch-Physikalisches Institut, Friedrich-Schiller-Universit{\"a}t Jena,}
	\mbox{\it D-07743 Jena, Germany}}



\begin{abstract}
Recent studies have provided evidence for the existence of new
asymptotically free trajectories in non-Abelian particle models without
asymptotic symmetry in the high-energy limit. 
We extend these results to a general 
$\SU(\NL)\times\SU(\Nc)$ Higgs-Yukawa model
that includes the non-Abelian sector of the standard model, 
finding further confirmation for such scenarios 
for a wide class of regularizations that account for
threshold behavior persisting to highest energies. 
We construct
these asymptotically free trajectories within conventional
$\overline{\text{MS}}$ schemes and systematic weak coupling expansions. 
The existence of these solutions
is argued to be a scheme-independent phenomenon, as
demonstrated for 
mass-dependent schemes based on general momentum-space
infrared regularizations. 
A change of scheme induces a map of the
theory's coupling space onto itself,
which in the present case also translates into a reparametrization 
of the space of asymptotically free solutions.

\end{abstract}

\maketitle

\section{Introduction}
\label{sec:intro}

Universality in physics characterizes the fact that long-range
effective properties of a system can be largely independent of its
microscopic details. In particle physics, where microscopic details,
say at the Planck scale, are neither known nor currently
experimentally accessible, universality is often quantified in terms
of observables which should be independent of the choice of the
regularization and the renormalization scheme.

On a technical level, universality can also be visible in properties
of renormalization group (RG) functions such as $\beta$ functions
specifying the behavior of couplings under a change of scale. A
standard textbook result is the scheme independence of the
perturbative one-loop $\beta$ function coefficient; in a
mass-independent scheme, also the two-loop coefficient is
universal. These results form the basis of classifying theories
according to their weak-coupling behavior with a prominent example
being asymptotic freedom (AF) towards high energies for the case of a
negative one-loop coefficient \cite{Gross:1973id,Politzer:1973fx,Gross:1973ju,Cheng:1973nv,Gross:1974cs,Politzer:1974fr,Chang:1974bv,Chang:1978nu,Fradkin:1975yt,Salam:1978dk,Bais:1978fv,Salam:1980ss,Callaway:1988ya}.

Though being universal, the one-loop $\beta$ function coefficent does
not necessarily provide a reasonable measure for the physical scale
dependence of couplings. A simple example is the running of the QED
fine-structure constant at, say, nano-electron-Volt scales: here the 
standard one-loop coefficient still assumes its standard value,
whereas the coupling (as, for instance, measured by Thomson
scattering) does not run at all, because the electron fluctuations
decouple below the electron mass threshold.

The reason for this apparent mismatch is that standard $\beta$
function definitions make implicit use of the deep Euclidean region
(DER), where all physical mass scales or external momenta are assumed
to be small with respect to the loop momenta of the fluctuations. By
contrast, definitions of RG functions that take mass or momentum
thresholds explicitly into account lead to $\beta$ functions that
describe the decoupling adequately. A famous example is given by RG
functions defined by the Callan-Symanzik equation
\cite{Symanzik:1970rt,Callan:1970yg,ZinnJustin:1989mi}.

The price to be paid for including physical threshold phenomena in an
RG description is that the corresponding $\beta$ functions become
scheme dependent even at one-loop order. This is natural, as this
dependence parametrizes the details of the physical decoupling of
massive modes; of course, such a scheme dependence cancels in physical
observables such as cross sections.

While threshold phenomena in RG functions are well-known and
controlled by standard procedures \cite{Collins:1978wz,Ovrut:1980dg,Ovrut:1979pk,Weinberg:1980wa,Bernreuther:1981sg,Marciano:1983pj}, their
potential role towards higher energies has been studied very
little. Here, the analysis in the DER seems only natural, as highest
momentum fluctuations are assumed to always exceed any mass scale. In
the case of mass generation through spontaneous symmetry breaking this
expectation is summarized as ``asymptotic symmetry''
\cite{Lee:1974gua}.

By contrast, new RG trajectories have recently been discovered in
non-Abelian gauge theories with various matter content that invalidate
the assumption of asymptotic symmetry \cite{Gies:2015lia}. Most
importantly, these trajectories give rise to new routes to AF
 and thus ultraviolet (UV) complete scenarios in \NAH\ models
\cite{Gies:2015lia,Gies:2016kkk} as well as gauged Yukawa models
\cite{Gies:2018vwk}, with large classes of models remaining to be
explored and used for model building. In fact, AF
theories still enjoy an unabated interest for the construction of UV complete models in particle physics \cite{Giudice:2014tma,Holdom:2014hla,Hetzel:2015bla,Pelaggi:2015kna,Pica:2016krb,Molgaard:2016bqf,Heikinheimo:2017nth,Einhorn:2017jbs,Hansen:2017pwe,Badziak:2017wxn}. 

Since the occurrence of symmetry-breaking-induced thresholds on all
scales is an essential ingredient for the corresponding RG flows, the
standard reasoning used for the DER and implying one-loop universality
is no longer applicable. This raises naturally the question of scheme
dependence: is the existence of these new AF UV
completions an universal statement? Can it be verified in a
scheme-independent fashion? Answering these questions is a goal of the
present work.

For this, we first generalize previous studies to a Yukawa model
with an $\SU(\NL)\times\SU(\Nc)$ gauge symmetry, covering the
non-Abelian part of the Standard Model (SM); also, the previously
considered \HTQCD\ and \NAH\ models represent limiting cases. In order
to make contact with the most widely used $\MSbar$ scheme of standard
perturbation theory, we elucidate the construction of AF
 trajectories on the basis of one-loop $\beta$ functions obtained
from dimensional regularization. The new UV-complete
trajectories become visible from these RG functions upon inclusion of
a running expectation value and higher dimensional operators, as is
familiar from an effective-field theory (EFT) approach.

A functional approach for the full Higgs potential can also be set up
within the $\MSbar$ scheme. We present several approaches to analyze
the resulting $\beta$ functional also including its global stability
features towards the UV limit. While the $\MSbar$ scheme -- though
widely used -- is a rather particular projection scheme, a more
comprehensive analysis can be performed on the basis of 
general mass-dependent schemes with momentum-space regularization, 
 as featured, e.g., by the 
functional RG (FRG). Here, we provide further evidence for the existence of these
AF trajectories for all admissible regulator
functions.  

In agreement with earlier findings
\cite{Gies:2015lia,Gies:2016kkk,Gies:2018vwk}, the new RG trajectories
occur as \textit{quasi-fixed points} (QFPs) of the $\beta$ functions in
the matter sector. These QFPs are driven by the AF
gauge couplings to the non-interacting Gau{\ss}ian fixed point (FP)
towards higher energies. The presence of an AF gauge
sector -- potentially also beyond the DER -- hence forms a crucial
ingredient in our construction. The important point, however, is that
this feature of AF can fully extend to further sectors of
the system which may not seem to be AF in the
conventional perturbative analysis restricted to the DER. 
For future work, an analysis going beyond the DER may also be
worthwhile for asymptotically safe particle-physics scenarios \cite{Litim:2014uca,Codello:2016muj,Bond:2016dvk,Bond:2017lnq,Dondi:2017civ} which
have recently attracted substantial attention for concrete model building \cite{Litim:2015iea,Esbensen:2015cjw,Bajc:2016efj,Mann:2017wzh,Bond:2017wut,Pelaggi:2017abg,Molinaro:2018kjz,Barducci:2018ysr,Abel:2018fls,Wang:2018yer}.

The paper is organized as follows: In
\Secref{sec:AF_in_standard_perturbation_theory}, we first introduce
the class of models featuring a local $\SU(\NL)\times\SU(\Nc)$ gauge
symmetry, highlight several relevant limiting cases, and review
the standard perturbative analysis also establishing our notation. In
order to transcend the limitations of the DER, \Secref{sec:EFT_MSbar}
presents a perturbative weak-coupling analysis within an
EFT approach, allowing for an inclusion of
higher-dimensional operators. Here, the analysis primarily relies on
the most widely used $\MSbar$ scheme based on dimensional
regularization, elucidating how the new AF
trajectories become visible in the most conventional scheme using
standard methods. Section \ref{sec:EFT_MSbar_FRG} is devoted to a
functional analysis of the flow of the effective potential, still
using the $\MSbar$ scheme. We construct various functional
approximations to the QFP trajectories; this includes also a
controlled weak-coupling expansion, illustrating how the perturbative
EFT emerge in the functional picture. In order to
discuss general classes of RG schemes, we set up the FRG
equations for the models in \Secref{sec:FRG}, employing a derivative
expansion of the effective action and also accounting for threshold
effects in the gauge sectors. The scheme-independent existence of the
new AF trajectories is then demonstrated in
\Secref{sec:weakhexpansion} using a weak-coupling analysis of the
FRG equations. As expected, a change of regularization
scheme induces a map of the coupling space onto itself, thereby
rearranging the space of initial conditions used for specifying
AF trajectories. As a new ingredient, this theory
space also includes rescaling parameters that distinguish between
different AF trajectories.

\section{Asymptotic freedom within perturbative renormalizability}\label{sec:AF_in_standard_perturbation_theory}

Ultimately aiming at the SM, we base our concrete studies in this work on a toy model which
comprises both non-Abelian sectors, the $\SU(\NL)$ gauge group 
as part of the electroweak interaction coupled to scalars and 
fermions, described in~\cite{Gies:2013pma}, and the strong-interaction-type
 $\SU(\Nc)$ gauge group coupled only to fermions as studied in~\cite{Gies:2018vwk}.	
Whenever feasible, we specialize to the SM matter content including its flavor and generation substructure.
We implicitly assume the existence of further sectors -- 
to be ignored for the purpose of this work -- 
that cancel perturbative gauge anomalies or a global Witten anomaly 
possibly occurring for certain $\NL$ and fermion content.

More explicitly, let us consider a complex scalar
\begin{align}
	\phi=\begin{pmatrix}
		\phi^1 \\
		\vdots\\
		\phi^{\NL}
	\end{pmatrix},\quad \phi'(x)= \E^{\I \bar{g} \alpha_i(x) t_i}\phi(x)
	\label{eq:SUL_trasformation_phi}
\end{align}
which transforms according to the fundamental representation of the local $\SU(\NL)$ gauge group.
The generators of the su$(\NL)$ Lie algebra are $t_i$ where $i=\{1,\dots,\NL^2-1\}$
and $\bar{g}$ is the charge associated to this Lie group.
Let us consider also a vector of $\NL$ Weyl fermions $\psiL$
belonging to the fundamental representation of $\SU(\NL)$.
Corresponding right-handed Weyl components transform trivially, 
\emph{i.e.}, as a singlet, under $\SU(\NL)$.
For $\NL=2$, we identify the components as top and bottom quark or their corresponding 
counter-parts of other generations.
With respect to the  $\SU(\Nc)$ color gauge group, each Weyl spinor transforms under the fundamental representation.
The corresponding gauge transformations for the fermions are
\begin{align}
\psiL'(x)&=\E^{\I \bar{g} \alpha_i(x)t_i} \otimes \E^{\I\bargs \alpha_{\mathrm{s}I}(x) T_I}\psiL(x),\\
\psiR'(x)&=\E^{\I\bargs \alpha_{\mathrm{s}I}(x) T_I}\psiR(x),
\label{eq:SULSUNc_trasformation_psiLR}
\end{align}
for arbitrary gauge functions $\alpha_i(x)$,
$\alpha_{\mathrm{s}I}(x)$. Here $T_I$ are the generators of the
su$(\Nc)$ Lie algebra with $I=\{1,\dots,\Nc^2-1\}$ and $\bargs$ is the
strong gauge coupling.

The essential part of the classical action that we address in
four-dimensional Euclidean spacetime reads
\begin{align}
	S_\text{cl}&=\int\! d^{4}x\left[\frac{1}{4}F_{i\mu\nu}^{\phantom{\mu}}F_i^{\mu\nu}+\frac{1}{4}G_{I\mu\nu}^{\phantom{\mu}}G_I^{\mu\nu}+(D_\mu\phi)^{\dagger a}(D^\mu\phi)^a\right.\nonumber\\
	&\quad+\bar m^2\trho+\frac{\bar{\lambda}}{2} \trho^2+\barpsiL^{aA}\I\slashed{D}^{abAB}\psiL^{bB}+\barpsiR^A\I\slashed{D}^{AB}\psiR^B\nonumber\\
	&\quad\left. +\I\bar h (\barpsiL^{aA}\phi^a\psiR^A+\barpsiR^A\phi^{\dagger a}\psiL^{aA})\right],
	\label{eq:S_cl_euclidean}
\end{align}
where the scalar field amplitude $\trho$ is the $\SU(\NL)$ invariant $\trho=\phi^{\dagger a}\phi^a$.
The indices $a,b,c,\cdots$ and $A,B,C,\cdots$ starting at the beginning of the alphabet are associated to the fundamental representations of $\SU(\NL)$ and $\SU(\Nc)$, respectively, $a=\{1,\dots,\NL\}$ and $A=\{1,\dots,\Nc\}$.
We explicitly introduce only one Yukawa coupling. For the limiting case of the SM, this Yukawa coupling will play the role of the top Yukawa coupling which is quantitatively the most relevant Yukawa coupling for the running of the Higgs potential. For $\NL=2$ it is also straightforward to introduce a bottom-like Yukawa coupling for the second component via the charge conjugated Higgs field. We suppresse possible lepton terms as well as generation indices which are implicitly understood and will be included in the counting of degrees of freedom whenever relevant.

The right-handed quarks are coupled to the gluons $G_I^\mu$ through the covariant derivative
\begin{align}
	D_\mu^{AB}=\delta^{AB}\de_\mu+\I\bargs G_{I\mu} T_I^{AB}.
\end{align}
The covariant derivative acting on the left-handed components involves also the gauge boson vector fields $W_i^\mu$,
\begin{align}
	D_\mu^{abAB}=\delta^{AB}\left(\delta^{ab}\de_\mu+\I \bar g W_{i\mu} t_i^{ab}\right)+\delta^{ab}\I\bargs G_{I\mu} T_I^{AB}.
\end{align}
The complex scalar is coupled only to the $W$ bosons,
\begin{align}
(D_\mu\phi)^{a}=\left(\delta^{ab}\de_\mu+\I \bar g W_{i\mu} t_i^{ab}\right)\phi^b.
\end{align}
The classical parameter space of this model is spanned by five bare couplings: the weak gauge coupling $\bar g$, the strong gauge coupling $\bargs$, the Yukawa coupling for the top quark $\bar h$, the scalar mass parameter $\bar m$ and the scalar quartic coupling $\bar{\lambda}$. While the mass parameter is power-counting relevant, all other couplings are marginal.

In the remainder of this section, we review the standard perturbative analysis for the above model at one loop and only for perturbatively renormalizable couplings in the DER.
In the latter approximation, we set any propagator masses to zero since they are supposed to be negligible with respect to the RG scale in the UV limit.
In particular we do not consider any contributions coming from the scalar mass parameter $\bar m$ as well as from a nontrivial vacuum expectation value in the case where the scalar potential is in the spontanously symmetry-broken (SSB) regime.
Moreover, we focus on the UV behavior of this toy model and look for totally AF trajectories.
In order to address this point, we need to study the RG flow equations for the renormalized dimensionless couplings $g$, $\gs$, $h$ and $\lambda$.
Their definitions in terms of the bare couplings and wave-function renormalizations are detailed later on in \Secref{sec:FRG}.

\subsection{Gauge sector}

Let us start by analyzing the RG flow equation for the gauge couplings.
The RG equation for the gauge coupling of the SU($\NL$) group is~\cite{Gross:1973ju} 
\be
\begin{split}
	\de_t g^2&=\etaW g^2,\\ \etaW&=-\frac{g^2}{48\pi^2}\left(22\NL-\dgammaL\NfL-\Nsc\right).
\end{split}
\label{eq:beta_g2_1loop_DER}
\ee
Here, $\dgammaL$ denotes the dimension of the Clifford-algebra representation (with $\dgammaL=2$ for the left-handed Weyl spinors of the SM). We also introduced
$\NfL$ as the number of fermionic $\NL$-tuples.
In the SM, we have 3 doublets for the leptons and $9$ doublets for the quarks, accounting for their $\SU(3)$ color, therefore $\NfL=12$.
The number of scalar $\NL$-tuples is counted by $\Nsc$, with $\Nsc=1$ for the SM.
The RG equation for the strong gauge coupling $\gs$ reads
\be
\begin{split}
	\de_t \gs^2=\etaG \gs^2,\quad \etaG=-\frac{\gs^2}{48\pi^2}\left(22\Nc-\dgammac\Nfc\right),
\end{split}\label{eq:beta_gs2_1loop_DER}
\ee
where $\dgammac$ denotes the dimension of the combined left- and right-handed Clifford algebra, i.e, $\dgammac=4$, and $\Nfc$ is the number of quark flavors. For the SM, we have in summary
\begin{align}
	\NL=2,\,\Nc=3,\,\NfL=12,\,\Nfc=6,\,\Nsc=1.
	\label{eq:SM_parameters}
\end{align}
In this case, both one-loop $\beta$ functions are negative such that $g$ and $\gs$ approach the AF
 Gau\ss ian FP in the UV limit.
In the present work, we use these $\beta$ functions for various specific models differing by their matter and gauge-symmetry content.

\subsection{Yukawa sector}
\label{subsec:Yukawa_sector}

In the present section, we retain the general $\Nc$, $\NL$ dependence as well as generic fermionic matter content specified by $\NfL$ and $\Nfc$,
while we set $\Nsc =1$.
The standard one-loop RG flow equation for the top-Yukawa coupling $h^2$ in the DER reads
\begin{align}
	\de_t h^2= (\eta_\phi+\etaL+\etaR) h^2-\frac{3}{8\pi^2}\frac{\Nc^2-1}{\Nc}h^2\gs^2,	\label{eq:beta_h2_1Loop_DER_general_NL_Nc}
\end{align}
where the anomalous dimensions for the scalar, 
the left- and right-handed Weyl spinors are
\begin{gather}
	\eta_\phi=\frac{\Nc}{8\pi^2}h^2-\frac{3}{16\pi^2}\frac{\NL^2-1}{\NL}g^2,\label{eq:eta_phi_1Loop_DER}\\
	\etaL=\frac{1}{16\pi^2}h^2,\quad\etaR=\frac{\NL}{16\pi^2}h^2\label{eq:eta_psi_1Loop_DER}.
\end{gather}
\begin{figure}[t!]
	\includegraphics[width=0.49\columnwidth]{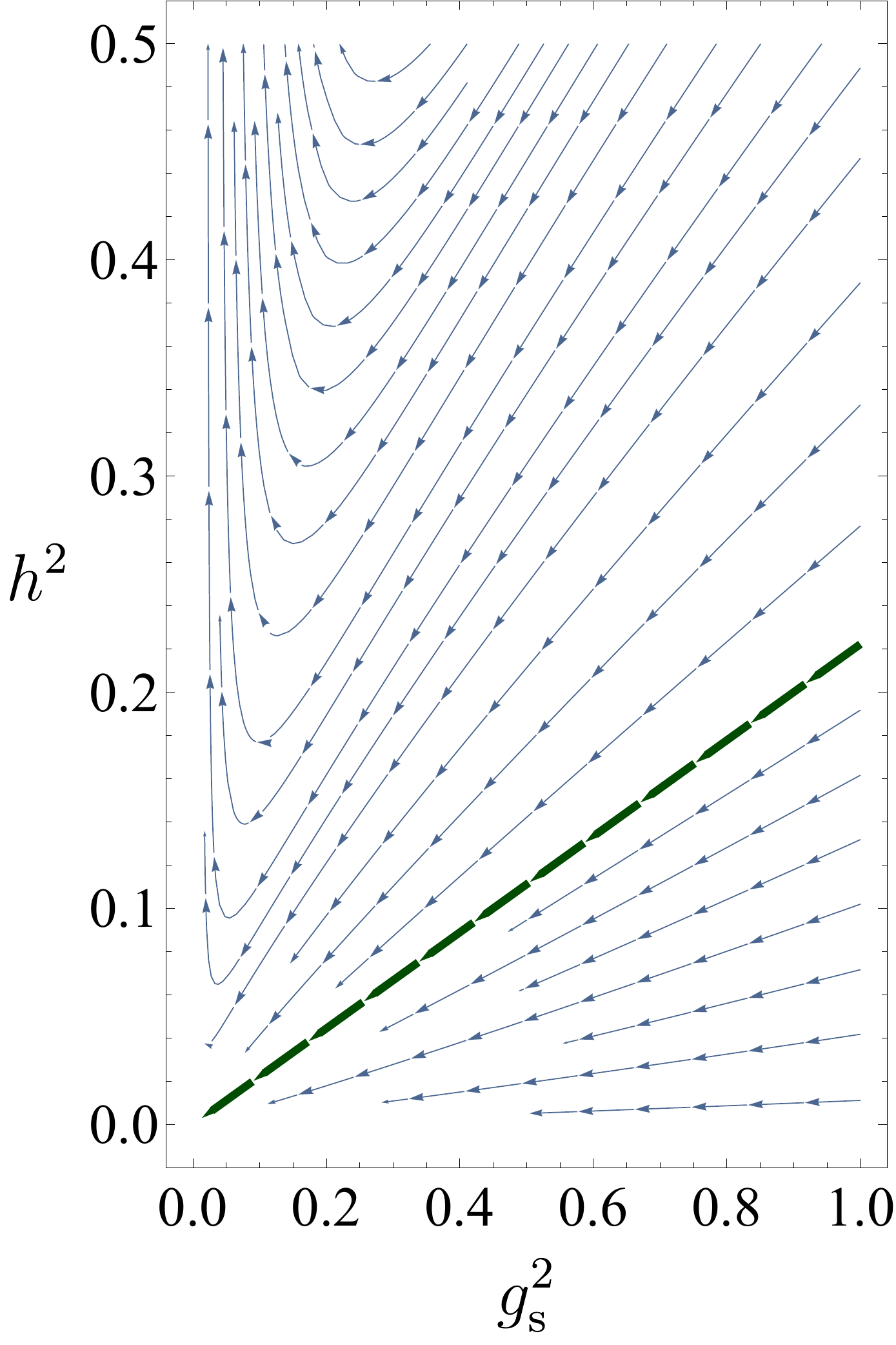}\hfill
	\includegraphics[width=0.49\columnwidth]{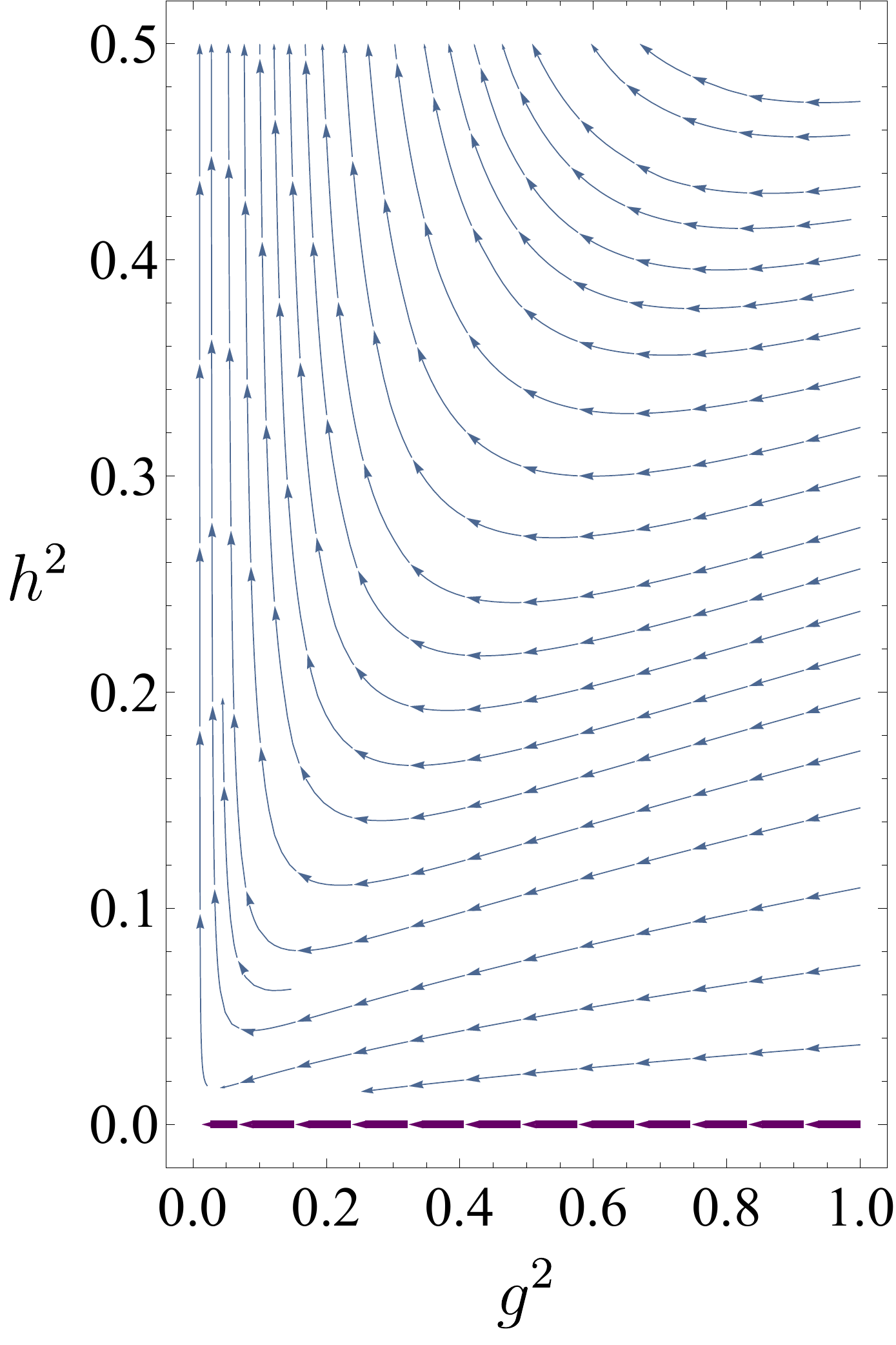}
	\caption{
		\emph{Left Panel}: the one-loop RG flow of the model
        with SM-like matter content in the DER
        projected onto the $(h^2,\gs^2)$ plane.  The UV
        repulsive QFP trajectory highlighted by the green line
        corresponds to the solution in
        \Eqref{eq:h2/gs2_QFP_YQCD_DER}.
        \emph{Right Panel}: the
        corresponding RG flow  projected onto the $(h^2,g^2)$ plane.
        The UV repulsive QFP trajectory highlighted by the purple
        line corresponds to the trivial solution $h^2=0$.  The
        arrows for both flows are pointing towards the UV.}
	\label{fig:StreamPlot_h2_of_g2_and_gs2}
\end{figure}
\subsubsection{\HTQCD\ model}
In order to obtain a better understanding of the RG trajectories towards the UV in the three dimensional
space $(h^2,g^2,\gs^2)$, let us start with the flow within the $(h^2,\gs^2)$ plane.
This corresponds to setting $g^2=0$ inside the RG flow equation for the top-Yukawa coupling.
Moreover we choose $\NL=2$ for illustration as in the SM.
In this case we expect a similar behavior as for the \HTQCD\ model analyzed in Ref.~\cite{Gies:2018vwk},
for which there is an AF region, bounded by a special AF trajectory along which $h^2$ is proportional to $\gs^2$.
This behavior can be characterized in terms of a rescaled Yukawa coupling,
\begin{align}
	 \hrescs^2=\frac{h^2}{\gs^2}. \label{eq:h2/gs2_def}
\end{align}
Its $\beta$ function is
\begin{align}
\de_t \hrescs^{2} = \frac{3+2\Nc}{16\pi^2}\gs^2\hrescs^{2}\left(\hrescs^{2} -  \chis^2\right),
	\label{eq:beta_h2/gs2_1Loop_DER}
\end{align}
where
\begin{align}
	\chis^2=\frac{1}{3+2\Nc}\left[\frac{4}{3}\left(\Nfc-\Nc\right)-\frac{6}{\Nc}\right].	\label{eq:h2/gs2_QFP}
\end{align}
The $\beta$ function in \Eqref{eq:beta_h2/gs2_1Loop_DER} has only one nontrivial zero for $\gs^2\neq 0$ which is $\hat{h}_*^2=\chis^2$.
A partial fixed point for a ratio of AF couplings such as in \Eqref{eq:h2/gs2_def} has been called \emph{quasi-fixed point} (QFP) in Ref.~\cite{Gies:2016kkk}.
It is a defining condition for AF scaling solutions and an useful tool to search for such trajectories~\cite{Gross:1973ju,Chang:1974bv,Callaway:1988ya,Giudice:2014tma,Gies:2018vwk}.
In Ref.~\cite{Gies:2018vwk} we observed also that AF requires the matter content parameter $\Nfc$  to stay within a finite window for fixed $\Nc$.
The upper bound of this window is given by the requirement that $\etaG<0$, while
the lower bound can be derived from Eq.~\eqref{eq:h2/gs2_QFP} by demanding
$\chis^{2}>0$. Thus, we obtain the criterion
\begin{align}
	\Nc + \frac{9}{2\Nc} < \Nfc < \frac{11}{2}\Nc,
\label{eq:AF_yukawa_window_Nfc}
\end{align}
which is fulfilled by the SM parameters of \Eqref{eq:SM_parameters}.
In this case, the ratio in~\Eqref{eq:h2/gs2_QFP} attains the value
\begin{align}
	\chis^2=\frac{2}{9}.
	\label{eq:h2/gs2_QFP_YQCD_DER}
\end{align}

\subsubsection{Non-Abelian Higgs model}

A similar analysis can be performed also by projecting the flow in~\Eqref{eq:beta_h2_1Loop_DER_general_NL_Nc} onto the $(h^2,g^2)$ plane, corresponding to taking the $\gs^2\to0$ limit.
Setting $\Nc=3$ for illustration, we can search for AF trajectories along which the Yukawa coupling becomes proportional to $g^2$.
Namely, we are interested in a QFP for the rescaled coupling
\begin{align}
	 \hrescg^2=\frac{h^2}{g^2}. \label{eq:h2/g2_def}
\end{align}
The corresponding RG flow equation is in this case
\begin{align}
	\de_t \hrescg^2=\frac{\NL+7}{16\pi^2} g^2\hrescg^2\left(\hrescg^2-\chig^2\right),
	\label{eq:beta_h2/g2_1Loop_DER}
\end{align}
where $\chig^2$ reads
\begin{align}
	\chig^2=\frac{2}{3(\NL+7)}\left(\NfL+\frac{1}{2}-\frac{13}{2}\NL-\frac{9}{2\NL}\right).	\label{eq:h2/g2_QFP}
\end{align}
The constraint on the matter content in order to have AF for both couplings $g^2$ and $h^2$ then is
\begin{align}
	\frac{1}{2}\left(13\NL+\frac{9}{\NL}-1\right)<\NfL<11\NL-\frac{1}{2}.	\label{eq:AF_yukawa_window_NfL}
\end{align}
However, the lower bound is not fulfilled for the SM, resulting
in a negative value for $\chig^2$,
\begin{align}
	\chig^2=-\frac{11}{54}.
	\label{eq:h2/g2_QFP_DER}
\end{align}
We can therefore conclude that nontrivial solutions for the QFP
equation $\de_t\hrescg^2=0$ do not exist in the positive part of the
$(h^2,g^2)$ plane, where the only possible solution for $g^2\neq 0$ is the trivial one $h^2=0$.

The two scenarios are illustrated for the SM case in \Figref{fig:StreamPlot_h2_of_g2_and_gs2}, where the flows projected onto the
$(h^2,\gs^2)$ (left) and $(h^2,g^2)$ (right) planes are depicted.
On the left panel, the special AF trajectory expressed in \Eqref{eq:h2/gs2_QFP_YQCD_DER} in the $g^2\to 0$ limit is highlighted by a
green line.
On the right panel, the flow in the $\gs^2\to 0$ limit is shown, where
the trivial solution corresponding to the axis $h^2=0$ is highlighted
by a purple line.
It is clear from the left panel that the trajectory $\hat{h}_*^2=\chis^2$
represents an upper bound for TAF.
In fact, if at some initializing RG scale
$\hresc_0^2>\chis^2$, the Yukawa
coupling hits a Landau pole at some finite energy scale towards the UV within this one-loop approximation.
On the other hand the Yukawa coupling becomes AF for those initial values which fulfill the constraint $\hresc_0^2\leq\chis^2$.
For the same reasons, there are no AF trajectories for the top-Yukawa coupling  due to the negative value of $\chig^2$ in the physical quadrant of the plane $(h^2,g^2)>0$.
The only possible way to have TAF is to set $h^2$ equal to the trivial null value.
The QFP nature of these two special trajectories can be better understood by looking at the
RG flow for the rescaled couplings $\hrescs^2$ or $\hrescg^2$ as a function of $\gs^2$ or
$g^2$, respectively.
They correspond to IR attractive trajectories, which govern the low-energy behavior
of the model, enhancing its predictive power.

\begin{figure}[t!]
	\includegraphics[width=0.9\columnwidth]{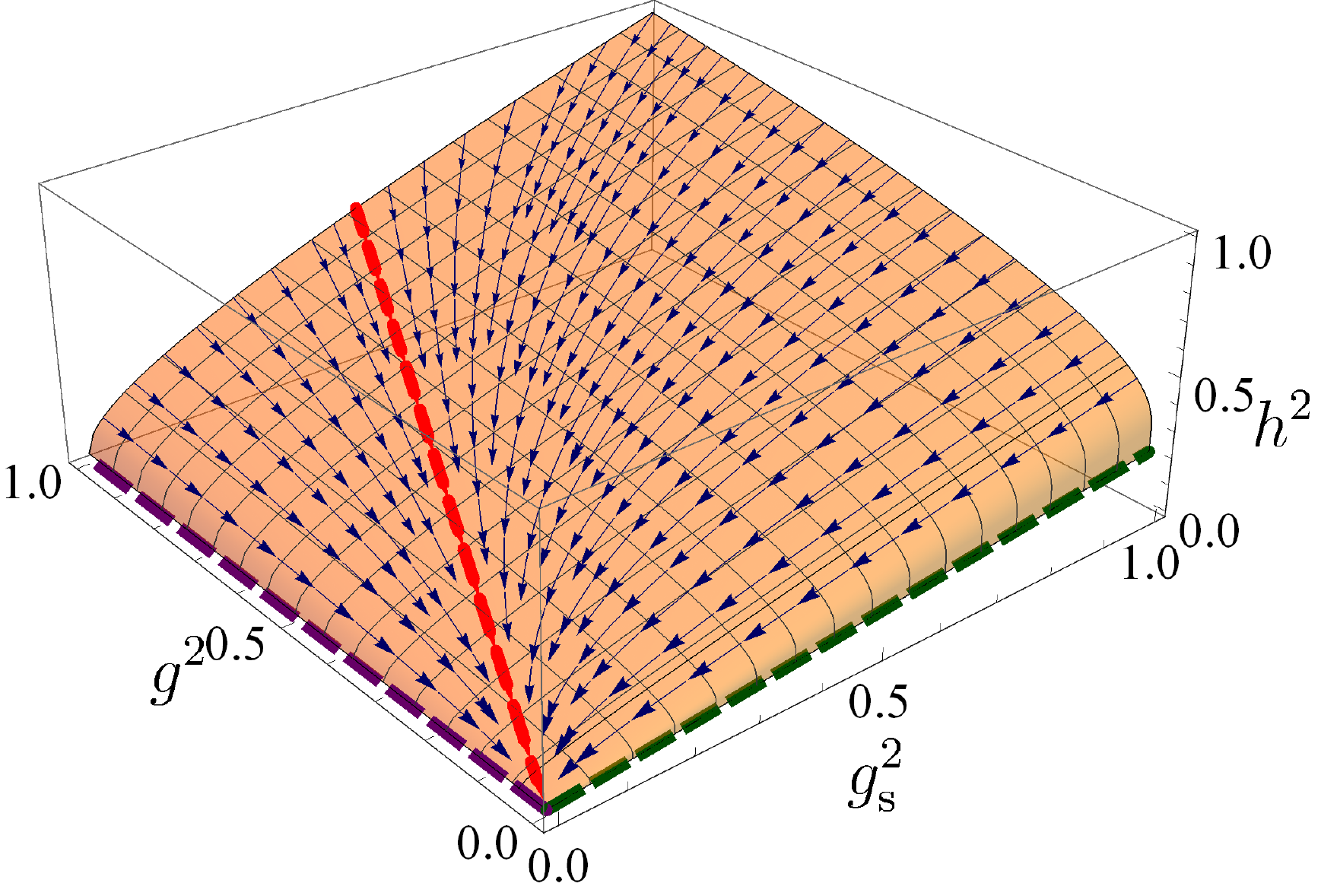}
	\caption{
		The upper critical surface $h^2=\Omega(\gs^2,g^2)$ of
		total AF for the perturbatively renormalizable model in the DER
		and for the SM set of parameters summarized in \Eqref{eq:SM_parameters}.
		The special trajectory in \Eqref{eq:QFP_g2/gs2_h2/gs2_generalscheme} along which
		$h^2$ is proportional to $g^2$ as well as $\gs^2$ is highlighted by a red line.
		It is an UV attractive (repulsive) trajectory along the directions tangent (orthogonal) to the critical surface.
		The intersection of $\Omega$ with the $g^2=0$ plane is
                highlighted by a green line with a slope given by
                \Eqref{eq:h2/gs2_QFP_YQCD_DER}; the intersection of
                $\Omega$ with the $\gs^2=0$ plane is shown as a purple
                line satisfying $h^2=\gs^2=0$.
		The arrows of the RG stream flow on the top of the critical surface are
		pointing towards the UV.
	}
	\label{fig:AF_surface}
\end{figure}
\subsubsection{\SULSUc\ model}

Let us next address the running of the Yukawa coupling in the presence of both gauge couplings. It is possible to analytically integrate the RG 
$\beta$ function of $h^2$ in \Eqref{eq:beta_h2_1Loop_DER_general_NL_Nc} together with the $\beta$ functions for $g^2$ and $\gs^2$.
As explained in \Appref{App:1loop_DER} the matter content parameters $\NfL$ and $\Nfc$ must fulfill the following necessary but not sufficient condition in order to feature total AF,
\begin{align}
	\chi^2=\frac{9(\Nc^2-1)}{\Nc (11 \Nc-2\Nfc)}+\frac{9(\NL^2-1)}{ \NL (22\NL-2\NfL-1)}-1>0,		\label{eq:TAF_necessary_condition}
\end{align}
which generalizes the two lower bounds $\chis^2>0$ and $\chig^2>0$  previously obtained for $\NL=2$ and $\Nc=3$.
In the SM case, the inequality \eqref{eq:TAF_necessary_condition} is satisfied, since $\chi^2=227/266$.
If the condition \eqref{eq:TAF_necessary_condition} holds,
we can identify a critical surface parametrized by a function $h^2=\Omega(\gs^2,g^2)$ which represents the upper bound for TAF.
In other words, for any initial condition such that $h_0^2\leq\Omega(g_\text{s0}^2,g_0^2)$ the top-Yukawa coupling becomes AF and approaches the Gau\ss ian fixed point in the UV limit.
As detailed in \Appref{App:1loop_DER}, this surface is an UV-repulsive surface along its
normal directions, while all the trajectories on the surface itself are in the UV limit attracted towards a special one where the top-Yukawa coupling and the gauge couplings are proportional to each other. In order to find the corresponding equation for this trajectory,
let us use the QFP criteria and consider first the ratio of the
two gauge couplings
\begin{align}
	\gresc^2=\frac{g^2}{\gs^2}.
	\label{eq:g2/gs2_def}
\end{align}
The flow equation for $\gresc^2$ is then
\begin{align}
	\de_t \gresc^2=\frac{g^2}{48\pi^2}\left[22\Nc-4\Nfc-(22\NL-2\NfL-1)\gresc^2\right],
\end{align}
which has a QFP solution for $g^2\neq 0$ at
\begin{align}
	\gresc_*^2=\frac{2(11\Nc-2\Nfc)}{22\NL-2\NfL-1}.
	\label{eq:g2/gs2_QFP_generalModel}
\end{align}
The solution $g^2=\gresc_*^2\gs^2$ identifies a plane in the three
dimensional space of parameters $(g^2,\gs^2,h^2)$ whose intersection with the
critical surface $\Omega$ is a trajectory along which
the top-Yukawa coupling is proportional to both gauge couplings.
Assuming $g^2=\gresc_*^2\gs^2$, we perform the same rescaling as in~\Eqref{eq:h2/gs2_def}, arriving at the $\beta$ function for the
rescaled top-Yukawa coupling
\begin{align}
	\de_t \hrescs=\hrescs^2\gs^2\frac{2\Nc+\NL+1}{16\pi^2}\left[\hrescs^2-\frac{2(11\Nc-2\Nfc)\chi^2}{3(2\Nc+\NL+1)}\right],
\end{align}
which has a nontrivial QFP solution at
\begin{align}
	\hresc_*^2=\frac{2(11\Nc-2\Nfc)\chi^2}{3(2\Nc+\NL+1)}. \label{eq:h2/gs2_QFP_generalModel}
\end{align}
Alternatively,  \Eqref{eq:h2/g2_def} could be used in the same way.

Equations~(\ref{eq:g2/gs2_QFP_generalModel}) and (\ref{eq:h2/gs2_QFP_generalModel})
are positive for the SM set of parameters and attain the QFP values
\begin{align}
	\hat{g}^2_*=\frac{42}{19},\qquad \hat{h}_*^2=\frac{227}{171}.
	\label{eq:QFP_g2/gs2_h2/gs2_generalscheme}
\end{align}
The RG flow in the
three dimensional space of couplings is plotted in \Figref{fig:AF_surface} exhibiting the
critical surface $\Omega(\gs^2,g^2)$ for the SM case and the RG flow on top of it.
Since this surface represents the upper bound for TAF, the directions
normal to it are UV repulsive.
On the surface itself however, all the trajectories are attracted
towards the special one described in
\Eqref{eq:QFP_g2/gs2_h2/gs2_generalscheme} in the UV limit, as is
highlighted by a red line in \Figref{fig:AF_surface}.  In the same
plot, the two trajectories in the $g^2=0$ (green) and $\gs^2=0$
(purple) planes are also highlighted, corresponding to those of
\Figref{fig:StreamPlot_h2_of_g2_and_gs2}.

\subsection{Scalar sector}

\begin{figure}[t!]
  \includegraphics[width=0.49\columnwidth]{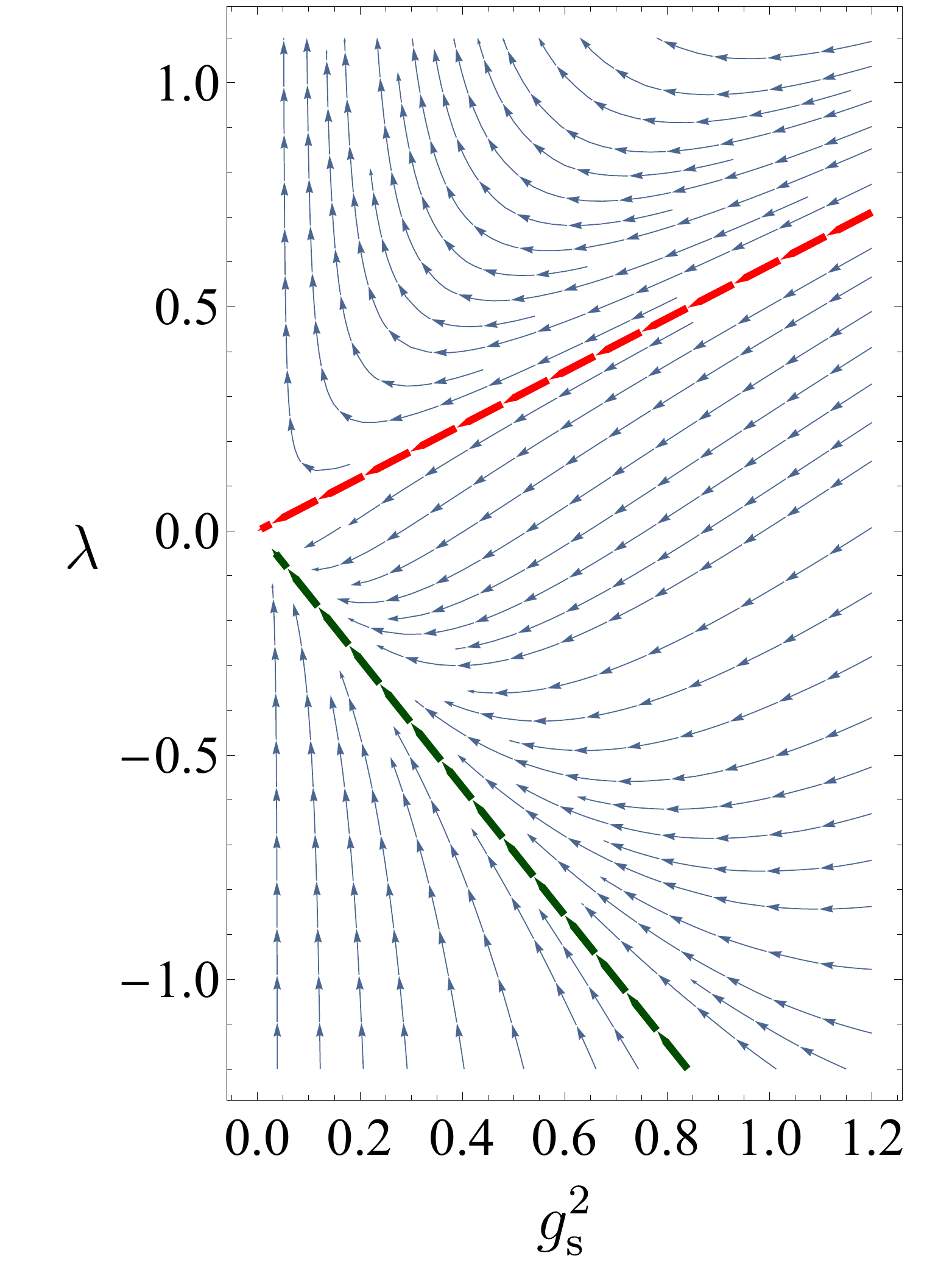}\hfill
  \includegraphics[width=0.49\columnwidth]{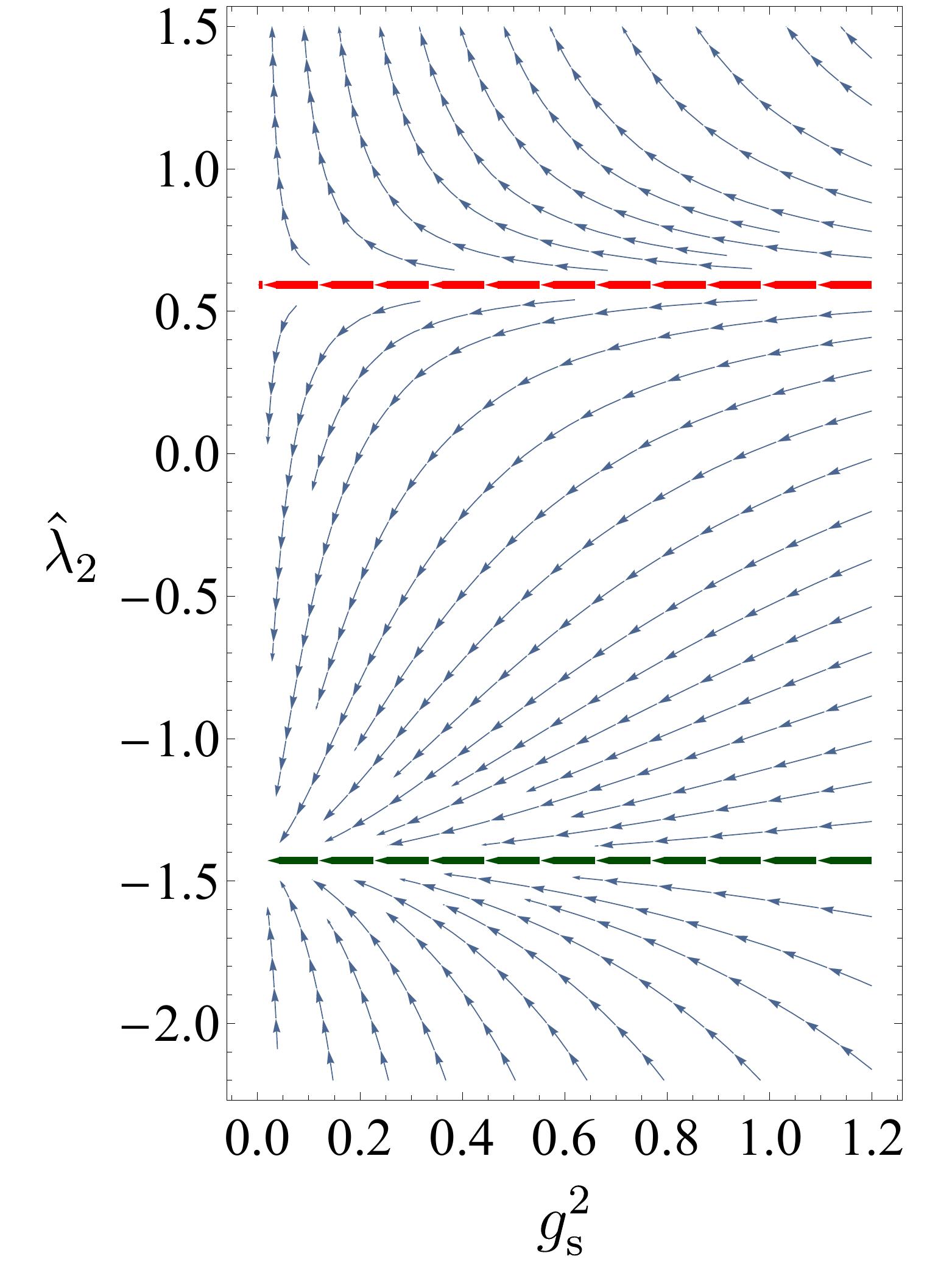}
  \caption{
  	\emph{Left Panel}: one-loop RG flow of the quartic scalar
    coupling $\lambda$ of the model with SM-like matter content as a
    function of $\gs^2$, using the special trajectory defined by
    \Eqref{eq:QFP_g2/gs2_h2/gs2_generalscheme}.
    \emph{Right Panel}: RG flow for the rescaled quartic coupling $\lresc_2$, defined in \Eqref{eq:L2def} with $P=1/2$, as a function of $\gs^2$.
    The red and green lines correspond to the UV-repulsive and attractive trajectories, respectively, corresponding to the roots
    $\lresc_2^\pm$ in \Eqref{eq:L2_pm_QFP_generalmodel}.  }
  \label{fig:StreamPlot_lamda2_of_gs2}
\end{figure}

Now we investigate the scalar sector and thus include also the running of the quartic scalar coupling $\lambda$.
Its $\beta$ function at one loop in the DER for our model defined by Eq.~\eqref{eq:S_cl_euclidean} is
\begin{align}
	\de_t\lambda&=2\eta_\phi\lambda+\frac{3(\NL-1)(\NL^2+2\NL-2)}{32\pi^2\NL^2}g^4\nonumber\\
	&\quad+\frac{\NL+4}{8\pi^2}\lambda^2-\frac{\Nc}{4\pi^2}h^4 ,	\label{eq:beta_lambda_1Loop_DER_general_NL_Nc}
\end{align}
where $\eta_\phi$ is given by~\Eqref{eq:eta_phi_1Loop_DER}.
Since we are interested in the special trajectory described by Eqs.~(\ref{eq:g2/gs2_QFP_generalModel}) and (\ref{eq:h2/gs2_QFP_generalModel}), along which the top-Yukawa and the gauge couplings are proportional, we can express $h^2$ and $g^2$ as a function of $\gs^2$.
Thus the beta function $\de_t \lambda$ turns out to be just a function of $\lambda$ and $\gs$.
Any AF solution must correspond to a particular scaling
of the quartic coupling with respect to the gauge coupling.
The latter is best revealed by inspecting the
flow for the ratio
\begin{align}	
	\lresc_{2}=\frac{\lambda}{\gs^{4P}}, \quad P>0 \label{eq:L2def} .
\end{align}
Here, the positive power $P$ is either fixed by the QFP 
condition for $\lresc_{2}$ at nonvanishing $\gs^2$,
or remains a free parameter.
The $\beta$ function for this rescaled Higgs coupling then receives an extra contribution coming from the running of $\gs$. Indeed
\begin{align}
	\de_t\lresc_2=&2\hatetaphi\lresc_2\gs^2+\frac{3(\NL-1)(\NL^2+2\NL-2)}{32\pi^2\NL^2}\hat{g}_*^4\gs^{4-4P}\nonumber\\
	&+\frac{\NL+4}{8\pi^2}\lresc_2^2\gs^{4P}-\frac{\Nc}{4\pi^2}\hat h_*^4\gs^{4-4P}+2 P\lresc_2 \hatetaG \gs^2.
\end{align}
Here, we have introduced the rescaled gluon and scalar anomalous dimensions
\begin{align}
	\hatetaG&=\frac{\etaG}{\gs^2}= - \frac{22\Nc-4\Nfc}{48\pi^2},
		\label{eq:hat_eta_G}\\
	\hatetaphi&=\frac{\eta_\phi}{\gs^2}=\frac{\Nc}{8\pi^2}\hat{h}^2_*-\frac{3(\NL^2-1)}{16\pi^2\NL}\hat g_*^2,
	\label{eq:hat_eta_phi}
\end{align}
which assume constant values on the QFP.
Close inspection reveals that  a nontrivial finite QFP solution for $\lresc_2$ in the UV limit 
requires $P=1/2$.
By choosing the SM set of parameters, see \Eqref{eq:SM_parameters},
we find the two roots
\begin{align}
	\lresc_2^\pm=\frac{1}{342} \left(-143\pm\sqrt{119402}\right),\quad P=\frac{1}{2}.
	\label{eq:L2_pm_QFP_generalmodel}
\end{align}
The stability properties of these two QFPs can be deduced by plotting the RG flow of $\lambda_2$ or $\lresc_2$ as a function of $\gs^2$, as shown in \Figref{fig:StreamPlot_lamda2_of_gs2}.
The positive root (red line) corresponds to an UV-repulsive trajectory.
By contrast, the negative root (green line) characterizes an UV-attractive trajectory.
For any initial condition with $\lresc_2<\lresc_2^+$, 
the rescaled quartic scalar coupling is attracted towards the negative root
in the UV, and the perturbative potential appears to become unstable.
On the other hand for an initial value bigger than $\lresc_{2}^+$,
the scalar coupling hits a Landau pole at some finite energy scale towards the UV.
Therefore $\lresc_2=\lresc_{2}^+$ corresponds to the only trajectory along which
the theory is UV-complete. The perturbatively renormalizable potential is automatically  stable then.
This trajectory is IR attractive and the low-energy behavior is governed by the QFP
value $\lresc_{2}^+$ which means that the theory exhibits an high degree of predictivity.

Comparing our toy-model flow to that of the SM, current data suggests that the SM flow is governed by its vicinity to the analogue of the critical surface $\Omega$, with the gauge couplings, the top-Yukawa coupling $h$ and the scalar coupling $\lambda$ all exhibiting a flow to smaller values above the Fermi scale. As the strong coupling $\gs$ is larger than the weak coupling $g$, the gauge sector has not yet reached its QFP \eqref{eq:g2/gs2_QFP_generalModel}. Also, the top-Yukawa coupling is below its QFP value \eqref{eq:h2/gs2_QFP_generalModel}, and behaves AF, cf.~\Figref{fig:StreamPlot_h2_of_g2_and_gs2} (left panel). The scalar coupling appears to be near critical \cite{Buttazzo:2013uya,Bednyakov:2015sca,Andreassen:2017rzq,Alekhin:2017kpj,Chigusa:2018uuj,McDowall:2018tdg}, with $\lresc_2$ being slightly below (the analogue of) $\lresc_2^+$, such that $\lambda$ appears to approach zero or potentially drop below zero towards higher scales, cf.~ \Figref{fig:StreamPlot_lamda2_of_gs2} (left panel). Of course, the contribution of the hypercharge U(1) group that would dominate the flow far above the Planck scale are ignored in the present discussion.

As a last remark of this section, we observe from the one-loop $\beta$ functions
for the Yukawa coupling and the quartic scalar coupling in the DER,
Eqs.~(\ref{eq:beta_h2_1Loop_DER_general_NL_Nc}-\ref{eq:eta_psi_1Loop_DER})~and~(\ref{eq:beta_lambda_1Loop_DER_general_NL_Nc}) that it is possible to recover
the corresponding $\beta$ functions for various limiting models.
For instance, recovering the \NAH\ model from the general case is straightforwardly possible in the DER, by setting $h^2\to 0$ and $\gs^2\to 0$.
Naively, the flow equation $\de_t h^2$
reduces to the one for the \HTQCD\ case \cite{Gies:2018vwk} by taking the limits $g^2\to 0$ and $\NL\to 2$.
Whereas the flow $\de_t\lambda$
 would reduce to the \HTQCD\ model in the limits $g^2\to0$ and $\NL\to 1/2$.
This seeming contradiction can be resolved by taking the unitary-gauge limit before approaching the DER. In this way the Goldstone modes decouple from the theory
and do not propagate because of their infinite mass.
Thus, the \HTQCD\ model simply corresponds to the limit $g^2\to 0$ of the present model in the unitary gauge.
More details on this reduction are given in \Secref{sec:FRG}.

\section{Effective field theory approach to the scalar potential in $\MSbar$}\label{sec:EFT_MSbar}

Let us generalize  the previously
outlined construction to the inclusion of perturbatively
nonrenormalizable interactions which will result in the description of
new AF models.  In adding higher-dimensional operators to
the scalar potential in \Eqref{eq:S_cl_euclidean}, we follow
the EFT paradigm and start with the most widely used $\MSbar$ scheme for concreteness. Here, we concentrate on momentum-independent scalar self-interactions
which form the effective potential.  As detailed in the next sections, the
consistency of these solutions requires an infinite number of
higher-dimensional operators. The class of point-like
scalar self-interactions is such an infinite set that becomes manageable by functional methods, as discussed in the following.

The goal of the present section is to explain how to
reveal these solutions and to properly account for some of their properties 
in a parameterization where first only a finite number of couplings with higher dimension
is included.
These steps then generalize to the inclusion of \emph{all} interactions up to some given dimensionality
in the effective Lagrangian.
Still, the crucial ingredient in the construction is a treatment of the $\beta$ functions
of these operators that slightly differs from the standard EFT one:
the scale dependence of one coupling 
or Wilson coefficient in the EFT expansion has to be treated as free.
The subsequent sections then demonstrate that this additional freedom has to be present
in any rigorous definition of the RG flow of the model, because of the 
infinite dimensionality of the theory space. It plays the role of a boundary condition
in a functional representation of the quantum dynamics.

In the DER, where all mass parameters are
neglected, it is a well known fact that 
higher-dimensional scalar self-interactions do not influence the running of the lower dimensional
ones. This is because divergences giving rise to powers of the renormalization
scale are replaced by corresponding powers of the masses in the
$\MSbar$ scheme.  However, the DER does not exhaust all
possible asymptotic behaviors of a quantum field theory, as we show in the following.

We begin with a systematic polynomial expansion of the
scalar effective potential. For convenience, we now switch to
dimensionless renormalized quantities which are obtained by rescaling the
dimensionful ones with suitable powers of the RG scale and
wave function renormalizations. Precise
definitions will be given below in Sec.~\ref{sec:FRG}. Let us call the
dimensionless effective potential $u$. 
We expand the potential about
the scale-dependent minimum $\kappa$,
which is vanishing in the symmetric (SYM) regime,
and positive in the spontaneous-symmetry-broken 
(SSB) regime. For instance, 
in the latter case a polynomial approximation
of the effective potential reads
\begin{align}
	u(\rho)=\sum_{n=2}^{N_p}
	\frac{\lambda_n}{n!}(\rho-\kappa)^n,
	\label{eq:ExpansionAroundKappa}
\end{align}
where $\rho$ is the dimensionless renormalized analog of the squared scalar field amplitude $\trho=\phi^\dagger \phi$.
Generically, we expect all couplings to be generated by fluctuations. Truncating the sum at some finite $N_p$ corresponds
to a polynomial approximation of the potential.

For reasons of clarity, we first study the simpler limiting models, the \HTQCD\ and the \NAH\ model, separately.
Both models represent well-defined limiting cases of the general model with $\SU(\NL)\times\SU(\Nc)$ gauge symmetry. In either case, we choose the remaining matter content as in the SM, cf.~\Eqref{eq:SM_parameters}, for illustration, and perform the analysis in the massless $\MSbar$ scheme.
Most of our results will be generalized to the full 
	$\SU(\NL)\times\SU(\Nc)$ model
	and to more general RG schemes in \Secref{sec:weakhexpansion}.
However, already in \Secref{sec:EFT_MSbar_FRG} we unveil novel AF solutions for 
the general $\SU(\NL)\times\SU(\Nc)$ model in the $\MSbar$ scheme.

\subsection{{\boldmath $\mathbbm{Z}_2$}-Yukawa-QCD model}
\label{sec:EFT_MSbar_HTQCD}

For the \HTQCD\ model, the flow equations for the nontrivial minimum $\kappa$ and the quartic scalar coupling $\lambda_2$, obtained by dimensional regularization in the $\MSbar$ scheme, are
\begin{align}
\de_t\kappa&=\left[-2-\frac{3h^2}{8\pi^2}-\frac{3\lambda_2}{8\pi^2}
+\frac{3h^4}{4\pi^2\lambda_2}-\frac{\kappa\lambda_3}{4\pi^2}
\right]\kappa, \label{eq:kappadotEFT}\\
\de_t\lambda_2&=\frac{9\lambda_2^2}{16\pi^2}+\frac{3h^2\lambda_2}{4\pi^2}
-\frac{3h^4}{4\pi^2}\nonumber\\
&\quad+\kappa\lambda_3\left(\frac{\lambda_2}{\pi^2}+\frac{3h^4}{4\pi^2\lambda_2}\right)+\frac{\lambda_2}{4\pi^2}\kappa^2\lambda_4.   \label{eq:lambdadotEFT}
\end{align}
The $\beta$ function for the minimum $\kappa$ involves only the couplings $\lambda_2$ and $\lambda_3$, while the $\beta$ function of a general self-interaction $\lambda_j$ depends on all $\lambda_n$ up to $n=j+2$.
These $\beta$ functions follow straightforwardly from the functional $\MSbar$ flow of the effective potential discussed in
the next section.
Compared to the standard one-loop flow in the DER, which is contained in \Eqref{eq:lambdadotEFT} in the limit $\kappa\to 0$, it appears that  nonvanishing 
values of $\kappa\lambda_3$ or $\kappa^2\lambda_4$
can considerably influence the flow of the quartic coupling.
In fact, this has implications for the construction
of AF trajectories.

As in \Secref{subsec:Yukawa_sector},
we look for AF scaling solutions by means of a QFP 
condition for $\lresc_{2}$, as defined in \Eqref{eq:L2def}.
Beyond the restriction to perturbatively renormalizable
couplings, and in the parameterization of 
\Eqref{eq:ExpansionAroundKappa},
similar conditions can be imposed on
suitably defined rescaled 
couplings
\be
\lresc_{n}=\frac{\lambda_n}{\gs^{2P_n}}, 
\label{eq:lambdatoLn}
\ee
with $P_2=2P$, cf. \Eqref{eq:L2def}. 
Also the coupling $\kappa$
may scale asymptotically as a definite power of $\gs^2$,
\be
\kap=\gs^{2Q}\kappa,
\label{eq:kappatoxikappa}
\ee
where the real power $Q$ is {\it a priori} arbitrary.
Recursive solutions to the QFP
condition can be constructed by keeping 
one coupling of the scalar potential as a free
parameter.
See Sec.~V of Ref.~\cite{Gies:2018vwk} for
a general description of this recursive problem.
Various ways to search for scaling solutions and for performing the
recursive procedure are possible. 
In practice, we find it useful, to 
express all
$\lresc_n$  as a function of $\gs$ and
$\kap$, cf.~\cite{Gies:2016kkk}.

For definiteness, we concentrate in this work on solutions
exhibiting the property that $\lresc_{2}\neq 0$ at the QFP
(though this might be a scheme-dependent statement).
We now illustrate this process by considering $N_p=2$.
At this order, we set $\lambda_4=0$, such that the beta functions 
for the two ratios of \Eqref{eq:L2def} and \Eqref{eq:kappatoxikappa}
become functions of $\kap$, $\lresc_2$, $\lambda_3$, $h^2$ and $\gs^2$.
The dependence on the Yukawa coupling can be eliminated by considering the special
trajectory along which $h^2=\chis^2\gs^2$ where the QFP value for $\chis^2$ is given by \Eqref{eq:h2/gs2_QFP}.
For the SM parameters for the remaining matter content, the finite ratio $\chis^2$ takes
the value as in \Eqref{eq:h2/gs2_QFP_YQCD_DER}.
Thus the RG flow equations for $\kap$ and $\lresc_2$ within the $\MSbar$ renormalization scheme read
\begin{align}
\de_t\kap&=\left[-2-\frac{\gs^2}{12\pi^2}-\frac{3\gs^{4P}\lresc_{2}}{8\pi^2}
+\frac{\gs^{4(1-P)}}{27\pi^2 \lresc_{2}}-Q\frac{7 \gs^2}{8\pi^2}
\right.\nonumber\\
&\quad\left. -\frac{\gs^{-2Q}\kap\lambda_3}{4\pi^2}\right]\kap\, , \label{eq:xikappadotEFT}\\
\de_t \lresc_{2}&=\frac{9\gs^{4P} \lresc_{2}^2}{16\pi^2}+\frac{\gs^2\lresc_{2}}{6\pi^2}
-\frac{\gs^{4(1-P)}}{27\pi^2}+P\frac{7 \gs^2\lresc_{2}}{4\pi^2}\nonumber\\
&\quad+\left[\frac{\lresc_{2}}{\pi^2}+\frac{\gs^{4(1-2P)}}{27\pi^2\lresc_{2}}\right]
\gs^{-2Q}\kap\lambda_3\, .
\label{eq:L2dotEFT}
\end{align}
The two terms proportional to the rescaled powers $P$ and $Q$ are the contributions coming from the running of the strong gauge coupling $\gs^2$.
Its $\beta$ function in the $\MSbar$ scheme equals the flow equation within the DER, namely \Eqref{eq:beta_gs2_1loop_DER}.
Then we look for QFPs with nonnegative $\lresc_{2}$ and $\kap$ in the
$\gs^2\to 0$ limit, while leaving $\lambda_3$ as a free parameter.
It is straightforward to rediscover the \emph{Cheng--Eichten--Li} (CEL) solution, for $P=1/2$, $\kap=0$, and $\lambda_3=0$. In this case, we find two QFP solutions~\cite{Cheng:1973nv,Gies:2018vwk}
\begin{align}
	\lresc_2^\pm=\frac{1}{27}\left(-25\pm\sqrt{673}\right),\quad P=\frac{1}{2}.
	\label{eq:CEL_solution}
\end{align}
The resulting flow structure is similar to
\Eqref{eq:L2_pm_QFP_generalmodel}, and the same conclusions as
outlined below \Eqref{eq:L2_pm_QFP_generalmodel} apply.

Let us generalize our discussion by considering the case $P>1/2$ and focus on \Eqref{eq:L2dotEFT}.
For small $\gs^2$, we can neglect the terms proportional
to $\gs^{4P}$ and $\gs^2$ in comparison with $\gs^{4(1-P)}$, and we retain also $\lambda_3$.
Solving the QFP equation $\de_t \lresc_{2}=0$ for
$\lambda_3$, in the $\gs^2\to0$ limit, we obtain
\be
\lambda_3=\gs^{2(Q+2P)}\frac{\lresc_{2}}{\kap}.
\label{eq:lambda3EFTMSbar}
\ee
Inserting \Eqref{eq:lambda3EFTMSbar} into \Eqref{eq:xikappadotEFT}
and again keeping only the leading $\gs^2$ dependence
gives the QFP
\begin{align}
\lresc_{2}=\frac{1}{54\pi^2}\, , \quad\quad P=1,
\label{eq:L2EFTMSbsar}
\end{align}
and a free $\kap$.
Thus, there is a two-parameter family of AF solutions
labeled by $\kap>0$  and $Q>-2$. 
It is identified by \Eqref{eq:L2EFTMSbsar}
and
\begin{align}
\lresc_{3}=\frac{1}{54\pi^2\kap}\, , \quad\quad P_3=Q+2.
\label{eq:L3_MSbar}
\end{align}

For completeness, let us discuss the case $P<1/2$, still in the $N_p=2$ truncation.
By the same process we first solve the equation $\de_t\lresc_2=0$
for $\lresc_3$, and investigate which $\gs^2$ scaling
of this coupling might produce QFPs for $\lresc_{2}$.
Then we input such scaling with an arbitrary coefficient $ \lresc_{3}$ and search
for QFPs for $\kap$ and $\lresc_{2}$ where such couplings
are finite and nonnegative.
It turns out that, for any $P< 1/2$, there is no acceptable solution.

After having worked out the problem at order $N_p=2$, one might increase
$N_p$ and check the stability of the known solution upon inclusion of
more couplings. 
However, this returns the QFPs described by 
\Eqref{eq:L2EFTMSbsar} and \Eqref{eq:L3_MSbar}, where $\kap$ is free
and any $\lresc_{n}$ is a function of it, which separately arises as solutions
of the equation $\de_t\lresc_{n-1}=0$.
For instance, at $N_p=3$, one finds again the same solution as before,
that is \Eqref{eq:L3_MSbar}, complemented by
\begin{align}
	\lresc_{4}=\frac{\lambda_4}{\gs^{4(Q+1)}}=-\frac{1}{54\pi^2\kap^2}\, .
	\label{eq:L4_MSbar}
\end{align}
The simplest way to address the result of this recursive problem is by considering
all the $\lresc_{n}$'s at once. As we will show in the following \Secref{sec:EFT_MSbar_FRG},
this can be done by a functional approach where the full scalar potential $u(\rho)$ is accounted for.

\subsection{Non-Abelian Higgs model}\label{subsec:EFT_MSbar_NAH}

Let us apply the same strategy as before to the \NAH\ model, setting $\gs\to 0$ and $h\to0$; for simplicity, we work with $\NL=2$. We expand the dimensionless potential around a nontrivial minimum $\kappa$ as in \Eqref{eq:ExpansionAroundKappa}.
Let us choose the polynomial expansion parameter $N_p=2$ and retain the nonperturbatively renormalizable coupling $\lambda_3$ as a free parameter.
This leads to the RG flow equation for $\kappa$ and $\lambda_2$
obtained by dimensional regularization in the $\MSbar$ scheme
and for the $\SU(2)_\text{L}$ gauge group,
\begin{align}
\de_t\kappa&=\left[-2+\frac{9g^2}{32\pi^2}-\frac{3\lambda_2}{8\pi^2}-\frac{9g^4}{64\pi^2\lambda_2}-\frac{\kappa\lambda_3}{4\pi^2}\right]\kappa, 
    \label{eq:kappadotEFT_NAH}\\
\de_t\lambda_2&=\frac{3\lambda_2^2}{4\pi^2}-\frac{9\lambda_2g^2}{16\pi^2}+\frac{9g^4}{64\pi^2}+\frac{\kappa\lambda_3\lambda_2}{\pi^2}-\frac{9g^4\kappa\lambda_3}{64\pi^2\lambda_2}.     \label{eq:lambdadotEFT_NAH}
\end{align}
Since we are interested in looking for AF trajectories, we rescale the couplings
similar to those in Eqs.~(\eqref{eq:L2def}),~\eqref{eq:lambdatoLn}),~and~(\eqref{eq:kappatoxikappa}), where the strong gauge coupling $\gs$ is replaced by the gauge coupling $g$,
\begin{align}
	\lrescg_{2}=\frac{\lambda_2}{g^{4P}},\quad\quad \lrescg_{n>2}=\frac{\lambda_n}{g^{2P_n}},\quad\quad\kapg=g^{2Q}\kappa.
	\label{eq:lrescaled_NAH}
\end{align}
The corresponding RG flow equations for $\kapg$ and $\lrescg_2$ read
\begin{align}
\de_t\kapg&=\left[-2-\frac{9g^{4-4P}}{64\pi^2\lrescg_2}-\frac{\kapg\lrescg_3g^{2(P_3-Q)}}{4\pi^2}+\frac{9g^2}{32\pi^2}\right.\nonumber\\
	&\quad\left. -\frac{43Qg^2}{48\pi^2}-\frac{3\lrescg_2g^{4P}}{8\pi^2}\right]\kapg, 
    \label{eq:beta_kap_MSbar_NAH}\\
\de_t\lrescg_2&=\frac{9g^{4(1-P)}}{64\pi^2}-\frac{9\lrescg_2 g^2}{16\pi^2}+\frac{43P\lrescg_2 g^2}{24\pi^2}+\frac{3\lrescg_2^2g^{4P}}{4\pi^2}\nonumber\\
	&\quad +\frac{\lrescg_3\kapg}{64\pi^2\lrescg_2}\left[64\lrescg_2^2-9g^{4(1-2P)}\right]g^{2(P_3-Q)}.     \label{eq:beta_lresc2_MSbar_NAH}
\end{align}
Apart from the present use of the $\MSbar$ scheme, these equations generalize the ones discussed in \cite{Gies:2016kkk} by an independent $Q$ rescaling of the minimum $\kappa$.
The two terms proportional to the rescaled powers $P$ and $Q$ are the contributions coming from the running of the weak gauge coupling $g^2$.
Its $\beta$ function in the $\MSbar$ scheme equals the flow equation within the DER, namely \Eqref{eq:beta_g2_1loop_DER}.
Since we want to construct QFP solutions where $\kapg$ approaches a finite value in the UV limit, it is possible to see from \Eqref{eq:beta_kap_MSbar_NAH} that
the following three values for the rescaled powers are allowed: $P=1$, or $P_3=Q$ or $P_3=Q+2-2P$.

In the first case where $P=1$, only the first two terms in \Eqref{eq:beta_kap_MSbar_NAH}
contribute to the QFP equation $\de_t\kapg=0$ at leading order in $g^2$,
providing a constant solution for $\lrescg_2$.
Substituting $P=1$ in \Eqref{eq:beta_lresc2_MSbar_NAH}, the value of the rescaled power $P_3$ is fixed by the
relation $P_3=Q+2$ in order to have a finite $g^2\to 0$ limit for the $\beta$ function $\de_t\lrescg_2$.
To summarize this first possible solution, we have
\be
\begin{split}
	\lrescg_2&=-\frac{9}{128\pi^2},\quad &&P=1,\\
	\kapg&=-\frac{9}{128\pi^2\lrescg_3},\quad &&P_3=Q+2,
\end{split}
\label{eq:NAH_EFT_P=1}
\ee
where $\lrescg_3$ and $Q\geq -2$ remain two free parameters.
However, this solution has to be rejected since it is not compatible with our assumption to expand the potential at its minimum, i.e., $\lrescg_2>0$, in order to interpret the coefficients as couplings and mass parameters during the flow towards the UV.

Analogous considerations can be performed also for the second possibility
where $P_3=Q$. In this case, the QFP solution is
\be
\begin{split}
	\lrescg_2=\pm\frac{3}{8},\quad&&P=\frac{1}{2},\\
	\kapg=-\frac{8\pi^2}{\lrescg_3},\quad&&P_3=Q,
\end{split}
\label{eq:NAH_EFT_P=1/2}
\ee
which admits a suitable solution with a positive value for $\lrescg_2$.
In addition, the presence of a nontrivial minimum requires that $\lrescg_3<0$.
For completeness we stress that the third possibility with $P_3=Q+2-2P$
does not lead to any real solution since the QFP equation $\de_t\lrescg_2=0$ admits only complex roots
at leading order in $g^2$.

The construction generalizes to higher orders in the polynomial expansion $N_p$.
For instance, the $P=1$ solution, c.f.~\Eqref{eq:NAH_EFT_P=1}, survives and we have, for example, for $N_p=4$
\be
\begin{split}
	\lrescg_4&=\frac{9}{128\pi^2 \kapg^2},\quad&&P_4=2 Q+2,\\
	\lrescg_5&=-\frac{9}{64\pi^2 \kapg^3},\quad&&P_5=3 Q+2.
\end{split}
\label{eq:lambda_n_EFT_MSbar_NAH_P=1}
\ee
This still represents a two-parameter family of solutions with couplings $\lrescg_{n>2}$ having alternating signs for $\kapg>0$.
The solution in \Eqref{eq:NAH_EFT_P=1/2} acquires a different QFP value for $\lrescg_2$, as
its $\beta$ function receives leading-order contributions both from $\lrescg_3$ as well as from $\lrescg_4$.
For example for $N_p=4$, the solution reads
\be
\begin{split}
	\lrescg_2&=\pm\frac{\sqrt{3}}{2},\quad&&P=\frac{1}{2},\\
	\lrescg_4&=\frac{26\pi^2}{\kapg^2},\quad&&P_4=2 Q,\\
	\lrescg_5&=-\frac{187\pi^2}{2\kapg^3},\quad&&P_5=3 Q,
\end{split}
\label{eq:lambda_n_EFT_MSbar_NAH_P=1/2}
\ee
while $\lrescg_3$ and $P_3$ are still given by \Eqref{eq:NAH_EFT_P=1/2}.
This solution has again alternating signs for the higher order couplings
if the potential is in the SSB regime.

These findings motivate a full functional analysis beyond the polynomial expansion of the potential.

\section{Full effective potential in $\MSbar$}
\label{sec:EFT_MSbar_FRG}

The existence of nonpolynomial structures in the functional RG flow of the scalar potential can already be anticipated from
classic results of one-loop computations with field-dependent thresholds~\cite{Coleman:1973jx,Jackiw:1974cv}.
In this section, we stick to evaluating the loop integrals in dimensional regularization.
For examples of this procedure, see~\cite{Jack:1982hf,Ford:1992pn,Martin:2001vx}.
According to the $\MSbar$ prescription, the $\beta$ function equals the residue
of the $(d-4)^{-1}$ poles of these integrals, which can be singled out by taking
RG time derivatives followed by the $d\to 4$ limit.
Recent applications of these flow equations have shown several advantages
of dealing with functional perturbative beta functions,
see~\cite{ODwyer:2007brp,Codello:2017hhh}.

Let us begin with the general $\SU(\NL)\times\SU(\Nc)$ model with SM matter content, i.e., $\NL=2$ and $\Nc=3$. 
 The functional flow
equation for the dimensionless scalar potential at one loop in the
$\MSbar$ scheme is
\begin{align}
  \de_t u=-4 u+(2+\eta_\phi)\rho u^\prime+\frac{\omegaBr^2 +3
    \omegaBl^2 +9 \omegaW^2 -12\omegaF^2}{32\pi^2},
  \label{eq:betau_MSbar_generalscheme}
\end{align}
where we have used the Landau gauge, and $\omegaBr$ and $\omegaBl$ are
the bosonic thresholds associated to the radial Higgs fluctuation and
the three Goldstone fluctuations, 
\begin{align}
  \omegaBr=u^\prime(\rho)+2\rho u^{\prime\prime}(\rho), \quad
  \omegaBl=u'(\rho).
  \label{eq:omegaBrBl_def}
\end{align}
The arguments associated to the gauge boson and fermionic threshold
contributions are defined as
\begin{align}
  \omegaW=\frac{g^2\rho}{2},\quad
  \omegaF=h^2\rho.
  \label{eq:omegaWF_def}
\end{align}
The scalar anomalous dimension $\eta_\phi$ is given by \Eqref{eq:eta_phi_1Loop_DER}.

In the limiting case of the \HTQCD\ model, the scalar field $\phi$ is real,
therefore only the physical Higgs excitation contributes
to the scalar threshold function. Also the degrees of freedom associated
to the weak gauge bosons do not occur.
Then, the RG flow equation for $u(\rho)$ in this model reads
\begin{align}
  \de_t u=-4 u+(2+\eta_\phi)\rho u^\prime+\frac{\omegaBr^2 - 12\omegaF^2}{32\pi^2}.
  \label{eq:betau_phi^4_HTQCD}
\end{align}
On the other hand, the \NAH\ model is recovered simply by ignoring
the quantum effects arising from the fermions.
The $\beta$ function for the dimensionless potential then reads
\begin{align}
  \de_t u=&-4u+(2+\eta_\phi)\rho u^\prime+\frac{\omegaBr^2 +3 \omegaBl^2
    +9 \omegaW^2 }{32\pi^2}.
  \label{eq:betau_MSbar_NAH}
\end{align}
We already know from the previous sections that AF trajectories in the theory space
can be detected by simply looking for QFPs of the flow for rescaled couplings.
To implement this condition in a functional approach we define a rescaled field variable
$x$ and its potential $f(x)$ as
\begin{align}
  x=\gs^{2P}\rho \quad \text{or}\quad x=g^{2 P}\rho,\qquad f(x)=u(\rho).
  \label{eq:rhotox_utof}
\end{align}
The field amplitude $\rho$ is then multiplied by an appropriate power of an AF gauge
coupling, which is either the weak gauge coupling for the \NAH\ model or
the strong gauge coupling in the \HTQCD\ model. In the general model, we also use the strong gauge coupling for the rescaling.
Denoting the nontrivial minimum by $x_0$, we have
\begin{align}
  f'(x_0)&=0,		\label{eq:x0_def} \\
  f^{(n)}(x_0)&=\xi_n,\quad\text{for}\, n\geq 2. \label{eq:xin_def}
\end{align}
The arbitrary rescaling power $P$ has to be chosen as
the $P$ value corresponding to the scaling of the quartic scalar coupling such that $\xi_2=\lresc_{2}$, because we specifically look for QFPs where $\lresc_{2}$ approaches a finite value in the UV limit.
Notice that the relation
between $\xi_n$ and $\lresc_n$ (and between $x_0$ and $\kap$) at finite value of $g^2$ is a simple rescaling,
but in the $\{g^2,\gs^2\}\to 0$ limit these couplings might attain 
different fixed-point values.
Thus, the rescaling of \Eqref{eq:rhotox_utof} is
expected to be useful as long as the quartic scalar coupling
is the leading term in the approach of the scalar potential
to flatness.
According to the rescaling in \Eqref{eq:rhotox_utof}, the functional RG flow equation
for $f(x)$ is thus
\begin{align}
	\de_t f(x)\equiv \de_t f(x)|_x=\de_t u(\rho)|_\rho - P\,\eta_{\mathrm{G/W}}\, xf'(x),
\end{align}
depending on whether we use $\gs$ or $g$ to rescale the field amplitude $\rho$.
The anomalous dimensions $\eta_\mathrm{G/W}$, as well as $\eta_\phi$ and $\de_t h^2$,  in the $\MSbar$ renormalization scheme are the same
as in the DER, cf.~Eqs.~(\ref{eq:beta_g2_1loop_DER},~\ref{eq:beta_gs2_1loop_DER},~\ref{eq:beta_h2_1Loop_DER_general_NL_Nc})~and~(\ref{eq:eta_phi_1Loop_DER}), due to the vanishing of power-like
divergent diagrams.

\subsection{{\boldmath $\phi^4$}-dominance approximation}
\label{sec:phi4D_MSbar}

In order to get closer to a full functional description, we first use a simple approximation of the $\beta$ function $\de_t f(x)$ by asserting that the scalar fluctuations are dominated by the marginal quartic coupling in the UV limit.
More precisely, we assume that the scalar potential appearing in the threshold functions 
takes the form $u(\rho)=\lambda_2 \rho^2/2$.
Nevertheless, we still retain the full $u(\rho)$ dependence in the scaling term and on the left-hand side
of $\de_t u(\rho)$ as an unknown arbitrary function of $\rho$.
This assumption leads to the following approximation for the radial Higgs excitation and Goldstone fluctuations:
\begin{align}
  \omegaBr=3\lambda_2\rho,\qquad\omegaBl=\lambda_2\rho.
  \label{eq:omegaBs_phi^4}
\end{align}

\subsubsection{\SULSUc\ model}

We start with the general model, specifically considering the trajectories described
by \Eqref{eq:QFP_g2/gs2_h2/gs2_generalscheme} along which
the top-Yukawa and weak gauge coupling become
proportional to $\gs$ in the UV limit.
This yields a $\beta$ function for the
rescaled scalar potential $f(x)$ that depends only on the AF strong gauge coupling $\gs$,
\begin{align}
  \de_t f&= - 4 f+ d_x x f^\prime
  +\frac{3\, x^2}{128\pi^2}\biggl[16\xi_2^2\gs^{4P}\nonumber\\
        &\quad
    -\Bigl(16\hat{h}^4_*-3\hat{g}^4_* \Bigr)\gs^{4-4P}\biggr],
  \label{eq:dotf_MSbar_phi^4_generalmodel}
\end{align}
where the scaling dimension $d_x$ of the rescaled field includes also a contribution from the running of the strong gauge coupling, in fact
\begin{align}
  d_x=2+\eta_\phi - P\etaG \equiv 2+\eta_x,
  \label{eq:dx_def_generalmodel}
\end{align}
where $\etaG$ is given by \Eqref{eq:beta_gs2_1loop_DER}. 
The QFP solutions for the ratios $\hat{g}^2_*$ and $\hat{h}^2_*$
are given by \Eqref{eq:QFP_g2/gs2_h2/gs2_generalscheme}.
The QFP equation, which is obtained by the requirement that 
the left-hand side of \Eqref{eq:dotf_MSbar_phi^4_generalmodel} is vanishing,
is solved by
\begin{align}
  f(x)&= C_fx^{4/d_x} - \frac{3\, x^2}{256\pi^2\eta_x}\biggl[16\xi_2^2\gs^{4P}\nonumber\\
    &\quad-\Bigl(16\hat{h}^4_*-3\hat{g}^4_* \Bigr)\gs^{4-4P}\biggr],
  \label{eq:f(x)_phi^4_MSbar_generalmodel}
\end{align}
where $C_f$ is a free integration constant, parameterizing the 
general  solution for the associated homogeneous equation.
Setting $C_f=0$ and requiring the consistency condition $f''(0)=\xi_2$
singles out the same solution with $P=1/2$ and $\xi_2=\lresc_{2}^+$ as it was
found in \Secref{sec:AF_in_standard_perturbation_theory}, cf. \Eqref{eq:L2_pm_QFP_generalmodel}.
For any nonvanishing $C_f$ the QFP potential behaves as a nonrational power of $x$ at the origin;
therefore, its second order derivative at $x=0$ is singular for $\eta_{x}>0$. 
If the system is in the SYM regime, the anomalous dimension for the rescaled field is indeed positive in the \HTQCD\ model for all values of $P$. Hence, the singularity would affect large classes of correlation functions expanded about the symmetric ground state, such that we consider such solutions as unphysical. On the other hand, in the general model and the \NAH\ model, $\eta_x$ can be negative for small enough values of $P$, because of the negative gauge-loop contribution entering in $\eta_\phi$.
	
The problematic singular behavior at the origin might be avoided in all models if
there is at least one nontrivial minimum for $f(x)$, in the spirit of the
Coleman-Weinberg mechanism~\cite{Coleman:1973jx}.  In fact, the system
of two equations that arises by setting $n=2$ in \Eqref{eq:xin_def}
can be solved for $C_f$ and $\xi_2$ as functions of $x_0$.  The
additional requirement that $\xi_2$ is finite and positive in the
$\gs^2\to0$ limit can be fulfilled only when $P=1$.
The expressions for $C_f$ and $\xi_2$ at leading order in $\gs^2$ are
\be
\begin{split}
  C_f&=-\frac{3 (16 \hat{h}^4_*-3 \hat{g}^4_* )}{256\pi^2}\left[\frac{1}{\eta_x}+\frac{1+2\log x_0}{2}\right],\\
  \xi_2&= \frac{3 (16 \hat{h}^4_*-3 \hat{g}^4_* )}{128\pi^2}>0,\qquad P=1.
  \label{eq:phi4_MSbar_generalmodel}
\end{split}
\ee  
If $x_0$ attains a finite value in the $\gs^2\to 0$ limit,
this corresponds to a potential that has a finite minimum as well as finite derivatives at
this minimum, which are given by
\begin{align}
  \xi_n= (-1)^{n+1}\frac{3 (16 \hat{h}^4_*-3 \hat{g}^4_* )}{128\pi^2}\frac{(n-3)!}{x_0^{n-2}},\qquad n\geq 3.
  \label{eq:xin_MSbar_generalmodel}
\end{align}
We can thus construct a family of solutions parametrized by the nontrivial
minimum $x_0$ with the desired property that the rescaled quartic
coupling at $x_0$ is finite in the UV limit.
This is in fact a two-parameter family of solutions,
as \Eqref{eq:xin_MSbar_generalmodel}
is compatible with an arbitrary asymptotic scale 
dependence of $x_0$ of the form
\begin{align}
x_0=\gs^{2 (P-Q)}\kap=\gs^{2 (1-Q)}\kap.
\label{eq:defQ}
\end{align}
The appearance of the additional parameter $Q$
occurs as in the EFT analysis of the previous section.
More details are provided in the following
 for the specific case of the {\HTQCD} model.

Evidence for the global stability of the scalar potential $f(x)$ can be obtained by studying the asymptotic behavior for large amplitudes $x$. In fact, the flow equation allows to study two different asymptotic limits, both corresponding to large amplitudes and small gauge coupling, but differing by the product $\gs^{2P} x$ being either small or large.
The former asymptotic region is addressed by taking first the $\gs^2\to 0$ limit and then the $x\to\infty$ limit, where
we find the following asymptotic behavior
\begin{align}
  f(x)\widesim[2]{x\to\infty}x^2 \frac{3 (16 \hat{h}^4_*-3 \hat{g}^4_* )}{128\pi^2}\frac{1}{4} \left[-1+2\log\left(\frac{x}{x_0}\right)\right].
\end{align}
The latter asymptotic regime is obtained by the opposite order, yielding
\begin{align}
  f(x)\widesim[2]{x\to\infty} \frac{3 (16 \hat{h}^4_*-3 \hat{g}^4_*)}{128\pi^2} \frac{x^2}{2\eta_x}>0.
\end{align}
In both regimes, we find a stable potential, providing evidence for global stability.

\subsubsection{\HTQCD\ model}

Within the $\phi^4$-dominance approximation, we can address the limiting case of the \HTQCD\ model by substituting the expressions
in \Eqref{eq:omegaBs_phi^4} into the RG flow equation \eqref{eq:betau_phi^4_HTQCD} for the scalar potential
\begin{align}
	 \de_t f&= - 4 f+ d_x x f^\prime
	 +\frac{3\, x^2}{32\pi^2}\left[3\xi_2^2\gs^{4P} - 4\hat{h}^4_*\gs^{4-4P}\right],
	\label{eq:dotf_MSbar_phi^4_YQCD}
\end{align}
where the QFP solution for the rescaled top-Yukawa coupling assumes
the value as in \Eqref{eq:h2/gs2_QFP_YQCD_DER}.
The QFP equation $\de_t f=0$ is solved by
\begin{align}
	f(x)= C_fx^{4/d_x} - \frac{3\, x^2}{64\pi^2\eta_x}\left[3\xi_2^2\gs^{4P} - 4\hat{h}^4_*\gs^{4-4P}\right],
	\label{eq:f(x)_phi^4_MSbar_YQCD}
\end{align}
where $C_f$ is again a free integration constant, parameterizing the 
general  solution for the associated homogeneous equation.
Setting $C_f=0$ and requiring the consistency condition $f''(0)=\xi_2$ singles out the
CEL solution with $P=1/2$ and $\xi_2=\lresc_{2}^\pm$ as for \Eqref{eq:CEL_solution}.

As discussed in the general model, the potential has a log-type
singularity in the second derivative at the origin for any $C_f\neq 0$, 
as $\eta_{x}$ is always positive in this model. This problem can be avoided if $f(x)$
admits a nontrivial minimum $x_0$. Keeping $x_0$ as a parameter, it is possible to solve
for $C_f$ and $\xi_2$, which are
\be
\begin{split}
		C_f&= - \frac{3\hat{h}^4_*}{16\pi^2}\left[\frac{1}{\eta_x}+\frac{1+2\log x_0}{2}\right],\\
	 	\xi_2&= \frac{3  \hat{h}^4_*}{8\pi^2}>0,\qquad P=1,
	\label{eq:Cf_xi2_phi^4l_MSbar_YQCD}
\end{split}
\ee
to leading order in $\gs^2$.
As for the general model, the rescaled quartic coupling $\xi_2$ can be finite only for $P=1$.
Moreover, the higher-order couplings at the nontrivial minimum are
\begin{align}
	\xi_n= (-1)^{n+1}\frac{3  \hat{h}^4_*}{8\pi^2}\frac{(n-3)!}{x_0^{n-2}},\qquad n\geq 3.
	\label{eq:xin_phi^4_MSbar_YQCD}
\end{align}
We can thus construct a one-parameter family of solutions which enjoy all the desired
properties usually expected for a QFP potential. Their singular behavior at vanishing
field values makes them invisible in an expansion for small field amplitudes.
To recover the two-parameter family of solutions 
observed in~\Secref{sec:EFT_MSbar_HTQCD},
it is sufficient to notice that \Eqref{eq:xin_phi^4_MSbar_YQCD}
still holds if $x_0$
scales as in \Eqref{eq:defQ}.
Inserting the latter scaling into \Eqref{eq:xin_phi^4_MSbar_YQCD}, we would find precisely
the results shown in \Eqref{eq:L3_MSbar} and \Eqref{eq:L4_MSbar},
as well as the predictions for all higher-order couplings
\be
\lresc_{n}=\frac{\lambda_n}{\gs^{2(n-2)Q+4}},\qquad n\geq 3,
\ee
which can be verified within the EFT approach.

Furthermore, the large-field behavior for any $\gs^2>0$ is
\begin{align}
	f(x)\widesim[2]{x\to\infty} \frac{3\hat{h}^4_*}{16\pi^2\eta_x} x^2>0,
\end{align}
whereas in the other asymptotic regime where the limit $\gs^2\to 0$ is taken before considering the $x\to\infty$ limit, the large-field behavior reads
\begin{align}
  f(x)\widesim[2]{x\to\infty}x^2 \frac{3 \hat{h}^4_*}{32\pi^2} \left[-1+2\log\left(\frac{x}{x_0}\right)\right].
\end{align}
In both cases, the potential appears stable.

\subsubsection{Non-Abelian Higgs model}

The limiting case of the \NAH\ model can be recovered from \Eqref{eq:dotf_MSbar_phi^4_generalmodel} by the substitutions
\begin{align}
	\gs^2\to g^2 \implies \hat{g}^2_*=1,\quad \hat{h}^2_*\to 0.
	\label{eq:from_generalmodel_to_NAH}
\end{align}
The RG flow equation for the rescaled potential $f(x)$ then becomes
\begin{align}
	\de_t f&= - 4 f+ d_x x f^\prime
	+\frac{3 \left(16\xi_2^2 g^{4P}+3g^{4-4P}\right)}{128\pi^2} x^2.
	\label{eq:dotf_MSbar_phi^4_NAH}
\end{align}
The quantum dimension $d_x$ includes a contribution from
the anomalous dimension of the gauge vector fields, since the scalar amplitude $\rho$ is
rescaled with the weak gauge coupling $g^2$:
\begin{align}
	d_x=2+\eta_\phi - P\etaW \equiv 2+\eta_x.
	\label{eq:dx_def_NAH}
\end{align}
The QFP solution of \Eqref{eq:dotf_MSbar_phi^4_NAH} is
\begin{align}
	f(x)=C_f x^{4/d_x}-\frac{3\left(16\xi_2^2g^{4P}+3 g^{4-4P}\right) x^2}{256\pi^2\eta_x},
	\label{eq:QFP_phi^4_MSbar_NAH}
\end{align}
that features a log-type singularity in the second derivative
at the origin $f''(0)$, as long as the integration constant $C_f$
is different from zero and $\eta_x$ positive.

In contrast to the general \SULSUc\ model or the \HTQCD\ model,
there is no real solution compatible with the consistency
condition $f''(0)=\xi_2$ for $C_f=0$. This reflects the conventional conclusion of triviality as seemingly evidenced by Landau-pole singularities in perturbation theory.
A different situation occurs for $C_f\neq0$. In this case indeed, the second derivative at the nontrivial minimum is finite only for $P=1$ and takes the value
\begin{align}
	\xi_2=-\frac{9}{128\pi^2},\qquad P=1,
\end{align}
to leading order in $g^2$.
As this is negative, it contradicts one of our selection criteria.
We can moreover find a recursive formula for all the higher-order couplings which is
\begin{align}
	\xi_n=(-1)^{n}\frac{9}{128\pi^2}\frac{(n-3)!}{x_0^{n-2}},\qquad n\geq 3.	\label{eq:xi_n_phi^4_MSbar_NAH}
\end{align}
This is in agreement with the solution found within the EFT approximation in the $\MSbar$ scheme. In fact, if we express the latter equation in terms of the finite rescaled couplings $\lrescg_n$ and $\kapg$ as in \Eqref{eq:lrescaled_NAH}, we would find
\be
\begin{split}
	\lrescg_{n}&=(-1)^{n-2}\frac{9(n-3)!}{128\pi^2 \kapg^{n-2}},
	\quad\text{for}\quad n\geq 3,\\
	P_n&=(n-2)Q+2,
	\label{eq:lresc_n_phi^4_MSbar_NAH}
\end{split}
\ee
which coincide with
Eqs.~\eqref{eq:NAH_EFT_P=1} and \eqref{eq:lambda_n_EFT_MSbar_NAH_P=1}.

Working out the behavior of the potential in the two asymptotic regions, we find in the intermediate asymptotic regime,
taking first the $g^2\to 0$ and then the $x\to\infty$ limit
\begin{align}
  f(x)\widesim[2]{x\to\infty} x^2 \frac{9}{128\pi^2}\frac{1 - 2\log(x/x_0)}{4},
\end{align}
while in the opposite order yields the large-field asymptotics
\begin{align}
  f(x)\widesim[2]{x\to\infty} - \frac{9 }{256\pi^2\eta_x} x^2 < 0.
\end{align}
Both asymptotic regions reveal that the potential is not stable.

	We conclude this subsection on the \NAH\ model by comparing the present $\phi^4$-dominance approximation with the EFT analysis:
	Within both approximations we discovered the $P=1$ solution,
 which has to be rejected as it involves a negative quartic coupling which violates our assumptions.
 Only the EFT analysis can reveal the acceptable solution associated to $P=1/2$, 
 c.f. \Eqref{eq:lambda_n_EFT_MSbar_NAH_P=1/2}.
	In the latter case, we have indeed observed that the contributions from the higher-order couplings $\lrescg_{3}$ and $\lrescg_{4}$ were crucial for finding the QFP value for $\lrescg_{2}$.
	We expect that neglecting the presence of these interaction terms within 
		the loops results in the impossibility to reveal these additional AF solutions
		in the present context.
	This illustrates the limitations of the non-systematic but nevertheless useful $\phi^4$-dominance approximation.

\subsection{Weak coupling expansion}
\label{sec:weakh_exp_MSbar}

In order to abandon the assumption about the dominance of the local four-point interaction, we now perform a parametrically controlled functional weak-coupling analysis.
For this, we neglect the subleading corrections in powers of
the gauge couplings to the UV-asymptotic behavior of $f(x)$,
and expand its $\beta$ function for weak coupling.
We first write the full flow equations of $f(x)$ at one loop
in the $\MSbar$ renormalization scheme for all the models under investigation.
In the general case, the RG flow equation for the rescaled scalar potential reads, cf. \Eqref{eq:betau_MSbar_generalscheme},
\begin{align}
	\de_t f=-4 f+d_x x f^\prime+\frac{\omegafBr^2 +3 \omegafBl^2
		+9 \omegafW^2 -12\omegafF^2}{32\pi^2},
	\label{eq:betaf_MSbar_generalmodel}
\end{align}
where the arguments of the threshold functions can be obtained from Eqs.~(\ref{eq:omegaBrBl_def}) and (\ref{eq:omegaWF_def}) after having rescaled the field amplitude $\rho$ according to \Eqref{eq:rhotox_utof}. We thus have
\begin{subequations}
\begin{align}
  \omegafBr&=\gs^{2P}(f'+2xf''), & \omegafBl&=\gs^{2P} f',\\
\omegafF&=\hat{h}^2_*\gs^{2-2P}x, & \omegafW&=\hat{g}^2_*\gs^{2-2P}\frac{x}{2},
\end{align}
\label{eq:omegaf_def}
\end{subequations}
where as usual we consider the special trajectories along which $h^2$ and $g^2$
become proportional to $\gs^2$. These are characterized by the QFP values in \Eqref{eq:QFP_g2/gs2_h2/gs2_generalscheme}. The full quantum dimension of $x$ is given by \Eqref{eq:dx_def_generalmodel}.

In the \HTQCD\ model, the degrees of freedom associated to the gauge bosons
and the Goldstone mode are not present thus the beta function of $f(x)$ becomes
\begin{align}
\de_t f=-4 f+d_x x f^\prime+\frac{\omegafBr^2 -12\omegafF^2}{32\pi^2}.
\label{eq:betaf_MSbar_HTQCD}
\end{align}
Also the QFP for $\hat{h}^2_*$ has to be changed and takes the value as in \Eqref{eq:h2/gs2_QFP_YQCD_DER}.

The \NAH\ model can be recovered from the general case by simply performing
the substitutions in \Eqref{eq:from_generalmodel_to_NAH}, thus $\de_t f$ reduces to
\begin{align}
	\de_t f=-4 f+d_x x f^\prime+\frac{\omegafBr^2 +3 \omegafBl^2
	+9 \omegafW^2}{32\pi^2},
\label{eq:betaf_MSbar_NAH}
\end{align}
where $d_x$ is given by \Eqref{eq:dx_def_NAH}.

Since the scalar, fermion and gauge boson loops appear with different
powers of the gauge couplings, we distinguish several cases
corresponding to the classification of leading-order terms in the UV
limit where $\{\gs^2,g^2\}\to0$ but $x$, $f'(x)$, and $f''(x)$ stay
finite.

If $P>1/2$, the bosonic contributions arising from the radial and/or Goldstone
fluctuations are negligible with respect to the fermionic and/or gauge boson fluctuations.
Also the $\{\gs^2,g^2\}$ dependence of $d_x$ is negligible which can nevertheless be
easily accounted for.
Under these approximations, the QFP equations $\de_t f=0$ for the different models
reduce to the flow equations obtained within the $\phi^4$-dominance approximation
if we set $\xi_2=0$.
As a consequence, the weak-coupling expansions of Eqs.~(\ref{eq:betaf_MSbar_generalmodel},~\ref{eq:betaf_MSbar_HTQCD})~and~ (\ref{eq:betaf_MSbar_NAH}) for $P>1/2$
agree with the approximation made in Sec.~\ref{sec:phi4D_MSbar} 
as far as the UV limits $\{\gs^2,g^2\}\to 0$ are concerned.

For $P<1/2$, the effects from the fermion and/or gauge boson loops are negligible and
the flow of the scalar potential is the same as the flow of a purely scalar
quantum field theory, where the dimension of the field is externally driven towards the canonical one as the classical sources given by $\gs^2$ or $g^2$ vanish.
In this case, as well as in the $P=1/2$ case, 
the QFP equation remains nonlinear and of second order,
and it does not offer straightforward analytical solutions.
As such, the present approximation is not helpful, and does not offer
better perspectives with respect to the results of the EFT-like analysis
in \Secref{sec:EFT_MSbar_FRG}.

This concludes our first analysis of a possible existence of further
AF trajectories within the well-known perturbative
$\MSbar$ scheme. To summarize for example the \HTQCD\ model:
in addition to the well-known CEL solution, we have found evidence for a family of further AF trajectories.
Let us
substantiate these findings by a more comprehensive analysis also addressing the
question of scheme dependence further.

\section{Renormalization in mass-dependent IR schemes} \label{sec:FRG}

The RG equations for action functionals as obtained from a
masslike scale-dependent deformation of the Gau{\ss}ian part of the action have been known for a long time~\cite{Wilson:1973jj,Wegner:1972ih}.
For the purpose of
extending the analysis of the scheme
dependence of AF solutions, 
we use the exact RG flow equation for the one-particle irreducible
effective average action $\Gamma_k$~\cite{Wetterich:1992yh} at an RG scale $k$, given by the Wetterich equation~\cite{Wetterich:1992yh,Ellwanger:1993mw,Morris:1993qb,Bonini:1992vh}
\begin{align}
\de_t\Gamma_k[\Phi]=\frac{1}{2}\text{STr}\left(\frac{\de_t R_k}{\Gamma^{(2)}_k[\Phi]+R_k}\right) ,	\label{eq:WetterichEQ}
\end{align}
where $t=\log k$ is the RG time. Here, $\Gamma_k^{(2)}[\Phi]$ denotes the second functional derivative with respect to the collective field variable $\Phi$. The function $R_k$ encodes a
general IR regularization of momentum modes near the scale $k$.
The derivative $\partial_t R_k$ in the numerator provides for a  UV regularization. The detailed form of $R_k$ therefore defines a regularization scheme within the FRG approach. Results that hold for any physically admissible regulator therefore provide evidence for scheme independence.
A solution $\Gamma_k$ to the \Eqref{eq:WetterichEQ} interpolates between the initial condition at some UV scale $\Lambda$, $\Gamma_{k=\Lambda}=S_\text{cl}$ in the form of a classical action, and the effective action $\Gamma_{k=0}=\Gamma$ generating the 1PI correlation functions of the full quantum theory, see \cite{Berges:2000ew,Pawlowski:2005xe,Gies:2006wv,Delamotte:2007pf,Braun:2011pp,Nagy:2012ef} for reviews.

In the following, we focus on the beta functional for a general scalar potential. For this, we solve \Eqref{eq:WetterichEQ} on a projected theory space spanned by the truncated action:
\begin{align}
\Gamma_k&=\int_x\left[\frac{\ZW}{4}F^{\quad}_{i\mu\nu}F_i^{\mu\nu}+\frac{\ZG}{4}G^{\quad}_{I\mu\nu}G_I^{\mu\nu}\nonumber\right.\\
&\quad+Z_\phi(D_\mu\phi)^{\dagger a}(D^\mu\phi)^a+U(\phi^\dagger\phi)\nonumber\\
&\quad+\ZL\barpsiL^{aA}\I\slashed{D}^{abAB}\psiL^{bB}+\ZR\barpsiR^A\I\slashed{D}^{AB}\psiR^B\nonumber\\
&\quad\left. +\I\bar h (\barpsiL^{aA}\phi^a\psiR^A+\barpsiR^A\phi^{\dagger a}\psiL^{aA})+L_\text{gf}+L_\text{gh}\right].
\label{eq:Gamma_k}
\end{align}
All couplings, wave function renormalizations $Z$, and the effective potential $U$ are $k$ dependent. This truncated theory space can be viewed as a leading-order derivative expansion of the action in terms of local operators which has been proven useful, e.g., in the analysis of the RG flow of the Higgs potential \cite{Gies:2013pma,Gies:2013fua,Gies:2014xha,Eichhorn:2014qka,Eichhorn:2015kea,Gies:2015lia,Jakovac:2015iqa,Jakovac:2015kka,Vacca:2015nta,Borchardt:2016xju,Gies:2016kkk,Jakovac:2017nsi,Gies:2017zwf,Gies:2017ajd,Sondenheimer:2017jin,Held:2018cxd}.

For simplicity, we refer to the $\NL=2$ case for the remainder of this section,
but we will consider $\Nc$ and the spacetime dimension $d$ as arbitrary parameters.
We use a gauge-fixing Lagrangian $L_\text{gf}$ of the general form
\begin{align}
	L_\text{gf}=\frac{\ZW}{2\zeta}\mathsf{F}_{i}^{*}\mathsf{F}_i^{\phantom{*}}+\frac{\ZG}{2\zeta_\text{s}}\mathsf{F}_{\text{s}I}^{*}\mathsf{F}_{\text{s}I}^{\phantom{*}},
\end{align}
where $\mathsf{F}_i$ and $\mathsf{F}_{\text{s}I}$ are the gauge-fixing conditions for the weak and the strong gauge group, respectively.
The corresponding gauge-fixing parameters are $\zeta$ and $\zeta_\text{s}$; below, we mostly quote results obtained in the Landau gauge, $\zeta, \zeta_{\text{s}}\to 0$.
Moreover, the ghost Lagrangian $L_\text{gh}$
\begin{align}
	L_\text{gh}=-\bar{c}_i M_{ij}c_{j} - \bar{b}_{I} M_{\text{s}IJ}b_{J},
\end{align}
where $c_i,\bar{c}_i,b_{I}$, and $\bar{b}_{I}$ are the ghost fields, encodes the determinants of the Faddeev-Popov operators
\begin{align}
	\mathcal{M}_{ij}=\frac{\delta\mathsf{F}_i}{\delta\alpha_j},\quad \mathcal{M}_{\text{s}IJ}=\frac{\delta\mathsf{F}_{\text{s}I}}{\delta\alpha_{\text{s}J}},
	\label{eq:Faddeev-Popov_operator_def}
\end{align}
where $\alpha_j$ and $\alpha_{\text{s}J}$ are the local parameters for the finite gauge transformations in Eqs.~(\ref{eq:SUL_trasformation_phi}) and (\ref{eq:SULSUNc_trasformation_psiLR}).
In order to take into account also the threshold effects coming from the SSB regime, we decompose the scalar field into the bare vev $\bar{v}$ and the fluctuations around it.
Without loss of generality we choose the radial mode in the first real component such that~\cite{Gies:2013pma,Gies:2016kkk}:
\begin{align}
	\phi=\frac{1}{\sqrt{2}}
	\begin{pmatrix} 	\bar{v}\\0 \end{pmatrix}
	+
	\frac{1}{\sqrt{2}}\begin{pmatrix} H+\I\theta_3\\
						\theta_2+\I\theta_1 \end{pmatrix},
	\label{eq:phi_splitting_SSB}
\end{align}
where the radial fluctuation $H$ corresponds to the Higgs excitation and the Goldstones form a triplet.
We choose the gauge-fixing functional for the $\SU(2)_\text{L}$ gauge group such that no mixing terms between the Goldstone modes and the gauge bosons appear in the propagators,
\begin{align}
	\mathsf{F}_i=\de_\mu W^\mu_i - \I\bar{g}\bar{v}\zeta\frac{Z_\phi}{\ZW}\left[ t_i^{12}\theta_{2}+\I t_i^{12}\theta_{1} +\I t_i^{11}\theta_{3}\right].
	\label{eq:gauge_fixing_SU(NL)} 
\end{align}
As usual, the Higgs excitation $H$ is not included,
thus the gauge-fixing condition involves only the Goldstone bosons and not the radial mode.
From the definition in \Eqref{eq:Faddeev-Popov_operator_def}, we identify the Faddeev-Popov operator
\begin{align}
	\mathcal{M}_{ij}&= - \Bigl[ \square + \frac{1}{4} g^{2}\bar{v}^{2}\zeta\frac{Z_\phi}{\ZW} \Bigr] \delta_{ij} - \bar{g}f_{ijk}\de_\mu W^\mu_k\nonumber\\
	&\quad+\bar{g}^2\bar{v}\zeta\frac{Z_\phi}{\ZW}
	\Bigl[t_i^{12}t_j^{21}H +t_i^{12}t_j^{22}\theta_2\nonumber\\
	&\quad+\I t_i^{1a} t_j^{a1}\theta_3+\I t_i^{1a} t_j^{a2}\theta_1\Bigr].
	\label{eq:Faddeev-Popov_operator_SU(NL)}
\end{align}
In the gluon sector, we use standard Lorenz gauge, such that the gauge-fixing functional for the $\SU(\Nc)$ gauge group and the corresponding Faddeev-Popov operator read
\begin{align}
	\mathsf{F}_{\text{s}I}=\de_\mu G^\mu_{I},\quad \mathcal{M}_{\text{s}IJ}=-\square \delta_{IJ}-\bargs f_{IJK}\de_\mu G^\mu_{K}.
	\label{eq:gauge_fixing_SU(NC)}
\end{align}
Let us introduce also the mass parameters for the elementary fields of the Lagrangian in the SSB regime.
The unrenormalized mass matrix for the gauge boson fields is
\begin{align}
	\bar{m}_{\text{W}\, ij}^{2}=\frac{Z_\phi}{2}\bar{g}^2\bar{v}^2\{t_i,t_j\}_{11},
\end{align}
where $\{\cdot,\cdot\}$ denotes the anticommutator.
The generators for the $\SU(2)_\text{L}$ gauge group are $t^i=\sigma^i/2$, therefore all gauge bosons acquire the same mass,
\begin{align}
	\bar{m}_{\text{W}\,ij}^{2}=\frac{Z_\phi}{4}\bar{g}^2\bar{v}^2\delta_{ij}.
\end{align}
Introducing the decomposition as in \Eqref{eq:phi_splitting_SSB}, we recover the following formulas for the mass of the scalar fluctuations,
\begin{align}
	m_{\mathrm{H}}^2 &= \left[
	U'(\phi^\dagger\phi)+
	\bar{v}^2 U''(\phi^\dagger\phi)
	\right]_{\phi^\dagger\phi=\bar{v}^2/2}, \\
	m_{\theta_{i}}^2 &= \left[U'(\phi^\dagger\phi)\right]_{\phi^\dagger\phi=\bar{v}^2/2} + \frac{Z_{\phi}^{2}}{\ZW^{2}}\frac{\bar{g}^{2}\bar{v}^{2}}{4}\zeta^{2}.
\end{align}
In the SYM regime, where the minimum of the potential is $\bar{v}=0$, all the scalar fluctuations acquire the same mass.
By contrast, in the SSB regime where $\bar{v}\neq 0$ and by definition $U'(\bar{v}^2/2)=0$, only the radial fluctuation $H$ becomes massive
which corresponds precisely to the Higgs excitation.
The angular fluctuations are instead massless in the Landau gauge and correspond to the Goldstone modes,
\begin{align}
	&m_{\mathrm{H}}^2=\bar{v}^2 U''(\bar{v}^2/2),\qquad m_{\theta_{i}}^2=0.
\end{align}
Furthermore the unrenormalized mass for the top quark is given by
\begin{align}
	\bar{m}_\text{t}=\frac{\bar{h}\bar{v}}{\sqrt{2}}.
\end{align}

At this point we would like to emphasize that these parameters of the elementary fields do not necessarily have to coincide with observables of the theory in the IR. 
Also the term spontaneous symmetry breaking is misleading, although often used in this context, as a local gauge symmetry cannot be spontaneously broken \cite{Elitzur:1975im}. Moreover,  the vev of the Higgs field is not a reliable order parameter \cite{Osterwalder:1977pc,Fradkin:1978dv} as it depends on the gauge choice even if the potential has a Mexican hat-type form \cite{Maas:2012ct}. 
To formulate the spectrum of a theory with a Brout-Englert-Higgs (BEH) effect in a gauge-invariant manner is cumbersome on a nonperturbative level due to the Gribov-Singer ambiguity \cite{Gribov:1977wm,Singer:1978dk,DellAntonio:1991mms,vanBaal:1991zw,vanBaal:1997gu,Maas:2011se}. 
This ambiguity states that commonly used gauge-fixing conditions like \eqref{eq:gauge_fixing_SU(NL)} and \eqref{eq:gauge_fixing_SU(NC)} are insufficient to fully fix the gauge. 
However, how this problem affects a BEH theory is still under investigation \cite{Lenz:1994tb,Maas:2010nc,Capri:2012ah,Capri:2013oja,Capri:2013gha}. 

That the elementary fields are not observable quantities is a consequence of the Gribov-Singer problem. 
Nevertheless, gauge-invariant approaches have been developed to formulate the spectrum appropriately \cite{Chernodub:2008rz,Masson:2010vx,Ilderton:2010tf}. Describing the observables in terms of gauge-invariant bound states, proposed by Fr\"ohlich, Morchio, and Strocchi, is an useful procedure \cite{Frohlich:1981yi,Frohlich:1980gj}. First, it can easily be generalized to other gauge groups \cite{Maas:2016ngo,Maas:2016qpu,Maas:2017xzh,Maas:2017wzi,Maas:2018xxu}. Second, it explains why the perturbative description of the spectrum of the weak sector of the standard model is so successful by using a one-to-one mapping of the symmetry structures among the bound states to the weak symmetry group. 
Thus, we will stick with the standard nomenclature (SYM, SSB, mass, ...) throughout this paper as we will concentrate on $\NL=2$, keeping in mind that actually not the vev breaks the gauge symmetry but the gauge fixing term and that the gauge-variant objects can be used to describe gauge-invariant observables with high precision for the weak sector of the standard model.

Since we are interested in FPs where the model asymptotically features 
a self-similar behavior, we study the RG flow for
the renormalized dimensionless quantities.  
Let us introduce therefore
the dimensionless renormalized
U($\NL$)-invariant scalar field amplitude
\begin{align}
	&\rho = Z_\phi \frac{\phi^{\dagger a}\phi^a}{k^{d-2}},
\end{align}
and the dimensionless renormalized couplings
\begin{align}
	h^2=\frac{\bar{h}^2 k^{d-4}}{Z_\phi \ZL\ZR },\quad g^2=\frac{\bar{g}^2k^{d-4}}{\ZW},\quad \gs^2=\frac{\bargs^2k^{d-4}}{\ZG}.
\end{align}
Inserting our truncation of the effective average action \Eqref{eq:Gamma_k} into the Wetterich equation \eqref{eq:WetterichEQ} and projecting onto the scalar sector allows to extract the RG flow equation for the dimensionless potential
\begin{align}
u(\rho)=k^{-d}U(Z_\phi^{-1}k^{d-2}\rho).
\end{align}
In a similar manner, the $\beta$ function for the dimensionless renormalized top-Yukawa coupling can be extracted.
Similarly, we can obtain the anomalous dimensions for the fields which are defined as
\begin{eqnarray}\begin{split}
	\eta_\phi&=-\de_t\log Z_\phi,& \quad\etaW&=-\de_t\log \ZW,\\
	\etaL&=-\de_t\log \ZL,& \quad\etaR&=-\de_t\log \ZR,\\
	\etaG&=-\de_t\log \ZG,
\end{split}
\end{eqnarray}
encoding the running of the scale-dependent wave function renormalizations.
The functional flow equation for the dimensionless renormalized potential in the
Landau gauge is given by~\cite{Gies:2018vwk,Gies:2013pma}
\begin{align}
	\de_t u &= - du+(d-2+\eta_\phi)\rho u^\prime +2 v_d\left\{ l_0^{(\mathrm{H})d}(\omegaBr,\eta_\phi)\right. \notag\\
	&\quad+3l_0^{(\theta)d}(\omegaBl,\eta_\phi)+3(d-1)l_0^{(\mathrm{W})d}(\omegaW,\etaW)\nonumber\\
	&\quad\left. - 4\Nc l_0^{(\mathrm F)d}(\omegaF,\eta_\psi)\right\},		
	\label{eq:betau}
\end{align}
where $v_d^{-1}=2^{d+1}\pi^{d/2}\Gamma(d/2)$ and the arguments of the threshold functions
$\omegaBr$, $\omegaBl$, $\omegaW$ as well as $\omegaF$ have already been defined in Eqs.~(\ref{eq:omegaBrBl_def},~\ref{eq:omegaWF_def}).
Let us remark here that additional contributions coming from the ghost loop,
the gluon loop, and the 
bottom-quark loop contribute only to the running of the $\rho$-independent vacuum energy and thus can be ignored for our present purpose.
From  \Eqref{eq:betau}, we can extract also the flow equation for the nontrivial minimum $\kappa$ in the SSB regime,
\begin{align}
	\kappa=\frac{Z_\phi\bar{v}^2}{2k^{d-2}},\quad\de_t\kappa=-\frac{\de_t u'(\rho)}{u''(\rho)}\bigg{|}_{\rho=\kappa}.\label{eq:beta_kappa}
\end{align}
The threshold functions $l_0^{(\Phi)}(\omega)$, with $\Phi\in\{\mathrm{H,\theta,F,W}\}$, carry the dependence on the momentum-space regularization of loop integrals specified by the form of $R_k$. Physically, they quantify how  massive modes decouple from the flow, once the RG scale crosses the mass threshold.
For their general definitions see the discussion in \Appref{app:threshold}.
From the $\beta$ functional \eqref{eq:betau} for the scalar potential,
the RG flow for the scalar self-couplings can straightforwardly be
derived to any order by polynomial expansion. More generally, 
\Eqref{eq:betau}
encodes the flow of the global properties of the Higgs potential to be
studied below.

Similar FRG flow equations for the Yukawa coupling and for the gauge
coupling, as well as FRG expressions for the anomalous dimensions
 of the fields,
 are presented in \Appref{app:FRGEs}.

\section{Full effective potential in the weak-coupling expansion in a general scheme} \label{sec:weakhexpansion}

%

Let us discuss the analytic weak-coupling expansion of the full functional flow; i.e., we expand the full functional equation for the rescaled potential $f(x)$, defined in \Eqref{eq:rhotox_utof}, in powers of the gauge couplings.
Due to the strong assumptions on the asymptotics of
higher order couplings which are implicit in this expansion,
as explained in \Secref{sec:weakh_exp_MSbar},
this analysis allows us to account for
the general scheme dependence of the corresponding
solutions.
In fact, we can address the one-loop flow equation
of $f(x)$ in an arbitrary 
regularization and renormalization scheme,
and thus get access to scheme-independent properties of the flow equation for $f(x)$.
We focus on the case where $\Nc=3$, $\NL=2$, and $d=4$. In addition, we address also the two limiting cases of the \HTQCD\ and the \NAH\ models.

In the $\SU(2)_\text{L}\times\SU(3)_\text{c}$ model, we have decided to rescale the field amplitude $\rho$ with the strong gauge coupling, i.e., $x=\gs^{2P}\rho$.
Therefore the functional flow equation for the rescaled potential is
\begin{align}
	\beta_f &= - 4 f+d_x x f'+\frac{1}{16\pi^2}\left\{ l_0^{(\mathrm H)}(\omegafBr)+3l_0^{(\theta)}(\omegafBl)\right.\nonumber\\
	&\quad\left.+9 l_0^{(\mathrm W)}(\omegafW)- 12 l_0^{(\mathrm F)}(\omegafF)\right\},
	\label{eq:betafFull1Loop_generalmodel}
\end{align}
where the anomalous scaling dimension for the rescaled field $d_x$ is given by \Eqref{eq:dx_def_generalmodel} and the arguments of the threshold functions are
given in \Eqref{eq:omegaf_def}.
The QFP solutions $\hat{g}^2_*$ and $\hat{h}^2_*$ take the same values as in
\Eqref{eq:QFP_g2/gs2_h2/gs2_generalscheme}.
Let us remind the reader here that these values have been calculated by assuming that any mass contributions induced by a nontrivial minimum in the scalar potential
are negligible in the flow equations for the gauge couplings and the top-Yukawa coupling.
In other words we have considered the latter beta functions in the DER,
where the arguments $\omegafBl$, $\omegafBr$, $\omegafW$ and $\omegafF$ are assumed
to go to zero in the UV limit, as in Ref.~\cite{Gies:2018vwk}.
The consistency of this assumption has to be tested once a scaling solution for the
Higgs potential is found.

In the \HTQCD\ model, the flow equation for the rescaled scalar potential takes the form
\begin{align}
	\beta_f = - 4 f+d_x x f'+\frac{1}{16\pi^2}\left\{ 
	l_0^{(\mathrm H)}(\omegafBr)	- 12 l_0^{(\mathrm F)}(\omegafF)\right\}.
	\label{eq:betafFull1LoopYQCD}
\end{align}
Within this case, the QFP solution $\hat{h}^2_*$ takes the value given in \Eqref{eq:h2/gs2_QFP_YQCD_DER}.

In the \NAH\ model with $\SU(2)_\text{L}$ gauge group the $\beta$ function for $f(x)$ reads
\be
\begin{split}
	\beta_f &=-4f+d_x x f'+\frac{1}{16\pi^2}\left\{ l_0^{(\mathrm H)}(\omegafBr)\right.\\
	&\quad \left.+3l_0^{(\theta)}(\omegafBl)+9 l_0^{(\mathrm W)}(\omegafW)\right\}, 
\end{split}
\label{eq:betafFull1LoopNAHiggs}
\ee
where the quantum dimension $d_x$ is given by \Eqref{eq:dx_def_NAH}.
Since only the weak gauge coupling is involved in this model, the scalar field is rescaled via an appropriate power of $g^2$, namely $x=g^{2P}\rho$.

Although in Eqs.~(\ref{eq:betafFull1Loop_generalmodel}$-$\ref{eq:betafFull1LoopNAHiggs}) 
		we have used the same notation for the threshold functions 
as in the FRG case of \Secref{sec:FRG}, in this section we generalize
their scope and we interpret them as threshold functions
in a generic scheme.
In other words, the
 threshold functions in Eqs.~(\ref{eq:betafFull1Loop_generalmodel}$-$\ref{eq:betafFull1LoopNAHiggs})
represent the loop contributions of the scalars, fermions, and gauge bosons
in any arbitrary regularization and renormalization scheme.
To each of the loop-momentum integrals we can associate generic regularization schemes which can even be different for each field.
As an example,
the $\MSbar$ scheme discussed in \Secref{sec:EFT_MSbar} and 
\Secref{sec:EFT_MSbar_FRG}, without RG improvement, i.e.,~suppressing
the anomalous dimensions in the threshold functions,
 would correspond to
\begin{equation}
l_0^{(\MSbar)}(\omega)=\frac{\omega^2}{2}
\label{eq:l_0MSbar}
\end{equation}
in $d=4$.

By Taylor expanding for small gauge couplings, these loop integrals
to first order take the form
\begin{align}
	\beta_f=\left[\beta_f\right]_{0}+\delta \beta_f,
	\label{eq:beta_f_weak-h}
\end{align}
where the first term is the $\beta$ function in the UV limit where $\{\gs^2,g^2\}\to0$ and the second term is the leading $\{\gs^2,g^2\}$ contribution upon expanding the loops and the anomalous dimension $\eta_x$.
This last point requires an important comment:
since $\eta_x$ depends on the properties of the Higgs potential at the nontrivial minimum $\kappa$ which could be a general function of the gauge couplings,
a self-consistency check of the Taylor expansion has to be performed, once the analytic QFP solution for $f(x)$ is computed.

According to the rescaling in \Eqref{eq:rhotox_utof}, quantum fluctuations can contribute to the zeroth-order term $[\beta_f]_0$ only for $P=1$.
For example, in the two limiting  models we have
%
\begin{equation}
  \left[\beta_f\right]_{0}=- 4 f + 2 x f^\prime , \quad \text{for $P<1$},
\label{eq:beta0Pl1}
\end{equation}
for both models, and
\begin{equation}
  \left[\beta_f\right]_{0}=- 4 f + 2 x f^\prime  - \frac{3 }{4\pi^2} l_0^{(\mathrm F)}\!\left(\frac{2x}{9}\right), \quad	\text{for $P=1$,}
\label{eq:beta0P=1Z2}
\end{equation}
in the \HTQCD\ model, and
\begin{equation}
  \left[\beta_f\right]_{0}=- 4 f + 2 x f^\prime  + \frac{9}{16\pi^2} l_0^{(\mathrm W)}\!\left(\frac{x}{2}\right), \quad	\text{for $P=1$,}
\label{eq:beta0P=1NAH}
\end{equation}
in the \NAH\ model.
Thus, for $P<1$ the zeroth order in the gauge couplings is trivial since no quantum fluctuations are retained.
On the other hand for $P=1$, we need a more detailed specification for the regulator
in order to address explicit properties of the QFP solutions, as discussed below.

Aiming at the leading $\{\gs^2,g^2\}$ corrections,
the values of $P\neq1$ are simpler to address in a generic scheme,
since the vertices of the theory, by assumption, scale like positive powers
of $\gs^2$ or $g^2$, depending on the model under consideration.
The leading contribution to $\delta\beta_f$ is produced by Taylor expanding the threshold functions to first order in the gauge couplings.
This gives rise to several coefficients which account for all the scheme dependence of the QFPs, namely
\be
	\mathcal{A}_\Phi=-\frac{1}{16\pi^2}\left[\de_z l_0^{(\Phi)}(z)\right]_{z=0},
\label{eq:defA_BF}
\ee
with $\Phi\in\{\mathrm{H,\theta,F,W}\}$ labeling the fluctuation modes.

Within an FRG scheme these coefficients
can also be written as
\be
	\mathcal{A}_\Phi
    =\frac{1}{2k^{2}}\int\!\! \frac{\mathrm{d}^4p}{(2\pi)^4}
	\frac{\tilde{\partial}_t P_\Phi(p^2)}{\left[P_\Phi(p^2)\right]^2}.
\label{eq:FRGdefA_BF}
\ee
Here, the operator $\tilde\partial_t$ denotes
differentiation with respect to $t=\log k$ 
acting only on the regulators, and
$P_{\Phi}$ 
is the inverse regularized propagator of 
each field. 
In the FRG formalism $P_{\Phi}$ depends on the regularization kernel
 $R_k$ in the Wetterich equation, cf.~\Eqref{eq:WetterichEQ}.
For more details and explicit expression see \Appref{app:threshold}.
The framework of \Eqref{eq:WetterichEQ} has been derived with 
$\Gamma_k$ as a 1PI effective action in the presence of an IR regularization
at the scale $k$, rather then an UV one~\cite{Morris:1993qb}.
As such, the requirement by which the shape functions
provide a physical coarse-graining is that they should diverge
for $k\to+\infty$ for a given $p^2$, and vanish for $k\to 0$.
For monotonic shape functions, the RG time derivative is always positive,
therefore $\mathcal{A}_\Phi>0$.
For example the piecewise linear regulator~\cite{Litim:2000ci,Litim:2001up},
discussed in \Appref{app:threshold}, leads to  $\mathcal{A}_\Phi=1/(32\pi^2)$.

Let us now address the case with $P\leq 1$, in the general
$\SU(2)_\text{L}\times\SU(3)_\text{c}$ model and also in the two limits
of the \HTQCD\ and \NAH\ models.

\subsection{{\boldmath $P\in(0,1/2)$}}
\label{sec:weak-coupling-P<1/2}

In this window of $P$ values and for all models under consideration, only the scalar loops contribute to the first correction in the $\beta$ function for $f(x)$.
In the general model, the leading correction scales as $\gs^{2P}$,
\begin{align}
	\delta \beta_f= - \gs^{2P}\left[\ABR(f'+2xf'')+3\ABL f'\right],
	\label{eq:deltabetaf_P(0,1/2)_generalmodel}
\end{align}
where we have distinguished the contributions coming from the Goldstone or radial modes with the labels $\theta$ and $\mathrm{H}$, respectively.
For the \HTQCD\ model, the leading correction $\delta\beta_f$ can be recovered by simply setting $\mathcal{A}_\theta=0$.
On the other hand, for the \NAH\ model $\delta\beta_f$ is the same as in \Eqref{eq:deltabetaf_P(0,1/2)_generalmodel} with the substitution $\gs^2\leftrightarrow g^2$.

By including the leading-order correction in $\gs^2$, the QFP equation for $f(x)$ becomes a second order ordinary differential equation (ODE) which can be analytically solved and leads to two different solutions.
The first one is given by a special case of the Kummer function which reduces to a quadratic polynomial,
\begin{align}
	f(x)=c\left[x^2-3(\ABR+\ABL) \gs^{2P}x\right].
	\label{eq:fCf_weakh_Pless1/2}
\end{align}
The second one grows exponentially for large field amplitudes.
However, we are only interested in solutions that obey power-like scaling for $x\to\infty$, since
a scalar product can then be defined on the space of eigenperturbations of these
solutions~\cite{Morris:1996nx,ODwyer:2007brp,Bridle:2016nsu}. 
Thus, we set the second integration constant to zero.

By imposing the defining properties for the nontrivial minimum and the rescaled quartic
scalar coupling, namely $f'(x_0)=0$ and $f''(x_0)=\xi_2$ respectively,
we find
\begin{align}
	\xi_2&=2 c ,\\
	x_0&=\frac{3}{2}(\ABL+\ABR) \gs^{2P}.
\end{align}
Therefore we can infer that, in the general model as well as in the \NAH\ model, the condition for having a nontrivial positive minimum is
\begin{align}
	\ABR>-\ABL,
	\label{eq:weak_coupling_P<1/2_condition-for-positive-minimum-generalmodel}
\end{align}
whereas in the \HTQCD\ model the latter expression becomes simply
\begin{align}
	\ABR>0.
\end{align}
As these conditions are satisfied for all admissible 
	FRG regularization schemes, 
within the latter framework
the existence of these solutions is a scheme-independent result.
A particular limiting case is the one of $\overline{{\text{MS}}}$,
	where these solutions are not present as $\ABR=\ABL=0$.
	We provide an interpretation  of this fact at the end
	of this section.

Let us perform the consistency check mentioned above, testing if the gauge and Yukawa coupling QFP values remain unaffected by threshold effects.
The solution in \Eqref{eq:fCf_weakh_Pless1/2} has been found by expanding the beta function in \Eqref{eq:betafFull1Loop_generalmodel} for small $\gs^2$ while keeping $x$, $f(x)$, and its derivatives finite.
	By inserting the QFP solution for $f(x)$ into the expression for the anomalous dimension $\eta_x$, we find that there is no contribution to $\delta\beta_f$ coming from $\eta_x$ for any $P<1/2$.
	In fact, the leading terms in $\eta_x$ scale as either $\gs^2$ or $\gs^{8P}$, and thus are negligible with respect to the scalar contributions in \Eqref{eq:deltabetaf_P(0,1/2)_generalmodel} scaling as $\gs^{2P}$.

As a matter of fact, it is also true that contributions proportional to $\gs^{8P}$ in the scalar anomalous dimension modify also the asymptotic UV behavior of the top-Yukawa coupling for $P\leq 1/4$, eventually leading to a different QFP value for $h^2$. A more detailed explanation of this fact can be found in \Appref{App:weak-coupling-expansion-P<1/4}.
	Thus, the consistency test is passed 
	only by the $1/4< P<1/2$ scaling solutions.
	For $P=1/4$ this does not require a change of the QFP solution for
	$f(x)$ we have just discussed, as only the QFP value of $\hresc_*^2$
	changes.
	For $P<1/4$ instead the current solutions for $f(x)$ are no
	longer valid for the \HTQCD\ model and for the general model,
	and they survive only in the \NAH\ model, where the 
	Yukawa coupling is not present.
	A different QFP for $P<1/4$ might still be possible in Yukawa models,
	if $\kappa$ and $h^2$ exhibit asymptotic scaling powers different
	from the ones discussed in this section.
	This behavior might require strong threshold phenomena and decoupling
	of some degrees of freedom.
	We leave this analysis for future investigations.

\subsection{{\boldmath $P=1/2$}}

For this value of $P$, the contributions from the scalar loops mix with the contribution from the fermionic loop and/or the gauge boson loop.
Let us consider first the general $\SU(2)_\text{L}\times\SU(3)_\text{c}$ model. In this case the first leading correction to $\beta_f$ is
\begin{align}
  \delta\beta_f&=-\gs\Bigl[\ABR (f^\prime+2xf^{\prime\prime})+3\ABL f'\nonumber\\
    &\quad +\frac{9}{2}\AW \hat{g}^2_* x - 12\AF\hat{h}^2_* x\Bigr],
    \label{eq:delta_beta_f_P=1/2_generalmodel}
\end{align}
where the QFP values $\hat{g}^2_*$ and $\hat{h}^2_*$ are given in \Eqref{eq:QFP_g2/gs2_h2/gs2_generalscheme}.
The QFP equation $[\beta_f]_0+\delta\beta_f=0$ is a second-order inhomogeneous linear ODE. Its solution consists of the general solution of the homogeneous part, c.f. \Eqref{eq:fCf_weakh_Pless1/2},
plus a particular solution of the nonhomogeneous equation, which is a quadratic polynomial in $x$.
We find therefore the QFP solution
\begin{align}
  f(x)&=\frac{\xi_2}{2}x^2-\frac{3\gs x }{4}\Bigl[2 \xi_2 (\ABL+\ABR)\nonumber\\
    &\quad +3\AW \hat{g}^2_* - 8\AF\hat{h}^2_*\Bigr],
    \label{eq:f(x)_weakcoupling_P=1/2_generalmodel}
\end{align}
where the rescaled quartic coupling $\xi_2$ remains a free parameter.
The rescaled potential has a nontrivial minimum at
\begin{align}
  x_0=3\gs\frac{ 2\xi_2 (\ABL+\ABR)+3\AW \hat{g}^2_* - 8\AF\hat{h}^2_* }{4\xi_2},
\end{align}
whose positivity requires
\begin{align}
  \xi_2>\frac{8\AF\hat{h}^2_* - 3\AW \hat{g}^2_*}{2(\ABL+\ABR)}.
\end{align}
Let us now investigate the two limiting cases.
In the \NAH\ model, we have 
\begin{align}
  \AF=0,\qquad \gs^2\to g^2\,\implies\,\hat{g}^2_*=1,
  \label{eq:limit_to_NAH}
\end{align}
implying the QFP potential
\begin{align}
  f(x)=\frac{\xi_2}{2}x^2-\frac{3g x}{4}\left[3\AW+2\xi_2(\ABL+\ABR)\right],
\end{align}
which has a nontrivial minimum at
\begin{align}
  x_0=3 g \frac{2\xi_2 (\ABL+\ABR)+3\AW}{4\xi_2}.
\end{align}
The condition for having a positive $x_0$ implies
\begin{align}
  \xi_2>-\frac{3\AW}{2(\ABL+\ABR)}.
\end{align}
Also the results for the \HTQCD\ model are easily recovered
from the general model by setting
\begin{align}
  \AW=0,\quad\ABL=0,\quad \hat{h}^2_*=\frac{2}{9},
  \label{eq:limit_to_HTQCD}
\end{align}	
where the QFP value of $\hat{h}^2_*$ is exactly the one in \Eqref{eq:h2/gs2_QFP_YQCD_DER}.
For example, the expression for the nontrivial minimum and the condition for its positivity become
\begin{align}
  x_0=\gs\frac{ 6\xi_2\ABR - 24\AF\hat{h}^2_* }{4\xi_2}>0\Leftrightarrow \xi_2>\frac{4 \AF\hat{h}^2_*}{\ABR}.
\end{align}

We conclude that, in all models under consideration and for $P= 1/2$,
the QFP equation $\beta_f=0$ admits scaling solutions which are in the SSB regime.
For any FRG scheme, the scheme-dependent coefficients $\mathcal{A}$ assume positive values and thus define a range of values for the rescaled quartic coupling $\xi_2$ that parametrizes the new AF solutions. The changes of this range for different FRG schemes, corresponds to the expected mapping of the coupling space onto itself induced by a scheme change.
Again $\overline{\text{MS}}$ appears special.
The vanishing of the $\mathcal{A}$'s signals
the fact that the leading order contribution in 
a weak coupling expansion is in fact quadratic in $f$,
such that at the present order only the
canonical scaling term contributes.
The latter is expected to be purely classical
and the corresponding QFP solutions unphysical.
Hence, also for $P=1/2$ and within the leading weak-coupling
approximation we find no new AF scaling solution in 
$\overline{\text{MS}}$.
This can be seen, as already stated in \Secref{sec:weakh_exp_MSbar},
as a deficiency of the present approximation strategy in the
$\overline{\text{MS}}$ case, since at leading order it
suppresses all interaction effects and at next-to-leading order
it returns the full nonlinear second-order ODE for the QFP
potential. Let us recall however, that analyzing the latter
equation with an EFT-like approximation in \Secref{sec:EFT_MSbar},
we did observe novel AF trajectories for $P=1/2$ in
the \NAH\ model, within the $\overline{\text{MS}}$ scheme.

Also for $P=1/2$, the contribution due to the anomalous dimension $\eta_x$ is subleading with respect to the correction in \Eqref{eq:delta_beta_f_P=1/2_generalmodel}.
	In addition, these solutions pass the consistency check that the $\beta$ function for the top-Yukawa coupling still has the same QFP solution as given in \Eqref{eq:QFP_g2/gs2_h2/gs2_generalscheme} for the general model.
	Indeed, this can be verified straightforwardly by substituting the solution in \Eqref{eq:f(x)_weakcoupling_P=1/2_generalmodel} into the $\beta$ function for $h^2$, c.f.~\Eqref{eq:betahsquared}.

\subsection{{\boldmath $P\in (1/2,1)$}}

In this case, the leading contribution to the $\beta$ function for the rescaled potential $f(x)$ is only due to the fermionic loop and/or the gauge boson loop, depending on the model under investigation.
For example in the general $\SU(2)_\text{L}\times\SU(3)_\text{c}$ model, the first leading correction is proportional to $\gs^{2-2P}$ and reads
\begin{align}
  \delta\beta_f = \gs^{2-2P} x \left[-\frac{9}{2}\AW\hat{g}^2_*+12\AF \hat{h}^2_*\right],
  \label{eq:deltabetaf_P(1/2,1)_generalmodel}
\end{align}
where the QFP values $\hat{g}^2_*$ and $\hat{h}^2_*$ are given in \Eqref{eq:QFP_g2/gs2_h2/gs2_generalscheme}.
The FP equation $[\beta_f]_0+\delta\beta_f=0$ remains a first-order ODE whose analytical solution is
\begin{align}
  f(x)=\frac{\xi_2}{2}x^2-\frac{3\gs^{2-2P} x}{4}\left[3\AW\hat{g}^2_*-8\AF\hat{h}^2_*\right],
\end{align}
where the rescaled quartic scalar coupling $\xi_2$ remains an arbitrary integration constant.
The QFP potential has a nontrivial positive minimum at
\begin{align}
  x_0=3\frac{3\AW\hat{g}^2_*-8\AF\hat{h}^2_*}{4\xi_2}\gs^{2-2P}>0\Leftrightarrow\AW>\frac{8\AF\hat{h}^2_*}{3\hat{g}^2_*}.
  \label{eq:x0_P(1/2,1)_generalmodel}
\end{align}
If one adopts the same scheme for both loops, such that $\AF=\AW$,
the latter condition is not satisfied in the SM case, where
\begin{equation}
3\hat{g}^2_*-8\hat{h}^2_*<0\,.
\end{equation}

Let us then consider the two limiting models separately. For the \HTQCD\ model
obtained from the substitutions in \Eqref{eq:limit_to_HTQCD},
the expression for the nontrivial minimum becomes
\begin{align}
  x_0=-\gs^{2-2P}\frac{6\AF\hat{h}^2_*}{\xi_2},
\end{align}
which would be negative for any positive value of $\xi_2$ and general 
FRG scheme coefficient $\AF>0$.
	
The opposite situation occurs in the \NAH\ model. From the substitution in \Eqref{eq:limit_to_NAH}, the nontrivial minimum now reduces to
\begin{align}
  x_0=g^{2-2P}\frac{9\AW}{4\xi_2}.
  \label{eq:x0_P(1/2,1)_NAH}
\end{align}
For any positive value of $\AW$ and $\xi_2$ the scalar potential for the \NAH\ model is in the broken regime and features a nontrivial minimum.

Also for these values of $P$, we need to perform the consistency check that the QFP of the top-Yukawa coupling remains unaffected: we observe that the contribution to $\delta\beta_f$ from the anomalous dimension $\eta_x$ is subleading with respect to the fermionic and gauge boson contributions and that the top-Yukawa coupling scales as $\gs^2$ in the UV limit. It thus has the same QFP solutions as in \Eqref{eq:QFP_g2/gs2_h2/gs2_generalscheme}.

We emphasize that for all values of $P<1$
the QFP solutions we have obtained are analytic in $x$  in the present weak-coupling approximation.
In the previous \Secref{sec:EFT_MSbar}, this was implemented by construction,
since we have projected the functional flow equation onto a polynomial ansatz.
In the present analysis, this happens because the contributions to $\beta_f$ producing non-analyticities are accompanied by subleading powers of $\gs^2$ or $g^2$ for $P<1$.
Indeed, both the appearance of the anomalous dimension $\eta_x$ 
in the scaling term $(2+\eta_x)xf'(x)$, and the contributions from
the threshold functions proportional to $x^2$ would produce a 
singularity of $f''(0)$ for any $\{\gs^2,g^2\}\neq0$, 
as discussed before in \Secref{sec:EFT_MSbar_FRG}, see also below.
Knowing about the presence of this singularity for any $P$ at $\{\gs^2,g^2\}\neq 0$, 
we can accept the previous solutions only if $x_0>0$.
This appears to be possible in all models under investigation for $P\leq1/2$. In addition, it is possible in the \NAH\ model, and in  the general $\SU(2)_\text{L}\times\SU(3)_\text{c}$ model for $P\in (1/2,1)$,
in the family of FRG schemes where an IR regularization is provided.

Moreover, we want to stress that all the solutions obtained for $P<1$ are consistent
with the assumptions made at the beginning:
the arguments of the threshold functions $\omegafBl,\omegafBr,\omegafF$ and $\omegafW$ go to zero in the UV limit, and
the flow equations for the gauge couplings and the top-Yukawa coupling can be treated as in the DER.

A general observation that could be raised against the 
	validity of the present approximation, in the FRG framework,
	is that neglecting the nonlinear contributions to the 
	flow equation of $f(x)$ might miss crucial terms
	and produce spurious or unphysical QFP solutions.
	In particular, part of the universal one-loop
	contribution, the one which arises from the
	Taylor expansion of the threshold functions
	to second order in  $\omegafBl,\omegafBr,\omegafF$ and $\omegafW$,
	is not accounted for in the present discussion
	of the $P<1$ scaling solutions.
	However, the inclusion of the latter contributions
	as well as of further nonlinearities, up to the 
	full complexity of the 
	FRG flow equations of \Secref{sec:FRG},
	has been performed, with specific
	regulator choices,
	for the \NAH\ model~\cite{Gies:2015lia,Gies:2016kkk}
	as well as for the  \HTQCD\ model~\cite{Gies:2018vwk}
	in fact confirming 
	the results of this leading-order
	weak coupling expansion.

As for the fate of these solutions in $\overline{\text{MS}}$,
	a similar conclusion as for the $P\leq1/2$ cases holds.
	The vanishing of the quadratically divergent coefficients $\mathcal{A}$
	results in the vanishing of the interaction effects retained at
	the leading order of the present approximation scheme.
	Thus, no new AF trajectory is visible in the  scheme
	for $P<1$.
However, this does not exclude the existence of new AF scaling solutions
in $\overline{\text{MS}}$ altogether. In fact, as discussed in 
Secs.~\ref{sec:EFT_MSbar} and \ref{sec:EFT_MSbar_FRG}
as well as in the following, a two-parameter family of these 
solutions occur in
the general $\SU(2)_\text{L}\times\SU(3)_\text{c}$ model
and in the \HTQCD\ model,
at $P=1$.

\subsection{{\boldmath $P=1$}}   \label{sec:weak-h^2_P=1}

For $P\geq 1$ the
$\beta$ functional to zeroth order in the gauge
couplings $[\beta_f]_0$ accounts for 
the full nonlinearity of the gauge and the fermion loops.
In fact, let us consider first the general $\SU(2)_\mathrm{L}\times\SU(3)_\mathrm{c}$
model. The leading RG flow equation for the rescaled scalar potential reads
\begin{align}
	[\beta_f]_0= - 4 f+2 x f' +\frac{1}{16\pi^2}\left[9 l_0^{(\mathrm W)}(\omegafW) - 12 l_0^{(\mathrm F)}(\omegafF)\right].
	\label{eq:betaf_P=1_generalmodel}
\end{align}
For $P=1$, the QFP solutions $\hat{g}^2_*$ and $\hat{h}^2_*$ will not have
the same values as in \Eqref{eq:QFP_g2/gs2_h2/gs2_generalscheme}, since it is no
longer true that the RG flow equations for the gauge couplings and the top-Yukawa
coupling can be treated as being in the DER. Indeed, the arguments $\omegafW$ and $\omegafF$
are finite and do not approach zero in the UV limit.
As a consequence, we expect that $g^2$ and $h^2$ are still proportional to $\gs^2$
in the UV limit, but with QFP solutions which depend nontrivially on $x_0$.
We can nevertheless consider $\hat{g}^2_*$ and $\hat{h}^2_*$ as finite ratios.

The corresponding QFP equation $[\beta_f]_0=0$ can be solved analytically and
leads to the integral solution
\begin{align}
	f(x)&= c\, x^2 -\frac{9}{32\pi^2} \omegafW^2\!\int_1^{\omegafW} \!\!\! \mathrm{d}y\, y^{-3} l^{(\mathrm W)}_0(y)\nonumber\\
	&\quad + \frac{3}{8\pi^2}\omegafF^2\!\int_1^{\omegafF} \!\!\! \mathrm{d}y\,y^{-3} l_0^{(\mathrm F)}(y),
	\label{eq:QFPf_P=1_weakh_anyscheme_generalmodel}
\end{align}
where $c$ is an arbitrary integration constant.
For instance, within the FRG framework the piecewise linear  regulator of \Appref{app:threshold}
would give a Coleman-Weinberg-like potential,
\begin{align}
	f(x)&=c \,x^2 - \frac{9}{64 \pi^2} \left[\omegafW+\omegafW^2\log\frac{\omegafW}{1+\omegafW}\right]\nonumber\\
	&\quad+ \frac{3}{16 \pi^2} \left[\omegafF+\omegafF^2\log\frac{\omegafF}{1+\omegafF}\right]
\end{align}
which has a log-type singularity at the origin in the second derivative,
given by the term $\sim x^2\log x$. This remains true in any scheme.
In fact by Taylor expanding the threshold functions $l_0^{(F)}(y)$ and $l_0^{(\mathrm{W})}(y)$ in \Eqref{eq:QFPf_P=1_weakh_anyscheme_generalmodel} around $y=0$,
a logarithmic divergence of the integral arises precisely from the quadratic term of this expansion.
In other words, the appearance of this singular
behavior is as universal as the one-loop beta function
of the marginal couplings.
Thus we expect that this feature survives also in the full $\gs^2$-dependent solution.

The freedom of choosing the parameter $c$ allows us to
construct QFP solutions, which circumvent the problem of nonanalytic structures at the minimum of the potential by developing a nontrivial minimum away from the origin.
The defining equation for the minimum $f'(x_0)=0$,
involves an integral of two arbitrary threshold functions
and might be hard to solve analytically for $x_0$. Still,  it can 
straightforwardly be used to express $c$ as a function of $x_0$.
From the point of view where the latter is the free parameter
labeling the QFP solutions, the natural question then is
as to whether it can be chosen such that $f''(x_0)=\xi_2$ is positive and finite
in the $\gs^2\to 0$ limit.
As for $P<1$ the answer to this question involves some scheme dependence,
which is encoded in the coefficients
\be
\mathcal{A}_\Phi(x_0)=-\frac{1}{16\pi^2}
\lim_{x\to x_0}
\left[\de_z l_0^{(\Phi)}(z)\right]_{z=z_\Phi},
\label{eq:defA_BF0}
\ee
where $\Phi\in\{\mathrm{F,W} \}$.
For all schemes which can be embedded into the FRG,
the coefficients $\mathcal{A}_\mathrm{F,W}(x_0)$ of \Eqref{eq:defA_BF0} are similar to the 
ones defined in \Eqref{eq:defA_BF}.
The evaluation of these loop integrals at
nonvanishing values of $z_\mathrm{F,W}$
simply results in adding $k^2 z_\mathrm{F,W}$
to $P_\mathrm{F,W}$ in the denominator of the integrand function, 
\begin{align}
\mathcal{A}_\Phi(x_0) = \frac{1}{2k^2} \int\!\! \frac{\mathrm{d}^4p}{(2\pi)^4}
\frac{\tilde{\partial}_t P_\Phi(p^2)}{\bigl[P_\Phi(p^2)+k^2 z_\Phi \bigr]^2},
\label{eq:defA(x_0)}
\end{align}
whose sign is positive for any IR regularization scheme, since the shape function
is monotonically increasing.

For any threshold functions, the rescaled quartic scalar coupling simplifies to
\begin{align}
\xi_2&=\frac{9\hat{g}^2_*}{4x_0}
\mathcal{A}_\mathrm{W}(x_0)
-\frac{6\hat{h}^2_*}{x_0}
\mathcal{A}_\mathrm{F}(x_0),
\label{eq:xi2_weak-h_P=1_generalscheme}
\end{align}
and the condition that $\xi_2$ be positive requires
	\begin{align}
3\hat{g}^2_*\mathcal{A}_\mathrm{W}(x_0)
	>	8\hat{h}^2_*\mathcal{A}_\mathrm{F}(x_0).
\label{eq:condition_for_xi2>0_P=1_generalmodel}
	\end{align}
For small values of $x_0$ this condition reduces to the one
in \Eqref{eq:x0_P(1/2,1)_generalmodel}.
In the FRG framework, we already observed in the previous section that the latter is 
not fulfilled in the SM case, if the same regulator is chosen
for both fields.
We expect this conclusion holds for any value of $x_0$,
and for generic FRG schemes.
In fact, in the opposite limiting case of large $x_0$,
$\mathcal{A}_\Phi(x_0)$ reduces to an  $x_0$-independent
number, and \Eqref{eq:condition_for_xi2>0_P=1_generalmodel}
again is equivalent to \Eqref {eq:x0_P(1/2,1)_generalmodel}.
For the special case of 
 the piecewise linear  regulator of \Appref{app:threshold}
the coefficients of \Eqref{eq:defA(x_0)} read
\begin{equation}
\mathcal{A}^{(\mathrm{p.lin.})}_\Phi(x_0)=\frac{1}{32\pi^2}\lim_{x\to x_0} \frac{1}{(1+z_\Phi)^2}, 
\quad \forall \Phi.
\label{eq:A(x_0)Litim}
\end{equation} 
such that  \Eqref{eq:condition_for_xi2>0_P=1_generalmodel} is not fulfilled 
for all $x_0\geq 0$.

Let us now consider the two limiting cases.
For the \HTQCD\ model where the gauge-boson loop is absent,
the inequality in \Eqref{eq:condition_for_xi2>0_P=1_generalmodel} becomes

\begin{align}
\mathcal{A}_\mathrm{F}(x_0)<0
\end{align}
which cannot be fulfilled for any admissible IR
regularization scheme.
Conversely, in the \NAH\ model the fermion loop is absent, resulting in the condition
	\begin{align}
	\mathcal{A}_\mathrm{W}(x_0)>0
	\end{align}
which is satisfied by any admissible regulator in the FRG
framework.

In  $\overline{\text{MS}}$ these conclusions get
twisted. In fact, \Eqref{eq:l_0MSbar} results in
\begin{equation}
\mathcal{A}^{(\MSbar)}_\Phi(x_0)=-\frac{1}{16\pi^2}\lim_{x\to x_0} z_\Phi<0, 
\quad \forall \Phi.
\label{eq:A(x_0)MSbar}
\end{equation}
This reproduces the existence of $P=1$ solutions in
the \HTQCD\ model and their absence in the \NAH\ model,
as already observed in Secs.~\ref{sec:EFT_MSbar} and \ref{sec:EFT_MSbar_FRG}.
Concerning the general $\SU(2)_\text{L}\times\SU(3)_\text{c}$ model,
replacing \Eqref{eq:A(x_0)MSbar} and \Eqref{eq:omegaf_def} inside 
\Eqref{eq:condition_for_xi2>0_P=1_generalmodel},
the latter inequality in $\overline{\text{MS}}$ becomes
\begin{equation}
 16 \hat{h}^4_*> 3\hat{g}^4_*,
\end{equation}
which does indeed hold for the SM, as was
already observed in \Secref{sec:phi4D_MSbar},
e.g.,~c.f.~\Eqref{eq:phi4_MSbar_generalmodel}.

\subsection{Summary}   \label{sec:summary}

	Let us summarize the results obtained so far.
	In most cases, a simple leading-order weak-coupling expansion
	of the flow equations suffices to unveil a two-parameters family
	of novel AF trajectories.
	In a general FRG scheme based on an IR regulator,
	these solutions are effectively labeled by the
	position of the nontrivial minimum  $x_0$ of the 
	rescaled scalar potential $f$, or equivalently by
	its finite asymptotic ratio $\hat{\kappa}$,
	and by the power $P$ defining the approach of the quartic coupling
	$\lambda_2$ to the Gaussian FP.
	
	In the general $\SU(2)_\text{L}\times\SU(3)_\text{c}$ model,
	for the specific case of SM matter content and under the assumption that
	the same kind of regulator is chosen for all fields,
	we have successfully constructed solutions with $P\in[1/4,1/2]$.
	For $P<1/4$ we do not exclude the possible existence of
	solutions for which $h^2$ exhibits an asymptotic scaling
	different from the one supported within the DER.
	The same conclusions can be drawn for the \HTQCD\ model.
	In the \NAH\ model,
    we have recovered the $P\in[0,1]$ solutions already described in 
    \cite{Gies:2015lia,Gies:2016kkk}.
    For completeness, the $P>1$ case is discussed in \Appref{app:P>1},
    where we recover the known solutions for the \NAH\ model
    and we find no AF trajectories for the
    Yukawa models.
    
    Different IR regulators result in a change of range of
    the attainable values for the finite ratio $\lresc_2=\xi_2$.
    Special regulator choices which assign different regularizations
    to different degrees of freedom can also change the 
    range of attainable values of the asymptotic power $P$.
    For instance, they can make the $P\in(1/2,1]$
    solutions available also in the
    $\SU(2)_\text{L}\times\SU(3)_\text{c}$ model,
    c.f.~Eqs.~(\ref{eq:x0_P(1/2,1)_generalmodel}) and (\ref{eq:condition_for_xi2>0_P=1_generalmodel}).

	The $\overline{\text{MS}}$ scheme
	appears to correspond to a peculiar limit of this remapping
	of allowed parameter ranges. 
	In fact, the two-parameter family of new AF solutions in
	this case is labeled by the finite ratio $\hat{\kappa}$
	and by the power $Q$, which describes the 
	asymptotic scaling of the running dimensionless vev $\kappa$.
	The power $P$ instead is fixed to $P=1$ in the 
	general  $\SU(2)_\text{L}\times\SU(3)_\text{c}$ model
	as well as in the \HTQCD\ model, while the \NAH\ model
	exhibits these solutions for $P=1/2$.
	
	These results are schematically summarized in
	Tab.~\ref{tab:table}. Here, for each model under
	consideration, we list
	the attainable values of the asymptotic power $P$
	related to the scaling of the quartic coupling $\lambda_2$.
	Furthermore, we recall the two remaining continuous parameters 
	that play the role of coordinates on the space
	of AF scaling solutions in the different RG schemes:
	$Q$ and $\hat{\kappa}$ in the $\overline{\text{MS}}$ scheme,
	$P$ and $\xi_2$ in the FRG schemes.

\section{Conclusions} \label{sec:conclusions}

The recently discovered class of new asymptotically free RG trajectories in various non-Abelian particle models~\cite{Gies:2015lia,Gies:2016kkk,Gies:2018vwk} 
calls for a critical assessment of their scheme independence.
In contrast to standard perturbative investigations, these new solutions become visible beyond the deep Euclidean region, because threshold effects can play an important role on all scales. 
As one-loop universality is no longer guaranteed beyond the deep Euclidean region, we have investigated in this work whether the existence of these UV complete trajectories is universal. 

\begin{table}[t!] 
	\begin{tabular}{ l | c  | c | c |}
		&SU(2)${}_\mathrm{L}\times\,$SU(3)${}_\mathrm{c}$  & \HTQCD & Non-Abelian Higgs\\
		\hline
		$\overline{\text{MS}}$ & $P=1,\, Q, \hat{\kappa}$ & $P=1,\, Q, \hat{\kappa}$  &
		$P=\frac{1}{2},\, Q, \kapg$  \\
		FRG & $P\in \left[\frac{1}{4},\frac{1}{2}\right]\!,\, \xi_2$ & 
		$P\in\left[\frac{1}{4},\frac{1}{2}\right]\!,\, \xi_2$ & $P\in(0,+\infty),\, \xi_2$ \\
		\addlinespace[0.2ex]\hline
	\end{tabular}
	\caption{Summary of the family of new AF solutions constructed in this work, for the models
		described in \Secref{sec:AF_in_standard_perturbation_theory}, and in the RG schemes we 
		have analyzed.
		For each solution, we provide the value(s)
		of $P$ for which they occur,
		and the (remaining) variables
		that parametrize the space of solutions.
		In the FRG schemes, we refer to the setup where the same
		regulator is used for all fields, as detailed in \Secref{sec:weak-coupling-P<1/2}.
		For the most general model we specify the SM matter content in the flow of the gauge and Yukawa couplings. }
	\label{tab:table}
\end{table}

We have specifically studied an $\SU(\NL) \times \SU(\Nc)$ gauge theory coupled to fermions which are either charged under both gauge groups or only with respect to the $\SU(\Nc)$ group. 
Moreover, we consider a scalar field being in the fundamental representation of $\SU(\NL)$ but a singlet of $\SU(\Nc)$. 
Thereby, we cover the non-Abelian subsector of the standard model, and generalize previous investigations of {\HTQCD} and {\NAH} models. 
The latter models represent particular limits of our general $\SU(\NL) \times \SU(\Nc)$ model.

In order to make contact with the most widely used RG scheme, we have started with the one-loop $\beta$ functions in the $\MSbar$ scheme. 
The new solutions become visible by inclusion of higher-dimensional operators as well as a nonvanishing expectation value for the scalar field inducing mass thresholds for various fields. 
This becomes even more transparent from the flow equation for the entire scalar potential. 
The $\beta$ functional for the potential $u(\rho)$ or its rescaled form $f(x)$ allows to include arbitrary generalizations of the scalar self-coupling as well as to address the global stability properties of the fixed-point potential. 
In addition, the functional analysis reveals that the family of new asymptotically free trajectories can be classified by rescaling parameters (such as $P$) and the position of the rescaled potential minimum (such as $\kap$ or $x_0$). These parameters can in turn be related to the boundary conditions specified for the quasi fixed-point potential, and thus the boundary conditions for correlation functions \cite{Gies:2016kkk}.
For instance, the standard perturbative solution of the {\HTQCD} model, i.e., the 
\emph{Cheng--Eichten--Li} solution~\cite{Cheng:1973nv},
 can be recovered as a particular boundary condition, which is implicitly assumed within perturbative computations.

As a probe of scheme independence, we have investigated the flow equations also using the 
functional-renormalization-group framework, that facilitates to perform the momentum space regularization with a general class of regulator functions. 
We demonstrated that the new solutions exist for any physically admissible 
functional-renormalization-group regulator which encodes the details of the decoupling of massive modes.
On a line of constant physics, a change of the regularization scheme induces a map of the coupling space of initial conditions, e.g., given in terms of bare couplings at an initial UV scale $\Lambda$, onto itself. 
Our results for the new trajectories show that such a mapping in theory space also includes the boundary conditions, i.e., parameters such as the rescaling power $P$, which classify the new 
asymptotically free trajectories.

The present work exemplifies common features of those models
  studied so far which allow for the construction of new asymptotically free
  trajectories: (i) a model needs an asymptotically free sector with a
  coupling that can serve as a rescaling parameter. While the
  rescaling is merely a technical step, it helps introducing the quasi-fixed-point
  concept that can be extended to full functions such as quasi-fixed-point scalar
  potentials. (ii) the (asymptotic) boundary conditions for such full
  quasi-fixed-point functions need to be specified explicitly or implicitly,
  e.g., through a suitable choice of rescaling parameters or the
  scaling of higher-dimensional operators. (iii) threshold effects
  need to be accounted for as they can invalidate the conventional
  naive analysis in the deep Euclidean region. 
  Whereas we have demonstrated that the new
  asymptotically free trajectories are clearly visible in a weak-coupling analysis, the
  conventional perturbative treatment does not fully account for (ii)
  and (iii) but is confined to implicit trivial choices. We emphasize
  that ingredient (i) does not mean that the full perturbatively
  renormalizable model needs to be
  asymptotically free; in fact, the \NAH\ model is a counterexample,
  as generic perturbatively
  renormalizable models are not completely asymptotically free due to
  scalar triviality. Our construction demonstrates that the asymptotic
  freedom of the gauge sector can be sufficient to seed asymptotic
  freedom in the complete model using ingredients (ii) and (iii).

This last point inspires to use the new trajectories for more
  realistic model building and, eventually, phenomenology. Asymptotic
  freedom can render a particle physics model high-energy
  complete. Our construction thus has the potential to solve the
  consistency problems of the standard model, namely the triviality
  problem being manifest in the Higgs and the $\mathrm{U}(1)$ sector
of the standard model at high energies. Our results for the
  non-Abelian models with semi-simple gauge groups may already be
  useful for the construction of asymptotically free grand unified
  models. If the construction can also be extended to an U(1)
hypercharge sector, a more direct completion of the standard model
could exist as a fundamental particle physics theory, offering new
routes to model building. This is also attractive in the light of our
observation that quasi-fixed-point trajectories can exist with a higher degree of
predictivity than their conventional perturbatively renormalizable counterparts.

\section*{Acknowledgments}

We thank Omar Zanusso for insightful discussions.
This work received funding
support by the DFG under Grants No. GRK1523/2, No. Gi328/9-1, and No. SO1777/1-1. RS
and AU acknowledge support by the Carl-Zeiss foundation.

\appendix

\section{The upper critical surface for total asymptotic freedom}
\label{App:1loop_DER}

In this appendix, we summarize the analysis of the one-loop flow in the DER as required for our model in the main text. Similar analyses have widely been discussed in the literature, see, e.g., \cite{Callaway:1988ya,Giudice:2014tma}.
In \Secref{subsec:Yukawa_sector}, we have presented the one-loop $\beta$ function for the top-Yukawa coupling in the DER and in the presence of the two gauge couplings $g^2$ and $\gs^2$, cf.~\Eqref{eq:beta_h2_1Loop_DER_general_NL_Nc}, having the form
\begin{align}
	\de_t h^2=h^2 \left(a_h h^2-a_g g^2-\as \gs^2\right),
	\label{eq:det_h2_1loop_DER_2gaugecouplings}
\end{align}
where the $a$'s coefficients are all positive for the SM set of parameters, cf.~\Eqref{eq:SM_parameters}:
\begin{align}
	a_h=\frac{9}{16\pi^2},\quad a_g=\frac{9}{32\pi^2},\quad \as=\frac{1}{\pi^2}.
\end{align}
Equation \eqref{eq:det_h2_1loop_DER_2gaugecouplings} can be integrated in the RG time $t$ together with the $\beta$ functions for the gauge couplings, Eqs.~(\ref{eq:beta_g2_1loop_DER}) and (\ref{eq:beta_gs2_1loop_DER}).
The integration of the latter is straightforward, yielding
\begin{align}
	g(t)^2=\frac{g_0^2}{1+g_0^2\hatetaW t},\qquad \gs(t)^2=\frac{g_\mathrm{s0}^2}{1+g_\mathrm{s0}^2\hatetaG t},
\end{align}
where $g_0^2$ and $g_\mathrm{s0}^2$ are the initial conditions at $t=0$.
The ratio $\hatetaG$ is defined in \Eqref{eq:hat_eta_G} and similarly for $\hatetaW=\etaW g^{-2}$, cf. \Eqref{eq:beta_g2_1loop_DER}.
The general solution of \Eqref{eq:det_h2_1loop_DER_2gaugecouplings} reads
\begin{align}
	\frac{1}{h(t)^2}=\left(\frac{1}{h_0^2} - a_h I(t)\right)\left[\frac{g_0^2}{g^2(t)}\right]^{\frac{a_g}{\hatetaW}}\left[\frac{g_\mathrm{s0}^2}{\gs^2(t)}\right]^{\frac{\as}{\hatetaG}},
	\label{eq:integration_of_h^2(t)}
\end{align}
where the function $I(t)$ is defined as \cite{Giudice:2014tma}
\begin{align}
	I(t)=\int_0^t d\tau \left[\frac{g^2(\tau)}{g_0^2}\right]^{\frac{a_g}{\hatetaW}}\left[\frac{\gs^2(\tau)}{g_\mathrm{s0}^2}\right]^{\frac{\as}{\hatetaG}}.
	\label{eq:integral_I_def}
\end{align}
The conditions for having total AF are
\begin{align}
	\frac{a_g}{\hatetaW}+\frac{\as}{\hatetaG}>1,\qquad h^2_0\leq\frac{1}{a_h I(\infty)}.
	\label{eq:TAF_conditions}
\end{align}
The first requirement is necessary in order to provide for
the existence of AF trajectories in the space of parameters $(h^2,g^2,\gs^2)$. Incidentally, this criterion is fulfilled for the SM set of parameters.
In fact, if this condition is not satisfied the integral $I(\infty)$ diverges; consequently, there will be a finite RG time $t_\mathrm{critic}$ at which the right-hand side of
\Eqref{eq:integration_of_h^2(t)} is zero and the top-Yukawa coupling hits a Landau pole,
i.e., $I(t_\mathrm{critic})=(h_0^2a_h)^{-1}$.
On the other hand, whenever the integral $I(\infty)$ converges,
AF is guaranteed as long as the right-hand side of \Eqref{eq:integration_of_h^2(t)}
remains positive. This is precisely the second inequality in \Eqref{eq:TAF_conditions}.

By denoting $\Omega(g_\text{s0}^2,g_0^2)=(a_h I(\infty))^{-1}$, the equation
$h_0^2=\Omega(g_\text{s0}^2,g_0^2)$ represents an upper critical surface for TAF in the $(h^2,g^2,\gs^2)$ space.
In fact, if $h_0^2>\Omega(g_\text{s0}^2,g_0^2)$, the negative one-loop contributions to
$\de_t h^2$ involving the gauge boson fluctuations are suppressed
in the UV limit, and the dominant positive scalar fluctuations drive the top-Yukawa coupling
towards a Landau pole. By contrast, all couplings approach the Gau\ss ian FP towards the UV for all initial conditions such that
$h_0^2\leq\Omega(g_\text{s0}^2,g_0^2)$.

	\begin{figure}[t!]
		\includegraphics[width=0.5\columnwidth]{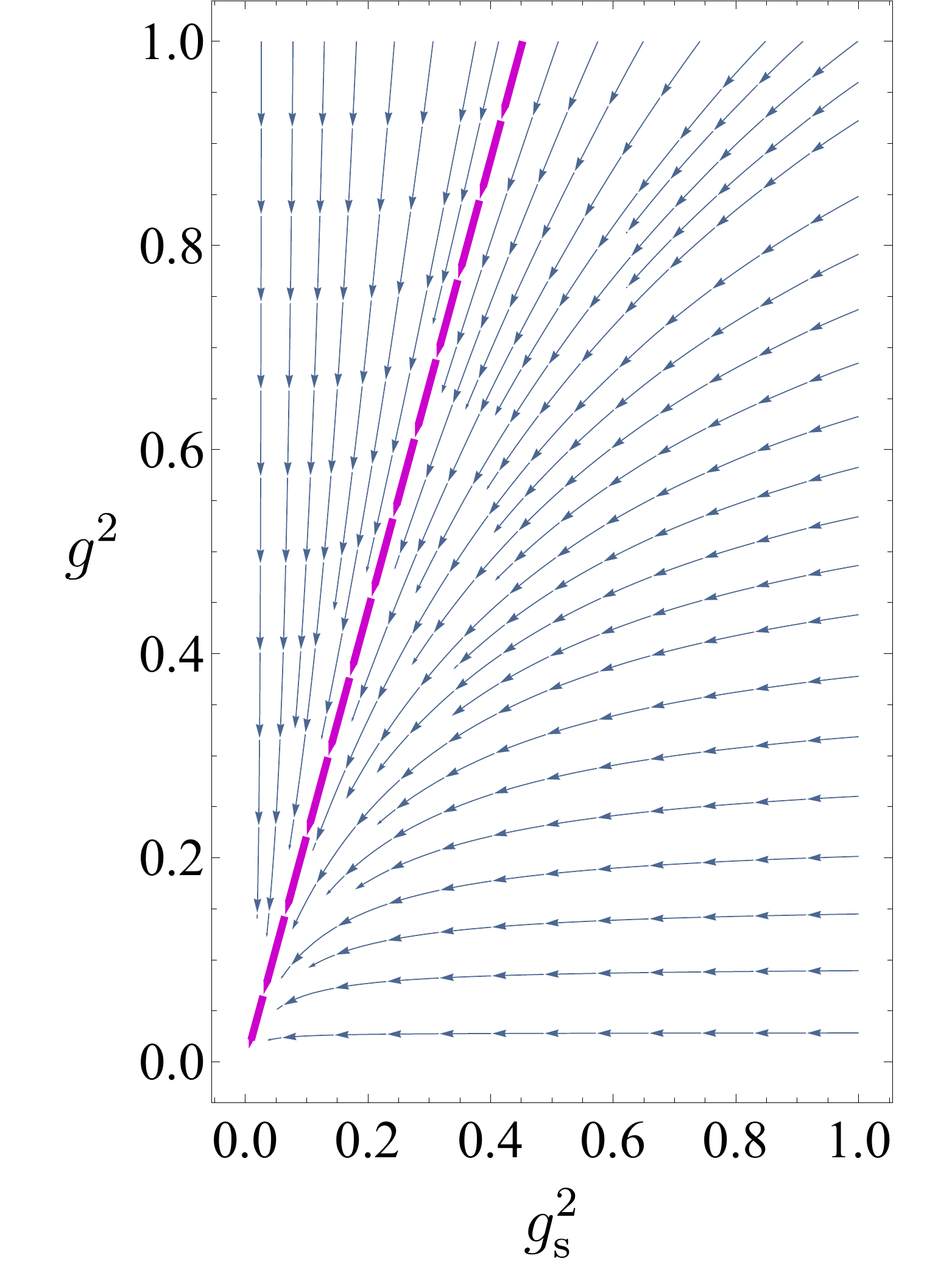}\hfill
		\includegraphics[width=0.5\columnwidth]{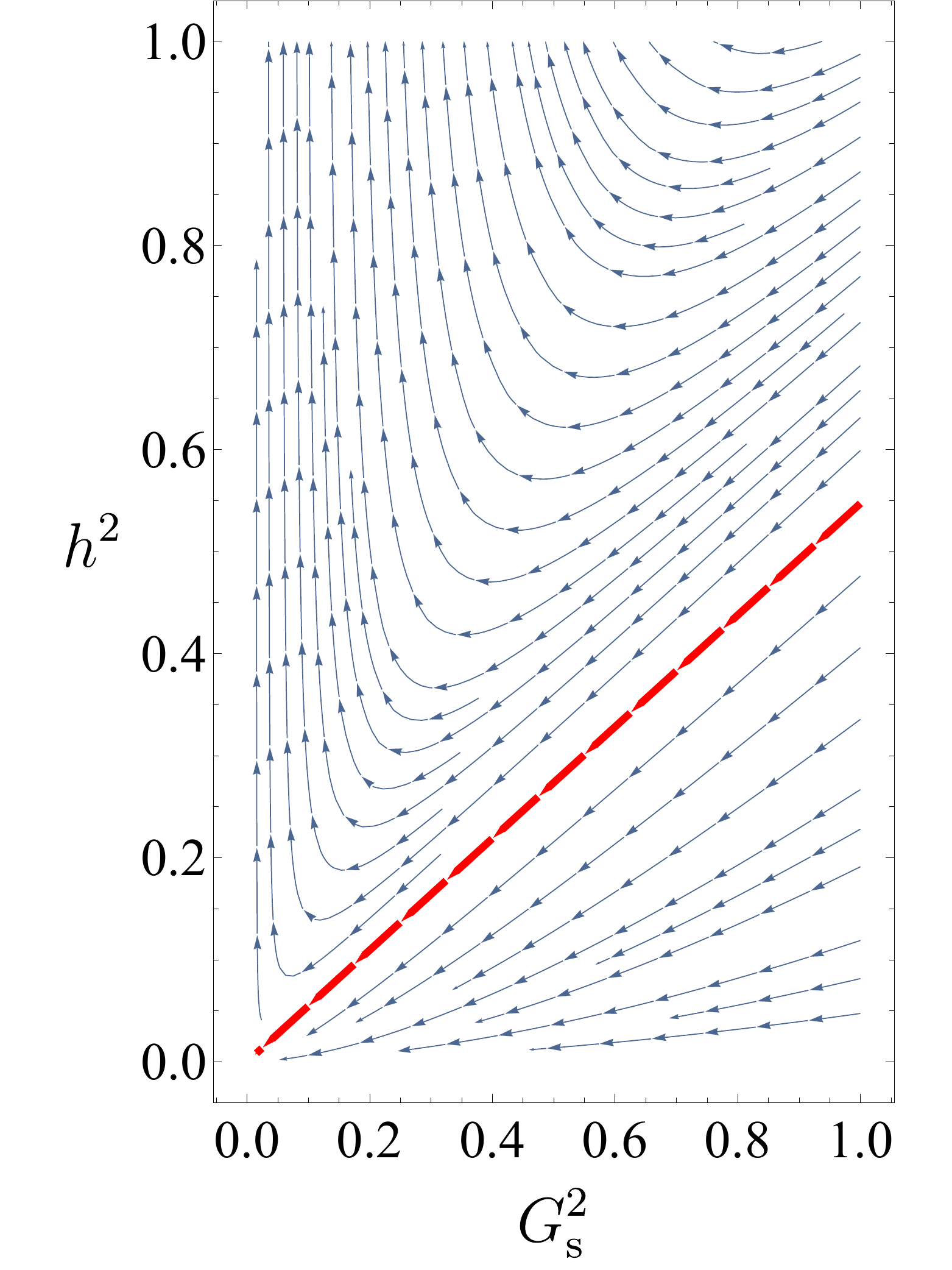}
		\caption{\emph{Left Panel}: phase diagram and flow of the gauge couplings
			in the $h^2=0$ plane. Both couplings exhibit an
			AF flow to the Gau\ss{}ian FP in the UV,
			being attracted by the trajectory where $g^2=\hat{g}^2_*
			\gs^2$ (magenta line).
			\emph{Right Panel}: phase diagram and coupling flow
			using the rotated gauge coupling $G_{\mathrm{s}}$
			(parametrizing the magenta-line trajectory of the left
			panel). AF trajectories require a
			sufficiently small initial Yukawa coupling $h_0^2\leq
			\Omega$ with the upper bound on the critical surface
			$\Omega$ denoted by the red line. The latter represents the special trajectory satisfying \Eqref{eq:A10}.}
		\label{fig:streamplot_g2gs2_plane_and_rotatedplane}
	\end{figure}

Let us concentrate in more detail on the RG flow on the surface $\Omega$:
By virtue of its critical nature, $\Omega$ represents an UV repulsive surface along its orthogonal directions.
Only for those initial conditions such that $h_0^2=\Omega(g_\text{s0}^2,g_0^2)$, the
integrated RG trajectories will remain on the critical surface itself for all $t>0$
and will approach the Gau\ss ian FP in the UV limit $t\to\infty$.
Next, we observe that the one-loop $\beta$ functions for the gauge couplings
are independent of the Yukawa-coupling in the DER.
As a consequence, the gauge-coupling flow on each slice of constant $h^2$ looks the same as within the
$(\gs^2,g^2)$ plane, see \Figref{fig:streamplot_g2gs2_plane_and_rotatedplane} (left panel).

Therefore, there will be a special AF trajectory also on the surface $\Omega$
along which the two gauge couplings are proportional to each other, representing an
UV attractive trajectory.
This special RG trajectory can be characterized more explicitly by introducing ``rotated'' gauge couplings
\be
\begin{split}
	G_\mathrm{s}^2&=\cos\theta\, \gs^2+\sin\theta\, g^2,\\
	G^2&=-\sin\theta\, \gs^2+\cos\theta\, g^2.
\end{split}
\ee
Choosing $\tan\theta=\hat{g}^2_*$, i.e., corresponding to the proportionality factor of the gauge couplings, it suffices to study the flow of $h^2$ for $G^2=0$.
Here, the $\beta$ functions for $h^2$ and $G_\mathrm{s}^2$ are
\begin{align}
	\de_t h^2&=h^2\left[a_h h^2-\cos\theta\, (a_g\hat{g}^2_*+\as)G^2_\mathrm{s}\right],\\
	\de_t G^2_\mathrm{s}&= - \hatetaG\cos\theta\, G^4_\mathrm{s},
\end{align}
exhibiting the phase diagram and RG flow as plotted in \Figref{fig:streamplot_g2gs2_plane_and_rotatedplane} (right panel).
Here the highlighted red line represents the QFP trajectory along which the top-Yukawa coupling
is proportional to $G_\mathrm{s}^2$, and the corresponding QFP value satisfies
\begin{align}
  \left(\frac{h^2}{G^2_\mathrm{s}}\right)_*=\left(\frac{h^2}{\gs^2}\right)_* \cos\theta.
  \label{eq:A10}
\end{align}
We conclude that the special trajectory in \Figref{fig:streamplot_g2gs2_plane_and_rotatedplane} along which all three perturbatively renormalizable couplings are proportional to each other corresponds exactly
to the QFP solution characterized in \Eqref{eq:QFP_g2/gs2_h2/gs2_generalscheme}.

\section{RG flow equations in mass-dependent IR schemes}
\label{app:FRGEs}

This appendix complements \Secref{sec:FRG} and
presents the FRG equations for the Yukawa coupling,
the fields anomalous dimensions, and the
beta functions of the gauge couplings.

\subsection{RG flow equations for the matter content}

Within the truncation of \Eqref{eq:Gamma_k}, different choices are possible for projecting on the scalar anomalous dimension, the anomalous dimension for the left-handed fermions, as well as for the Yukawa coupling~\cite{Gies:2013pma}.
For the present purpose, we compute $\de_t h^2$ and $\eta_\phi$ by projecting the flow onto the radial scalar operator $H$ in the SSB regime which represents the physical Higgs excitation.
Moreover, we concentrate on the fermionic wave function renormalizations associated with the massive top quark.
Accordingly, the flow equation for the Yukawa coupling in the Landau gauge reads~\cite{Gies:2018vwk}
\begin{widetext}
	\begin{align}
	\de_t h^2 &= (d-4+\eta_\phi+2\eta_\psi) h^2+ 4 v_d h^4\left\{	l_{11}^{(\mathrm{FH})d}(\omegaF,\omegaBr;\eta_\psi,\eta_\phi)-
	l_{11}^{(\mathrm{F}\theta)d}(\omegaF,\omegaBl;\eta_\psi,\eta_\phi)\right\}\nonumber\\
	&\quad 	- 8v_d\frac{\Nc^2-1}{2\Nc} (d-1)h^2\gs^2
	l_{11}^{(\mathrm{FG})d}(\omegaF,0;\eta_\psi,\etaG) \Big |_{\rho = \kappa},	
	\label{eq:betahsquared}
	\end{align}
\end{widetext}
where the spinor anomalous dimension $\eta_\psi$ is defined as the average of the anomalous dimension for the left and right-handed Weyl spinors,
\begin{align}
2\eta_\psi=\etaL+\etaR.
\end{align} 
There are no 1PI contributions from the weak gauge boson fluctuations to the flow of $h^2$ which is a special feature of the Landau gauge~\cite{Gies:2013pma,Sondenheimer:2016vko}. 
Thus, the weak gauge bosons contribute only via one-particle reducible graphs which are stored in the anomalous dimensions modifying the canonical scaling. 
The same conclusion also holds for unitary gauge in the SSB regime due to the decoupling of the involved Goldstone modes.
For the flow of the Yukawa coupling in the SSB regime, we use the projection prescription of \cite{Pawlowski:2014zaa,Gies:2017zwf}, which features a better convergence upon the inclusion of higher Yukawa operators, implying that \Eqref{eq:betahsquared} slightly differs from the flows used in \cite{Gies:2013pma,Gies:2013fua,Gies:2014xha}.

Finally, the scalar and spinor anomalous dimensions in the Landau gauge are
\begin{widetext}
	\begin{align}
	\eta_\phi &= \frac{8 v_d}{d} \left\{\rho(3 u''+2\rho u''')^2 m_2^{(\mathrm H)d}(\omegaBr,\eta_\phi)+3\rho(u'')^2 m_2^{(\theta)d}(\omegaBl,\eta_\phi)+ 2 \Nc h^2 \left[m_4^{(\mathrm F)d}(\omegaF,\eta_\psi) 
	- \rho h^2 m_2^{(\mathrm F) d} (\omegaF,\eta_\psi) \right]\right\}\nonumber\\
	&\quad+\frac{8 v_d(d-1)}{d} \left\{-\frac{3}{2}g^2 l^{(\theta \mathrm W)d}_{11}(\omegaBl,\omegaW,\eta_\phi,\etaW)+3\frac{\omegaW^2}{\rho}\left[ 2\, \widetilde{m}_{2}^{(\mathrm{W})d}(\omegaW,\etaW)
	+m^{(\mathrm W)d}_2(\omegaW,\etaW)\right] \right\}  \bigg{|}_{\rho = \kappa},		\label{eq:etaphi}
	\\
	%
	\etaR &= \frac{4 v_d h^2}{d}\left\{ m_{12}^{(\mathrm{ LH})d}(\omegaF,\omegaBr;\etaL,\etaR,\eta_\phi)
	+m_{12}^{(\mathrm L \theta)d}(\omegaF,\omegaBl;\etaL,\etaR,\eta_\phi)
	+2m_{12}^{(\mathrm L \theta)d}(0,\omegaBl;\etaL,\eta_\phi)\right\}\nonumber\\
	&\quad+\frac{8 v_d}{d} \frac{\Nc^2-1}{2 \Nc} \gs^2 (d-1) \left\{ m_{12}^{(\mathrm{RG})d}(\omegaF,0;\etaL,\etaR,\etaG)
	- \widetilde{m}_{12}^{(\mathrm{RG})d}(\omegaF,0;\etaL,\etaR,\etaG) \right\}\Big{|}_{\rho = \kappa}, \label{eq:etapsiR}
	\\
	%
	\etaL &= \frac{4 v_d }{d}h^2\left\{ m_{12}^{(\mathrm{RH})d}(\omegaF,\omegaBr;\etaL,\etaR,\eta_\phi)
	+m_{12}^{(\mathrm{RH})d}(\omegaF,\omegaBl;\etaL,\etaR,\eta_\phi)\right\}
	+\frac{2 v_d }{d}(d-1)g^2\left\{ m_{12}^{(\mathrm{LW})d}(\omegaF,\omegaW;\etaL,\etaR,\etaW)\right.\nonumber\\
	&\quad\left.
	-\widetilde{m}_{12}^{(\mathrm{LW})d}(\omegaF,\omegaW;\etaL,\etaR,\etaW)
	+2\left[m_{12}^{(\mathrm{LW})d}(0,\omegaW;\etaL,\etaW)
	-\widetilde{m}_{12}^{(\mathrm{LW})d}(0,\omegaW;\etaL,\etaW)\right]\right\}\nonumber\\
	&\quad+\frac{8 v_d}{d} \frac{\Nc^2-1}{2 \Nc} \gs^2 (d-1) \left\{ m_{12}^{(\mathrm{LG})d}(\omegaF,0;\etaL,\etaR,\etaG)
	-\widetilde{m}_{12}^{(\mathrm{LG})d}(\omegaF,0;\etaL,\etaR,\etaG) \right\}\Big{|}_{\rho = \kappa}. \label{eq:etapsiL}
	\end{align}
\end{widetext}
Notice that different labels in the threshold functions 
identify different propagators in the corresponding
one-loop integrals. For more details on this see
\Appref{app:threshold}.
For the rest of this paper we drop the index $d$ from the threshold functions,
as we work in $d=4$.

The universal one-loop $\beta$ functions can be straightforwardly obtained from the FRG results. For this, we obviously need to go into the DER by setting the mass-like arguments in the threshold functions to zero. Furthermore, we can also drop all anomalous dimensions inside the threshold functions as they correspond to higher-loop resummations. For instance, the scalar and spinor anomalous dimensions then take the same form as in \Eqref{eq:eta_phi_1Loop_DER} and \Eqref{eq:eta_psi_1Loop_DER}, for $\NL=2$.
Also the flow equation for the top-Yukawa coupling reduces to \Eqref{eq:beta_h2_1Loop_DER_general_NL_Nc} for $\NL=2$.

In order to rediscover the RG equations for the \HTQCD\ model, we need a gauge choice different from the Landau gauge  for the $\SU(2)_\text{L}$ sector, such that any contribution from the Goldstone modes decouple.
This goal can be achieved in the unitary gauge for the $\SU(2)_\text{L}$ gauge group, corresponding to the $\zeta\to\infty$ limit.
Then, all threshold functions associated to the Goldstone mode excitations, with $\omegaBl$ as argument, disappear.
For more details on the unitary gauge and $\beta$ functions in a generic $R_\zeta$ gauge in the context of the FRG,
we refer the reader to Ref.~\cite{Sondenheimer:2016vko}.

\subsection{RG flow equations for the gauge couplings}\label{subsec:RG_flow_gauge_couplings}

The functional treatment of the RG flow also goes along with generalizations of the $\beta$ functions for the gauge couplings $g^2$ and $\gs^2$.
Equations~\eqref{eq:beta_g2_1loop_DER} and \eqref{eq:beta_gs2_1loop_DER} represent 
these flows in the DER where all the mass parameters are ignored.
We list here instead the same flows including also the possibility that the scalar potential is in the SSB regime.
A nontrivial minimum for $u(\rho)$ gives rise to masses
for all fields which couple to the Higgs expectation value.
The most convenient way to compute these one-loop flows proceeds in the
background-field version of the gauge fixings used above, see~\cite{Gies:2013pma} for details in the present case.
The running of the weak gauge coupling at one loop is then determined
by the wave function renormalization. For the
$\SU(2)_\mathrm{L}$ gauge group and for one complex scalar as well as $\NfL$ left-handed fermion doublets, $\etaW$ reads
\begin{align}
\etaW &= - \frac{g^2}{48\pi^2}\biggl[44 L_\mathrm{W}(\mu_\mathrm{W}^2)-\dgammaL\sum^{\NfL}_{j=1}L_\mathrm{F}(\mu^2_{\mathrm{t}_j},\mu^2_{\mathrm{b}_j})\nonumber\\
&\quad - L_\phi(\mu^2_\mathrm{H})\biggr].
\end{align}
The threshold functions $L_\mathrm{W,F,\phi}$ guarantee the decoupling across 
mass thresholds in the SSB regime. They are normalized such that $L_\mathrm{W,F,\phi}(0)=1$, implying that we obtain the standard results in the DER.
In the SSB regime, the renormalized and dimensionless mass parameters are proportional to the vev $\kappa$ of the scalar potential,
\begin{align}
\mu_\mathrm{W}^2=\frac{g^2\kappa}{2},\quad
\mu_{\mathrm{t}_j,\mathrm{b}_j}^2=h_{\mathrm{t}_j,\mathrm{b}_j}^2\kappa,\quad
\mu_\mathrm{H}^2 =2 \lambda_2\kappa.
\end{align}
Here, we have also allowed for a bottom-type Yukawa coupling $h^2_{\mathrm{b}_j}$ in addition to the top-type Yukawas $h^2_{\mathrm{t}_j}$ associated to the $j$-th generation in order to model the decoupling of all quark mass thresholds.
Ignoring higher-loop resummations, the threshold functions read
\begin{align}
L_\mathrm{W}(\mu_\mathrm{W}^2)&=\frac{1}{44}\biggl(21+\frac{21}{1+\mu_\mathrm{H}^2}+2\biggr),\\
L_\mathrm{F}(\mu^2_{\mathrm{t}_j},\mu^2_{\mathrm{b}_j})&=\frac{1}{2}\biggl(\frac{1}{1+\mu_{\mathrm{t}_j}^2}+\frac{1}{1+\mu_{\mathrm{b}_j}^2}\biggr),\\
L_\phi(\mu^2_\mathrm{H})&=\frac{1}{2}\biggl(1+\frac{1}{1+\mu_\mathrm{H}^2}\biggr).
\end{align}
For the $\SU(3)_\mathrm{c}$ gauge group, the wave function renormalization $\etaG$
for the gluon field is
\begin{gather}
\etaG= - \frac{\gs^2}{48\pi^2}\biggl[22\Nc - 
\dgammac\sum^{\Nfc}_{j=1}L^\prime_\mathrm{F}(\mu^2_{\mathrm{Q}_j})\biggr],
\end{gather}
where in this case the fermionic threshold function $L^\prime_\mathrm{F}$ takes the form
\begin{align}
L^\prime_\mathrm{F}(\mu^2_{\mathrm{Q}_j})&=\frac{1}{1+\mu^2_{\mathrm{Q}_j}},
\end{align}
where $\mu^2_{\mathrm{Q}_j}=h^2_{\mathrm{Q}_j}\kappa$ is the mass for the $j$-th quark where $j$ has to be understood as a multiindex, labeling the position within the left-handed doublet as well as possible generation copies.

Let us briefly comment on the case of a rescaling power $P>1$ where further simplifications arise: 
as discussed in \Appref{app:P>1}, the gauge boson and
fermion fluctuations decouple from the dynamics for $P>1$, since their masses diverge
in the UV limit.
In this regime, all loop contributions from the massive modes drop out of the gauge coupling flows.
Depending on the nature of the neutrinos, they either decouple as well if they are Dirac neutrinos with a mass term generated via a Yukawa coupling to the vacuum expectation value. Or as Majorana neutrinos, they could essentially behave as nearly massless particles in the DER and thus would not decouple from $\etaW$. Counting the essentially massless neutrinos by $n_\nu$, we obtain
\begin{align}
\etaG= - \frac{11\Nc}{24\pi^2}\gs^2,\quad \etaW= - \frac{g^2}{48\pi^2}\left(23-\dgammaL\frac{n_\nu}{2}-\frac{1}{2}\right).
\end{align}
In this case, the ratio of the two gauge couplings,
defined in \Eqref{eq:g2/gs2_def}, takes the QFP value
\begin{align}
\hat{g}^2_*=\frac{44}{13}.
\label{eq:g2/gs2_QFP_decoupling}
\end{align}
On the other hand, if we treat the neutrinos as Dirac particles, their contribution decouples from $\etaW$ and the latter QFP value changes into
\begin{align}
\hat{g}^2_*=\frac{44}{15}.
\label{eq:g2/gs2_QFP_decoupling_Diracneutrinos}
\end{align}

\section{Threshold functions}
\label{app:threshold}

For the application of the functional-RG 
equation \eqref{eq:WetterichEQ},
we choose a regulator $R_k$ which is diagonal in field space.
We keep the freedom to have different regulators, specified by corresponding sub- or superscripts, for the Higgs scalar
(H) and for the Goldstone bosons ($\theta$), 
as this is possible in the SSB regime.
Notice however that we do not distinguish between different runnings of the wave function renormalization for the radial excitation $Z_{\mathrm{H}}$ and the Goldstone modes $Z_{\theta}$ in the SSB regime at the present level of our truncation. Thus we use $Z_{\phi}$ as a collective wave function renormalization. Its flow is stored in the anomalous dimension $\eta_{\phi}$ which coincides with the projection rule for the radial excitation.

Similarly, due to the choice of covariant gauges,
we can have different regulators 
for the transverse gluons (GT) or $W$ bosons (WT) and 
for the longitudinal gluons (GL) or $W$ bosons (WL).
In the Landau gauge used in this work, only
transverse gauge bosons propagate, hence we can
avoid the further specifications L and T,
and simply write G or W.
Finally, we account for
independent regularizations of the 
left-handed (L) and right-handed (R) spinors.
One of the left-handed Weyl spinors together with its right-handed partner becomes massive in the SSB regime.
The contributions of the corresponding Dirac field
are denoted with F.

The regularized kinetic (or squared kinetic) terms are
\begin{subequations}
\begin{eqnarray} 
	P_{\mathrm{H}} (x)\!&=&\! 
	x(1+r_{\mathrm{H}}(x))\\
	P_{\mathrm L}(x)\!&=&\! x(1 + r_{\mathrm L}(x))^2\\
	P_{\mathrm F}(x)\!&=&\! x(1 + r_{\mathrm L}(x))(1 + r_{\mathrm R}(x)) \,.
\end{eqnarray}
\end{subequations}
The definition of analogous terms for $\theta$, $\mathrm{G}$, and $\mathrm{W}$ 
are identical to the one for H.
Their RG time derivative is defined through the operator
\begin{equation}
\tilde{\partial}_t=\sum_{\Phi\in\{\mathrm{H},\theta,\mathrm{L,R,W,G}\}}
Z_\Phi^{-1} \partial_t\!\left(Z_\Phi r_\Phi\right)
\cdot \frac{\delta\phantom{r}}{\delta r_\Phi}\,.
\end{equation}
Recall however that we approximate 
$Z_\mathrm{H}=Z_\theta=Z_\phi$.

The loop momentum integrals appearing on the r.h.s.~of
\Eqref{eq:WetterichEQ} are classified by defining the
corresponding threshold functions. 
Most of the threshold functions used in this work can be 
found in App.~A of Ref.~\cite{Gies:2013pma}.
However, the present abbreviations differ from the ones
adopted there.
In the latter reference, any scalar contribution carries the label B,
and the letter G was used for the gauge bosons, corresponding to the $W$ bosons in our work.
This applies to the different versions of $l_0^d$ 
appearing in \Eqref{eq:betau};
to those of $l_{11}^d$ in \Eqref{eq:betahsquared} and \Eqref{eq:etaphi};
to the various forms of $m_4^d$ and $m_2^d$ in
\Eqref{eq:etaphi};
and finally to the $m_{1,2}^d$ in \Eqref{eq:etapsiR} and \Eqref{eq:etapsiL}.

Two kinds of threshold functions require a more
detailed discussion.
One threshold function is called $a_{3}^d$ 
in Ref.~\cite{Gies:2013pma}
and $\widetilde{m}_{1,2}^d$ in the present work, 
as well as in Ref.~\cite{Gies:2017zwf}.
Also, the function called $a_1^d$ in Ref.~\cite{Gies:2013pma}
is correspondingly renamed $\widetilde{m}_{2}^d$ in this work.
For clarity, we provide here the explicit 
definitions of these two kinds of threshold functions
\begin{widetext}
\begin{align}
	\widetilde{m}_{2}^{(\mathrm{W})d}(\omega;\etaW) &= -\frac{k^{6-d}}{16v_d} \int \!\!
	\frac{d^dp}{(2\pi)^d} \frac{1}{p^2}	\,\tilde\partial_t \left( \frac{1}{P_\mathrm{W} + \omega k^2} \right)^2 ,\\
	\widetilde{m}_{1,2}^{(\mathrm{LW})d}
	(\omega_1, \omega_2;\etaL,\etaR,\etaW) &= 
	- \frac{k^{4-d}}{4v_d} \int \!\!
	\frac{d^dp}{(2\pi)^d}\, \tilde\partial_t \left( \frac{1+r_{\mathrm R}}{P_\mathrm F +\omega_1 k^2} \frac{1}{P_\mathrm{W} + \omega_2 k^2} \right) ,\\
	\widetilde{m}_{1,2}^{(\mathrm{LG})d}
	(\omega_1, \omega_2;\etaL,\etaR,\etaG) &= 
	- \frac{k^{4-d}}{4v_d} \int \!\!
	\frac{d^dp}{(2\pi)^d}\, \tilde\partial_t \left( \frac{1+r_{\mathrm R}}{P_\mathrm F +\omega_1 k^2} \frac{1}{P_\mathrm{G} + \omega_2 k^2} \right) ,\\
	\widetilde{m}_{1,2}^{(\mathrm{RG})d}
	(\omega_1, \omega_2;\etaL,\etaR,\etaG) &= 
	- \frac{k^{4-d}}{4v_d} \int \!\!
	\frac{d^dp}{(2\pi)^d}\, \tilde\partial_t \left( \frac{1+r_{\mathrm L}}{P_\mathrm F +\omega_1 k^2} \frac{1}{P_\mathrm{G} + \omega_2 k^2} \right) .
\end{align}
\end{widetext}
Here, the operator $\tilde\partial_t$ denotes
 differentiation with respect to $t=\log k$ 
acting only on the regulators.

As an example, 
the piecewise linear regulator~\cite{Litim:2000ci,Litim:2001up},
\begin{align}\label{eq:cutoff}
r_{\mathrm{W}}(x)&=\left(\frac{1}{x}-1\right)\theta(1-x),\\
r_{\mathrm{L}/\mathrm{R}}(x)&=
\sqrt{1+r_{\mathrm{W}}(x)}-1,
\end{align}
where $x=q^2/k^2$,
yields the following results for these threshold functions
\begin{align}
	\widetilde{m}_{2}^{(\mathrm{W})d}(\omega;\etaW) &= \frac{1- \tfrac{\eta_\mathrm W}{d}}{d-2}\frac{1}{(1 + \omega)^3} \, , \\
	\begin{split}
	\widetilde{m}_{1,2}^{(\mathrm{LW})d}
	(\omega_1, \omega_2;\etaL,\etaR,\etaW)  &= \frac{1}{d-1}\frac{1}{(1+\omega_1)(1+\omega_2)} \\
	\times\bigg[
	2\frac{1-\tfrac{\eta_\mathrm W}{d+1}}{1+\omega_2} 
	+&\frac{\left( 1-\tfrac{\eta_\mathrm L}{d} \right) - \omega_1\left( 1-\tfrac{\eta_\mathrm R}{d} \right)}{1+\omega_1} 
	\bigg]\, .
	\end{split}
\end{align}
For any regulator, we note that  $\omega_1=0$ renders
$\widetilde{m}_{1,2}^{(\mathrm{LW})d}$
independent of $\etaR$ (and similarly 
$\widetilde{m}_{1,2}^{(\mathrm{RW})d}$ becomes independent of $\etaL$).
Correspondingly, we can drop the associated $\eta$ argument, as has been used in \Eqref{eq:etapsiR} and \Eqref{eq:etapsiL}.

\section{Weak gauge-coupling expansion for {\boldmath $P<1/4$}}
\label{App:weak-coupling-expansion-P<1/4}

At the end of \Secref{sec:weak-coupling-P<1/2} we have found indications that the UV behavior of the top-Yukawa coupling might change for $P\leq 1/4$ due to leading terms proportional to $\gs^{8P}$ in the scalar anomalous dimension.
Even though we demonstrated that this does not modify the QFP solution for the rescaled potential $f(x)$, the persistence of the QFP value of the top-Yukawa coupling is an important consistency check of our construction.

We start with QFP solution for $f(x)$, as given in \Eqref{eq:fCf_weakh_Pless1/2} for $P<1/2$, which has a nontrivial minimum $x_0$ and substitute it into the RG flow equation for $h^2$, c.f. \Eqref{eq:betahsquared}.
Within a weak-coupling expansion, we recover the same expression 
as for the DER, c.f. \Eqref{eq:beta_h2_1Loop_DER_general_NL_Nc},
plus an extra term proportional to $\gs^{8P}$,
\begin{align}
\de_t h^2=\de_t h^2\Big{|}_\text{DER}+\frac{9(\ABL+\ABR)\xi_2}{8\pi^2}h^2\gs^{8P}.
\label{eq:det_h2_p<1/4_weak-coupling}
\end{align}
The extra term arises from the scalar loop threshold function $m_2^{(B)d}$ in $\eta_\phi$, c.f.~\Eqref{eq:etaphi},
and contributes only in the SSB regime where $x_0\neq 0$.

For $P=1/4$, the extra term is of the same order as the DER limit for $\de_t h^2$.
Therefore the QFP value for the ratio among the top-Yukawa coupling and the strong-gauge coupling becomes $x_0$-dependent and reads
\begin{align}
\hresc^2_*=\frac{1}{18}\left(4-12(\ABL+\ABR)\xi_2+9\hat{g}_*^2\right),
\end{align}
where $\hat{g}_*^2$ is given by \Eqref{eq:QFP_g2/gs2_h2/gs2_generalscheme}. Moreover, the QFP value stays positive as long as
\begin{align}
\ABL+\ABR<\frac{4+9\hat{g}_*^2}{12\xi_2}.
\end{align}

The situation is different for $P<1/4$ where the last term in \Eqref{eq:det_h2_p<1/4_weak-coupling} becomes leading.
In order to capture the (in)existence of possible different scaling solutions for the top-Yukawa coupling with respect to the strong-gauge coupling, we look for QFP solutions for the ratio
\begin{align}
\hat{h}^2=\frac{h^2}{\gs^{2E}}
\label{eq:h2/gs2E-def}
\end{align}
with $E>0$.
With such a rescaling, the only possible QFP value is
\begin{align}
\hresc_*^2=-\frac{2\xi_2}{3}(\ABL+\ABR),\qquad \text{for}\,\, E=4P.
\end{align}
In view of the condition in \Eqref{eq:weak_coupling_P<1/2_condition-for-positive-minimum-generalmodel}, this solution is, however, negative and thus unphysical. In other words, the presence of a nontrivial minimum for the scalar potential prevents the existence of scaling solutions for the top-Yukawa coupling for all $P<1/4$.
Nevertheless scaling solutions for $f(x)$ do exist also for all $P<1/4$ and do not depend on the asymptotic behavior of the top-Yukawa coupling.

\section{Weak gauge-coupling expansion for {\boldmath $P>1$} }
\label{app:P>1}

It is worthwhile to study the possibility of new AF
trajectories for the case $P>1$ in the general model. Previous studies
found that they exist in the \NAH\ model \cite{Gies:2015lia,Gies:2016kkk},
whereas no valid solutions have been found in the
\HTQCD\ model \cite{Gies:2018vwk}. As the general model interpolates between
the two limiting cases, a search for their scheme-independent
(in)existence is particularly instructive. Our result confirms their
existence in the \NAH\ model as a special limiting case, whereas the general
model does not feature the same mechanism.

For $P>1$, the arguments in the fermionic loop $\omegafF$ and in the gauge boson loop $\omegafW$ defined in \Eqref{eq:omegaf_def}
diverge in the $\{\gs^2,g^2\}\to 0$ limit.
Therefore, in order to capture the correct UV behavior,
we are led to Taylor expand the threshold functions $l_0^{(F)}(\omegafF)$ and $l_0^{(\mathrm{W})}(\omegafW)$ in powers of $z_\mathrm{F,W}^{-1}$.
Let us define the new scheme-dependent coefficients
\be\begin{split}
	\mathcal{B}_\Phi&=\frac{1}{16\pi^2}\left[\de_{(z^{-1})} l_0^{(\Phi)}(z) \right]_{z^{-1}=0},\\
	&=\frac{1}{2k^{6}}\int\!\! \frac{\mathrm{d}^4p}{(2\pi)^4}
	\tilde{\partial}_t P_\Phi(p^2),
	\label{eq:BFW_def}
\end{split}\ee
with $\Phi\in\{\mathrm{F,W}\}$,
such that $\mathcal{B}_\Phi>0$ for general RG schemes providing an IR regularization.
For instance, the piecewise linear regulator yields the positive value $\mathcal{B}_\Phi=1/(32\pi^2)$.
On the other hand, the bosonic thresholds associated to the
radial Higgs fluctuation and the three Goldstone fluctuations are always subleading in the UV for $P>1$.
Moreover, the anomalous dimension $\eta_x$ can contribute to leading order in the $\beta$ function for $f(x)$ for these values of $P$.
For example, it has been observed in the \NAH\ model that $\etaW$ becomes leading for $P>2$ since it is proportional to $g^2$ \cite{Gies:2016kkk}.
The same conclusion holds also for the \HTQCD\ model: the anomalous dimension $\etaG$, being proportional to $\gs^2$, contributes to leading order for $P>2$.
In order to discuss also the possibility to have an anomalous dimension for the rescaled field,
we therefore solve a QFP differential equation where $\eta_x$ is retained as a parameter which has to go to zero in the UV limit.

Let us start by investigating the general $\SU(2)_\text{L}\times\SU(3)_\text{c}$ model.
By keeping the terms linear in $z_\mathrm{F,W}^{-1}$,
the $\beta$ function for the rescaled potential becomes
\begin{align}
  \beta_f=-4f+(2+\eta_x)xf'+\left[\frac{3\BG}{\hat{g}^2_*}-\frac{2\BF}{\hat{h}^2_*}\right]\frac{6\gs^{2P-2}}{x}.
  \label{eq:betaf_P>1_generalmodel}
\end{align}
The presence of a singular term at the origin induces a corresponding pole in the QFP solution which is obtained from integrating the QFP condition $\beta_f=0$ (at fixed $\gs$),
\begin{align}
  f(x)=c\, x^\frac{4}{2+\eta_x} +\left[\frac{3\BG}{\hat{g}^2_*}-\frac{2\BF}{\hat{h}^2_*}\right]\frac{6\gs^{2P-2}}{(6+\eta_x)x},
  \label{eq:QFP_P>1_generalscheme}
\end{align}
where $c$ is the integration constant of the first-order ODE.
Additionally, there is also a log-type singularity in the second derivative at the origin.
In fact by Taylor expanding the scaling term for small $\eta_x$, we get a contribution proportional to $x^2\log x$.
This singularity can be avoided if the potential admits a nontrivial minimum $x_0$ such that $f'(x_0)=0$.
The latter condition can be solved for $c$, yielding a function $c(\gs^2,x_0)$,
and by substituting it into the definition of the rescaled quartic scalar coupling $\xi_2=f''(x_0)$. This provides the following expression
\begin{align}
  \xi_2= \left[\frac{3\BG}{\hat{g}^2_*}-\frac{2\BF}{\hat{h}^2_*}\right]\frac{6\gs^{2P-2}}{x_0^3(2+\eta_x)}.
  \label{eq:xi2_general_scheme_P>1_generalmodel}
\end{align}
We observe that the QFP potential has a nontrivial minimum for positive $\xi_2$ only if
\begin{align}
  \BG>\frac{2\BF\hat{g}^2_*}{3\hat{h}^2_*}.
  \label{eq:B5}
\end{align}
The anomalous dimension $\eta_x$ depends nontrivially on $x_0$,
but the consistency criterion holds that $\eta_x\to 0$ in the UV limit.
From this property we can infer that for any finite value of $\xi_2$ the behavior of the
nontrivial minimum as a function of the strong gauge coupling is
$x_0\sim \gs^{2(P-1)/3}$ in the $\gs^2\to 0$ limit.
By substituting this scaling inside the definitions for $z_\mathrm{W,F}$
we observe that $z_\mathrm{W,F}\to\infty$ in the UV limit, as stated above.

For the \NAH\ model, there are no fermion fluctuations, so the
  right-hand side of \Eqref{eq:B5} vanishes and the criterion is
  satisfied in any scheme. The evidence found in
  \cite{Gies:2015lia,Gies:2016kkk} for the existence of new
  AF trajectories is thus confirmed in a
  scheme-independent manner. By contrast, there are no weak gauge
  contributions in the \HTQCD\ model, implying that the left-hand side of
  \Eqref{eq:B5} is zero in this limiting case. Hence, the criterion
  cannot be satisfied. 
  
  For the $\SU(2)_\text{L}\times\SU(3)_\text{c}$ model,
  a diagonal choice of regulators $\BG=\BF$ inside \Eqref{eq:xi2_general_scheme_P>1_generalmodel}
  would result in $\xi_2<0$ for SM matter content.
  It seems that the \Eqref{eq:B5} still leaves room for legitimate models in the general case. However, this is not the case as detailed in the following.

In writing \Eqref{eq:betaf_P>1_generalmodel}, we have also assumed that
$h^2$ and $g^2$ scale with $\gs^2$. This is true in the DER but
has to be verified  outside this regime.
For $P>1$, the arguments $\omegafF$ and $\omegafW$ diverge in the UV limit,
corresponding to a divergence of the gauge-boson and the top-quark thresholds.
Physically, this means that they decouple from the theory and do not propagate.
In \Secref{subsec:RG_flow_gauge_couplings}, we have seen that the anomalous dimensions $\etaG$ and $\etaW$ within the
decoupled regime are still proportional to $\gs^2$ and $g^2$, respectively.
Therefore, the scaling $g^2\sim\gs^2$ is still valid also outside the DER but the
constant of proportionality depends on the number of decoupled degrees of freedom.
For the SM case, the QFP value for $\hat{g}^2$ is given by \Eqref{eq:g2/gs2_QFP_decoupling}.
By contrast, the $\beta$ function for the top-Yukawa coupling changes drastically
beyond the DER.

Thus the scaling
$h^2\sim\gs^2$ might no longer be valid outside the DER. This leaves a loophole in the argument presented in 
Ref.~\cite{Gies:2018vwk}, where we had assumed that the approximation of the $\beta$\ function
$\de_t h^2$ in the DER holds also for $P>1$. This loophole will be closed by the following analysis:
as an example, let us assume that $\{\omegafW,\omegafF\}\to\infty$ and
$\omegafBr\to 0$.
Since we are looking for solutions with a nontrivial minimum, we can set
$\omegafBl=0$. By expanding \Eqref{eq:betahsquared}
and retaining only the leading terms in $\gs^2$, the $\beta$ function for the rescaled
top-Yukawa coupling, defined in \Eqref{eq:h2/gs2E-def},
reduces to
\begin{align}
  &\de_t \hat{h}^2\simeq 2 v_d \gs^{2E}\hat{h}^4+\hat{h}^2\left\{ - \etaG E-v_d\left[\frac{7}{x_0}\gs^{2P} - 24x_0\xi_2^2\gs^{6P}\right.\right.\nonumber\\
    &\quad \left.\left.-\frac{36}{\hat{g}^2_* x_0^2}\gs^{4P-2}\right]\right\}+\frac{v_d}{3x_0^2}\Bigl[3\gs^{4P-2E}-16x_0\gs^{2P-2E+2}\Bigr].
\end{align}
Let us assume the nontrivial minimum to scale with a power $(P-Q)$ of the strong
gauge coupling,
\begin{align}
  x_0=\gs^{2(P-Q)}\hat{\kappa},\quad\text{with}\quad P\leq Q\quad \text{or}\quad P>Q.
\end{align}
A careful analysis among all the possible combinations between the scaling powers
$P>1$, $E>0$, and $Q$, leads to the conclusion that one or more of the assumptions above are
violated  for any combination which allows
to have a QFP solution for $\hat{h}^2$.

As an example, let us consider the case where $P=E$ and $P=Q$. The rescaled
top-Yukawa coupling has a QFP solution depending on the nontrivial minimum which reads
\begin{align}
  \hat{h}^2_*= - \frac{16\gs^2}{3\hat{\kappa}\etaG P}.
\end{align}
However this solution is not compatible with our assumptions, since $\omegafF$ would stay
finite and does not diverge in the UV limit. All other cases can be analyzed analogously.

We conclude that the general model does not feature new
AF trajectories for $P>1$ similarly to the \HTQCD\ model,
whereas they do exist in the \NAH\ model in any scheme covered by our analysis.


\bibliography{bibliography}

\end{document}